\documentclass[11pt,a4paper,english]{article}
\usepackage{graphicx}

\usepackage{amscd,eucal}
\usepackage{xypic}
\usepackage{mathrsfs}

\usepackage{amssymb,latexsym}
\usepackage{amsmath}
\usepackage{epsf}
\usepackage{pdfsync}
\usepackage{epsfig}
\usepackage[parfill]{parskip}
\usepackage[matrix,arrow,color]{xy}
\usepackage[usenames]{color}
\usepackage{hyperref}

\input{xy}
\xyoption{all}

\input{epsf}

\makeatletter

\catcode`@=11
\renewcommand{\section}
{\@startsection{section}{1}{0pt}{\medskipamount}{\medskipamount}{\large\bf}}
\makeatletter\renewcommand{\subsection}
{\@startsection{subsection}{2}{\z@}{-3.25ex plus -1ex minus -.2ex}
{1.5ex plus .2ex}{\it }}

\numberwithin{equation}{section}
\catcode`@=12

\newcommand{\ba}{\begin{eqnarray*}}
\newcommand{\ea}{\end{eqnarray*}}
\newcommand{\ban}{\begin{eqnarray}}
\newcommand{\ean}{\end{eqnarray}}

\newcommand{\Tr}{{\rm Tr\,}}

\newcommand{\IZ}{\mathbb{Z}}
\newcommand{\IC}{\mathbb{C}}

\newcommand{\IR}{\mathbb{R}}
\newcommand{\IQ}{\mathbb{Q}}
\newcommand{\IF}{\mathbb{F}}

\newcommand{\frM}{\frak{M}}

\newcommand{\Xscr}{{\mathscr{X}}}

\newcommand{\Dscr}{{\mathscr{D}}}

\newcommand{\Ascr}{{\mathscr{A}}}
\newcommand{\Gscr}{{\mathscr{G}}}
\newcommand{\Vscr}{{\mathscr{V}}}
\newcommand{\Wscr}{{\mathscr{W}}}
\newcommand{\Escr}{{\mathscr{E}}}
\newcommand{\Oscr}{{\mathscr{O}}}
\newcommand{\Hscr}{{\mathscr{H}}}

\newcommand{\cW}{{\mathcal W}}
\newcommand{\cN}{{\mathcal N}}
\newcommand{\cM}{{\mathcal M}}
\newcommand{\cS}{{\mathcal S}}

\newcommand{\cE}{{\mathcal E}}
\newcommand{\cO}{{\mathcal O}}

\newcommand{\cQ}{{\mathcal Q}}
\newcommand{\Rcal}{{\mathcal R}}
\newcommand{\cZ}{{\mathcal Z}}
\newcommand{\Lcal}{{\mathcal L}}
\newcommand{\cF}{{\mathcal F}}
\newcommand{\Dcal}{{\mathcal D}}
\newcommand{\cT}{{\mathcal T}}

\newcommand{\Line}{{\mathcal L}}
\newcommand{\hil}{{\mathcal H}}
\newcommand{\cU}{{\mathcal U}}

\newcommand{\frg}{\mathfrak{g}}

\newcommand{\frh}{\mathfrak{h}}
\newcommand{\frgl}{\mathfrak{gl}}
\newcommand{\frsl}{\mathfrak{sl}}

\newcommand{\frsu}{\mathfrak{su}}

\def\Hilb{{\sf Hilb}}
\def\Sym{{\sf Sym}}
\def\ad{{\sf ad}}

\newcommand{\sfA}{{\mathsf{A}}}

\newcommand{\sfQ}{{\mathsf{Q}}}

\newcommand{\sfR}{{\mathsf{R}}}

\newcommand{\sfC}{{\mathsf{C}}}

\newcommand{\sfW}{{\mathsf{W}}}
\newcommand{\sfN}{{\mathsf{N}}}
\newcommand{\sfT}{{\mathsf{T}}}
\newcommand{\Pic}{{\mathsf{Pic}}}
\newcommand{\sfq}{\mathsf{q}}

\newcommand{\rmc}{\mathrm{c}}

\newcommand{\Module}{{\mathscr{M}}}

\newcommand{\DT}{{\tt DT}}
\newcommand{\BPS}{{\tt BPS}}

\def \min{\rm{min}}

\newcommand{\mbf}[1]{{\boldsymbol {#1} }}
\newcommand{\complex}{{\mathbb C}} 
\newcommand{\zed}{{\mathbb Z}} 
\newcommand{\real}{{\mathbb R}} 
\newcommand{\torus}{{\mathbb T}}

\def\e{{\,\rm e}\,}

\newcommand{\alg}{{\mathcal A}}

\newcommand{\Ecal}{{\mathcal E}}
\newcommand{\ch}{{\rm ch}}

\def\ii{{\,{\rm i}\,}}
\def\dd{{\rm d}}

\newcommand{\Hom}{\mathrm{Hom}}

\newcommand{\End}{\mathrm{End}}

\newcommand{\Euler}{{\sf Eu}}
\newcommand{\Ch}{{\sf Ch}}
\newcommand{\Thom}{{\sf Th}}

\def\beq{\begin{equation}}
\def\bee{\begin{equation}}
\def\eeq{\end{equation}}
\def\bea{\begin{eqnarray}}
\def\eea{\end{eqnarray}}
\def\bd{\begin{displaymath}}
\def\ed{\end{displaymath}}

\newcommand{\Cint}{\int\kern-10.5pt-\kern7pt}

\newcommand{\PP}{{\mathbb{P}}}

\newcommand{\be}{\begin{equation}}
\newcommand{\ee}{\end{equation}}

\newcommand\fverbit{\egroup\item[\fbox{\unhbox\pippobox}]}
\newbox\pippobox

{\def\cn{{\mathcal N}}

\def\be{\begin{equation}}
\def\ee{\end{equation}}
\def\bea{\begin{eqnarray}}
\def\eea{\end{eqnarray}}

\newtheorem{theorem}[equation]{Theorem}

\newtheorem{proposition}[equation]{Proposition}
\newtheorem{definition}[equation]{Definition}

\newtheorem{conjecture}[equation]{Conjecture}

\newcommand{\Proof}[1]{\noindent\underline{\textsf{Proof}}: #1 \hfill
  $\blacksquare$\\}

\begin{document}


\setcounter{page}{1}

\begin{flushright}
EMPG--15--09
\end{flushright}

\vskip 1cm

\begin{center}

{\Large\bf $\mbf{\cN=2}$ gauge theories, instanton moduli spaces
  \\[3mm] and geometric representation theory}\footnote{Contribution to the Special Issue ``Instanton Counting: Moduli
  Spaces, Representation Theory and Integrable Systems'' of the {\sl
    Journal of Geometry and Physics}, eds. U.~Bruzzo and F.~Sala.}

\vspace{10mm}

{\large\bf Richard~J.~Szabo}
\\[5mm]
\noindent{\em Department of Mathematics, Heriot--Watt
  University\\
Colin Maclaurin Building, Riccarton, Edinburgh EH14 4AS, UK}\\ and
{\em Maxwell Institute for Mathematical Sciences, Edinburgh, UK}\\ and
{\em The Higgs Centre for Theoretical Physics, Edinburgh, UK}\\
Email: \ {\tt
  R.J.Szabo@hw.ac.uk}

\vspace{10mm}

\begin{abstract}
\noindent
We survey some of the AGT relations between $\cN=2$ gauge theories in four dimensions and geometric representations of symmetry algebras of two-dimensional conformal field theory on the equivariant cohomology of their instanton moduli spaces. We treat the cases of gauge theories on both flat space and ALE spaces in some detail, and with emphasis on the implications arising from embedding them into supersymmetric theories in six dimensions. Along the way we construct new toric noncommutative ALE spaces using the
general theory of complex algebraic deformations of toric varieties, and indicate how to generalise the construction of instanton moduli spaces. We also compute the equivariant partition functions of topologically twisted six-dimensional Yang--Mills theory with maximal
supersymmetry in a general $\Omega$-background, and use the construction to obtain novel reductions to theories in four dimensions.
\end{abstract}

\end{center}


\bigskip


\tableofcontents

\allowdisplaybreaks

\bigskip

\section{Introduction and summary}

This paper is a survey of some recent mathematical results concerning the ever evolving topic of instanton counting in four-dimensional gauge theories with $\cN=2$ supersymmetry. For brevity of the exposition we consider only gauge theories without matter fields, with gauge group $U(N)$, and in the absence of complex codimension one singularities. Our survey is mostly technical and phrased in mathematical language, presenting details for the most part only where firm mathematical statements can be made. In particular, throughout this paper we exploit the geometric interpretation of instantons as moduli of torsion-free sheaves together with their algebraic realisations as modules over a noncommutative deformation of the pertinent space. We pay special attention to ways in which four-dimensional theories can be realised as reductions of certain supersymmetric theories in six dimensions, and discuss some of their implications; other surveys devoted to the interplay between supersymmetric gauge theories in four and six dimensions, with different emphasis, can be found in~\cite{Szabo:2009vw,Szabo:2011mj,art:ciraficiszabo2012}. We consider three such six-dimensional perspectives that we now explain in detail as a summary of the contents of the present contribution.

We begin in \S\ref{sec:braidedsym} with a mathematical introduction to topologically twisted $\cN=2$ gauge theories in four dimensions, which can be defined as a naive dimensional reduction of $\cN=1$ Yang--Mills theory in six dimensions. After explaining how the principle of supersymmetric localization reduces the partition functions to integrals of certain characteristic classes over the instanton moduli spaces, we consider a particular twisted dimensional reduction from six dimensions which defines Nekrasov's \emph{$\Omega$-deformation} of the four-dimensional $\cN=2$ gauge theory~\cite{art:nekrasov2003}. This recasts the partition function as a generating function for certain integrals in equivariant cohomology, to which powerful localization techniques can be applied. Our main examples in this paper concern gauge theories on $\IC^2$, its cyclic orbifolds, and its resolutions, i.e. ALE spaces. We provide a detailed description of the moduli spaces of instantons on ALE spaces, and the computations of their partition functions, elucidating various details. We also report on preliminary steps towards a new reformulation of the instanton partition functions in terms of contributions from modules over a toric noncommutative deformation of ALE spaces, extending the constructions of~\cite{CLSII,CLSIII} which dealt with gauge theories on $\IC^2$; this provides rigorous justification to some heuristic calculations in the literature within the context of noncommutative gauge theory, and moreover produces new instanton moduli spaces which are (commutative) deformations of those of the classical case.

In \S\ref{sec:20theory} we consider the four-dimensional $\cN=2$ gauge theory as a reduction of the elusive $\cN=(2,0)$ theory in six dimensions which is the trigger for the celebrated duality between gauge theories in four dimensions and conformal field theory in two dimensions that was discovered by Alday, Gaiotto and Tachikawa~\cite{art:aldaygaiottotachikawa2010}. We survey some of the results that have been obtained in this direction thus far, focusing again on instances where precise mathematical statements can be made concerning the geometric realisations of the equivariant cohomology of instanton moduli spaces as modules for $\cW$-algebras. We also summarise some of the results that relate these representations to the spectra of quantum integrable systems underlying the $\cN=2$ gauge theories. However, we stress that this article is \emph{not} meant to serve as a review of the AGT correspondence, and it is beyond its scope to survey all aspects and literature on the subject. A review of the AGT correspondence with emphasis on physical aspects can be found in~\cite{Tachikawa:2014dja} where a more exhaustive survey and list of references is presented, together with a pedagogical introduction to instanton moduli spaces and the computations of equivariant partition functions; another recent pedagogical review for physicists with different emphasis and vast literature survey can be found in~\cite{Itoyama:2015xia}.

In \S\ref{sec:N26D} we study the cohomological gauge theory in six dimensions with maximal ($\cN=2$) supersymmetry; in a certain sense, that we discuss, it can be regarded as sitting somewhere between the topologically twisted pure $\cN=2$ gauge theory and the cohomological gauge theory in four dimensions with maximal ($\cN=4$) supersymmetry. Its partition function localises to a weighted sum over generalised instantons which calculates the Donaldson--Thomas theory of toric Calabi--Yau threefolds. Here we compute the Coulomb branch partition function for an arbitrary $\Omega$-deformation of $\IC^3$, extending the calculation of~\cite{Cirafici:2008sn} which considered only the Calabi--Yau specialisation; similar treatments in the framework of equivariant K-theory can be found in~\cite{NekJapan,Awata:2009dd}. We use this general deformation to observe two reductions of the six-dimensional $\cN=2$ gauge theory to a four-dimensional theory; we do not understand fully the implications of these reductions yet, but we feel they are novel and worthy of mention. First, we show that instantons in four dimensions can be embedded as generalised instantons in six dimensions which are invariant under a certain $\IC^\times$-action; this is analogous to the construction of~\cite{Hanany:2003hp} which shows that the moduli space of vortices
in two-dimensional $\cN = (4, 4)$ gauge theories can be embedded as a $U (1)$-invariant subspace of the instanton moduli space. When the gauge theory is viewed as the worldvolume theory of D6-branes (as is required in the definition of the $\Omega$-deformation in this case), this reduction seems to be consistent with previously noted reductions in the literature; however, we do not understand what six-dimensional theory would reduce exactly to the $\cN=2$ gauge theories in four dimensions. Second, in the general $\Omega$-background we can consider the analog of the Nekrasov--Shatashvili limit~\cite{Nekrasov:2009rc} which clearly reduces the gauge theory to some exactly solvable four-dimensional theory whose properties remain to be understood.

We conclude with three appendices at the end of the paper which contain pertinent technical details used in the main text. In~\S\ref{sec:Young} we summarise the relevant combinatorial background on
Young diagrams and partitions which are used throughout this paper; the standard reference is the book~\cite{Andrews} in which further details and concepts may be found. In~\S\ref{sec:symmetric} we survey aspects from the theory of symmetric functions; the standard reference is the book~\cite{book:macdonald1995} which can be consulted for further general details, though we give a fairly detailed exposition of the Uglov functions as they are perhaps the less familiar class of symmetric functions and are particularly relevant to the discussion of \S\ref{sec:20theory}. Finally, in~\S\ref{app:Walgebras} we summarise some standard results about $\cW$-algebras; further particulars can be found in the review~\cite{Bouwknegt:1992wg}.

\bigskip

\section{$\cN=2$ gauge theories in four dimensions\label{sec:braidedsym}}

In this section we study supersymmetric gauge theories in four dimensions using their relation to the cohomology of moduli spaces of torsion-free sheaves. Our main focus is on formulating instanton counting problems. In particular, we provide a fairly detailed account of moduli of instantons on ALE spaces, and the connections between noncommutative instantons and framed sheaves. We elucidate details of some calculations below, while the material in the last two subsections is new.

\subsection{Supersymmetric localization\label{sec:SUSYloc}}

The main player in this paper is the moduli space of finite energy charge $n$ instantons in 
gauge theory on a Riemannian four-manifold $X$ (with Hodge operator $*$). This is
the space of gauge equivalence classes of connections on a principal bundle
on $X$, which are flat
at infinity, of second Chern class $n$
whose curvature $F$ satisfies the anti-self-duality equation
$F=-*F$. The modern interest in these moduli varieties is the important role
they play in supersymmetric quantum gauge theories. Gauge theories with $\cN=2$
supersymmetry can be topologically twisted and, via the principle of
supersymmetric localization, the formal functional integrals defining
their partition functions receive non-trivial contributions from only
supersymmetry-preserving configurations. These states are precisely the
instanton solutions, and hence the partition function and BPS
observables can be rigorously \emph{defined} as finite-dimensional integrals of
certain characteristic classes of natural vector bundles over
the instanton moduli spaces which
are induced by integrating out the bosonic and fermionic field degrees
of freedom. We begin by sketching how some of these formal arguments work; see e.g.~\cite{Cordes:1994fc} for
a review.

Let $G$ be a compact semi-simple Lie group with Lie algebra $\frg$, and let $P\to
X$ be a principal $G$-bundle. Let $\Ascr(P)$ be the space of
connections $A$ on $P$ which are flat at infinity. Since $\Ascr(P)$ is an affine space, the tangent space $T_A\Ascr(P)$ at any point $A\in\Ascr(P)$ can be canonically identified with $\Omega^1(X,\ad\,P)$. The \emph{twisted
$\cN=2$ supersymmetry algebra} is the algebra of odd
derivations of the differential graded algebra
$\Omega^\bullet\big(\Ascr(P)\big)$ of differential forms on
$\Ascr(P)$; it is generated by supercharges $\delta$ with
$\delta^2=0$ and carries a super-Lie algebra structure. The space of bosonic fields $\Vscr(P)$ of twisted $\cN=2$ gauge
theory on $X$ can be identified by regarding it as the
dimensional reduction of six-dimensional $\cN=1$ supersymmetric
Yang--Mills theory on the trivial complex line bundle $X\times\IC$
over $X$ in the limit where the fiber collapses to a point; by identifying the vertical components of connections in this limit
with sections $\phi\in\Omega^0(X,\ad\,P)$, one has
\bea
\Vscr(P) = \Ascr(P)\times \Omega^0(X,\ad\,P) \ .
\nonumber \eea
The superpartners of these fields are sections of the cotangent bundle
$T^*\Vscr(P)$, so that the total field content of the $\cN=2$ gauge
theory on $X$ lies in the infinite-dimensional space
\bea
\Wscr(P) = \Vscr(P) \times \Omega^\bullet\big(\Vscr(P) \big) \ .
\nonumber \eea
One can define a family of noncommutative associative products on these fields, parameterized locally by points $p\in X$, which turns $\Wscr(P)$ into
an algebra that generalizes the structure of a vertex operator
algebra in two-dimensional conformal field theory. The supersymmetry
generators $\delta$ act on $\Wscr(P)$ and so one can define the
cohomology $H^\bullet(\Wscr(P),\delta)$. The product on fields induces
a standard superalgebra structure on $H^\bullet(\Wscr(P),\delta)$,
which gives the \emph{chiral ring} of the $\cN=2$ gauge theory on $X$.

Let $\Gscr(P)$ be the group of automorphisms of $P$ which are trivial at infinity; its Lie algebra can be identified with the space of sections $\Omega^0(X,\ad\, P)$. Let
\bea
\frM_{G,P}(X):=\big\{(A,\phi)\in\Vscr(P)\, \big|\, F^+=0 \ , \ \nabla_A\phi=0 \big\}\,\big/\,
\Gscr(P) 
\nonumber \eea
denote the moduli space of pairs of anti-self-dual connections and covariantly constant sections, where $F^+:=\frac12\,(F+* F)$ denotes the
self-dual part of the curvature two-form $F\in\Omega^2(X,\ad\,P)$ of $A$ and $\nabla_A$ is the covariant derivative associated to $A$. There is an infinite rank vector bundle
$\pi:\Escr\to \Vscr(P)/\Gscr(P)$ given by
\bea
\Escr:=\Vscr(P)\times_{\Gscr(P)}\big(\Omega^{2,+}(X,\ad\,P)\oplus\Omega^1(X,\ad\,P)\big)
\nonumber \eea
where the superscript $+$ denotes the self-dual part of a two-form, on which the Hodge operator $*$ acts as the identity; it is associated to the principal $\Gscr(P)$-bundle $\Vscr(P)\to
\Vscr(P)/\Gscr(P)$. Let $\Thom(\Escr):= \pi_*^{-1}(1) \in
H^\bullet(\Escr)$ be the \emph{Thom class} of $\Escr$, where
$\pi_*:H^\bullet(\Escr)\to H^\bullet\big(\Vscr(P)/\Gscr(P)\big)$ is
the Thom isomorphism. The vector bundle $\Escr$ has a natural section
\bea
s(A,\phi) = \big(F^+,\nabla_A\phi\big)
\nonumber \eea
whose zero locus coincides with the moduli space
$\frM_{G,P}(X)$. Then the pullback of the Thom class
$s^*\big(\Thom(\Escr)\big)$ is a closed form representing the {Euler
class} $\Euler(\Escr)$ of the vector bundle $\Escr$. It obeys the
localization property
\bea
\int_{\Vscr(P)/\Gscr(P)}\, s^*\big(\Thom(\Escr)\big) \wedge \Oscr =
\int_{\frM_{G,P}(X)}\, i^*\Oscr 
\nonumber \eea
for any gauge invariant observable $\Oscr$ on $\Vscr(P)$, i.e. $s^*\big(\Thom(\Escr)\big)$ is the Poincar\'e dual to the embedding
$i:\frM_{G,P}(X)\hookrightarrow \Vscr(P)/\Gscr(P)$.

The partition function of cohomological
Yang--Mills theory can be interpreted geometrically as constructing the Mathai--Quillen representative of
the Thom class $\Thom(\Escr)$, which is accomplished in the twisted
$\cN=2$ gauge theory on $X$ via supersymmetric localization; this is the cohomological gauge theory that gives rise to the Donaldson invariants of four-manifolds~\cite{Witten:1988ze}.
Using standard localization arguments applied to the $\Gscr(P)$-equivariant cohomology of the space of fields $\Wscr(P)$, the weighted integration over the domain $\Wscr(P)$ localizes to an integral over the $\delta$-fixed points, i.e. to BPS observables representing cohomology classes in the chiral ring $H^\bullet(\Wscr(P),\delta)$, which under favourable conditions consists of irreducible connections with $\phi=0$. By taking a weighted sum over all topological classes of principal $G$-bundles $P\to X$, the partition function of the $\cN=2$ gauge theory on $X$ is then given \emph{exactly} by
\bea
\cZ_X(\sfq) = \sum_{n=0}^\infty \, \sfq^n \ \int_{\frM_{G,n}(X)}\, \cZ(P\to X) \ ,
\nonumber \eea
where $\sfq=\e^{2\pi\ii\tau}$ with $\tau$ the complexified gauge
coupling constant, $\frM_{G,n}(X)$ is the moduli space of anti-self-dual $G$-connections on a principal $G$-bundle $P\to M$ with $c_2(P)=n$, and the characteristic class $\cZ(P\to X)$ is a topological invariant depending only on the smooth structure of $X$. The instanton moduli space $\frM_{G,n}(X)$ has real dimension $4\,h\,n$, where $h$ is the dual Coxeter number of $G$. However, even though in some instances the classes $\cZ(P\to X)$ are given by tractable expressions, it is not always the case that the moduli space integrals can be evaluated explicitly.

In this paper we are interested in cases where $X$ is a complex surface and the structure group $G$ is the unitary group $U(N)$. By the seminal work of Donaldson~\cite{Donaldson84},
which translated the problem into the setting of algebraic geometry,
$U(N)$ instantons correspond to rank $N$ holomorphic vector bundles on
$X$. A partial compactification which serves as a 
resolution of singularities of this moduli space is provided by the
moduli space $\frM_{N,n}(X)$ of rank $N$ torsion-free sheaves
$\cE$ on a compactification $\overline{X}=X\cup D_\infty$ with
$c_2(\cE)=n$ and a
framing isomorphism (trivialization) $\phi_\cE:\cE|_{D_\infty}\to \cO_{D_\infty}^{\oplus r}$ over the compactification divisor $D_\infty$. The moduli integrals are still somewhat tricky to handle, and
need not even be well-defined due to the non-compactness of
$\frM_{N,n}(X)$. This problem was resolved in the seminal work of
Nekrasov~\cite{art:nekrasov2003} who showed that with a further
deformation of the gauge theory, called \emph{$\Omega$-deformation},
one can make mathematical sense of these moduli space integrals by regarding them
as integrations in equivariant cohomology. Nekrasov's theory applies
to the instances where $X$ is a \emph{toric} surface, i.e. $X$ carries an action of the algebraic torus $(\IC^\times)^2$ which is associated to a fan in $\IR^2$; in this case there
are extra supercharges and, in addition to the gauge group, the
equivariance with respect to the lift of the torus action on $X$ to
the space of fields is a powerful tool for analysing the $\cN=2$ gauge theory.

\subsection{$\cN=2$ gauge theory on $\IC^2$\label{sec:N2C2}}

We begin with the case where $X$ is the complex affine space $\IC^2$. Then
$\overline{X}$ is the projective plane $\PP^2=\IC^2\cup\ell_\infty$ with
$D_\infty=\ell_\infty\cong\PP^1$ a projective line, and the fine
moduli space $\frM_{N,n}(\IC^2)$ is a smooth quasi-projective complex
algebraic variety
of dimension $2\,N\,n$. In the rank one case $N=1$, $\frM_{1,n}(\IC^2)
\cong
\Hilb^n(\IC^2)$ is the Hilbert scheme of $n$ points in the plane
$\IC^2$ which is the moduli space parameterizing ideals in the polynomial ring
$\IC[z_1,z_2]=\cO_{\IC^2}$ of codimension $n$. The larger family of
moduli varieties $\frM_{N,n}(\IC^2)$ of torsion-free $\IC[z_1,z_2]$-modules of
rank $N>1$ are in this sense higher rank generalizations of Hilbert
schemes.

The moduli space $\frM_{N,n}(\IC^2)$ arises in other settings. For example,
it is the moduli space of $n$ D0-branes bound to $N$ D4-branes in
Type~IIA string theory~\cite{Douglas:1995bn}. It is also the moduli
space of $U(N)$ noncommutative instantons on a Moyal deformation of
Euclidean space $\IR^4$~\cite{NS}; see e.g.~\cite{Hamanaka} for an application of the connection between these two moduli spaces. In this paper we shall
frequently exploit these convenient realisations of instanton moduli as
torsion-free modules over a noncommutative deformation of the algebra $\IC[z_1,z_2]$~\cite{Kapustin:2000ek,CLSII,CLSIII}.

The maximal torus $\torus^2:= \IC^\times\times\IC^\times$ of the group $GL(2,\IC)$ acts by
complex rotations of $\IC^2$, while the group of constant gauge
transformations $GL(N,\IC)$ acts on $\frM_{N,n}(\IC^2)$ by rotating
the framing at infinity $\phi_\Ecal\mapsto
g\circ\phi_\Ecal$. Framed instantons are invariant only under
local 
rotations, so we consider the $\Omega$-deformed gauge theory in which 
rotational invariance holds only up to gauge equivalence. This implies
that the parameters of the framing rotations are given by the
expectation values of the scalar field $\phi$ in the $\cN=2$ vector
multiplet (``Higgs vevs''),
which can always be rotated into the Cartan
subalgebra of $\frgl_N$. Correspondingly, the torus
$\widetilde{\torus}=\torus^2\times(\IC^\times)^N$ acts on the moduli
space $\frM_{N,n}(\IC^2)$.

The $\Omega$-deformed $\cN=2$ gauge theory on $\IC^2$ can be defined as the
twisted dimensional reduction of six-dimensional $\cN=1$ gauge theory
on the total space of a flat $\IC^2$-bundle $M_{\tau_0} \to T^2$
in the limit where the real two-torus $T^2$, of complex structure
modulus $\tau_0$, collapses to a point; the total space of the
affine bundle $M_{\tau_0}$ is defined as the quotient of $\IC^2\times \IC$ by the
$\IZ^2$-action
$$
(n_1,n_2)\triangleright (z_1,z_2,w)=\big(t_1^{n_1}\, z_1,t_2^{n_2}\,
z_2,w+(n_1+\tau_0 \, n_2)\big)
$$
where $(n_1,n_2)\in\IZ^2$, $(z_1,z_2)\in\IC^2$, $w\in\IC$, and $(t_1,t_2)\in
\torus^2$. We
further introduce a flat $\torus^2$-bundle
over $T^2$; in the collapsing limit, fields of the gauge theory which are charged under the
$SL(2,\IC)$ R-symmetry group are sections of the pullback of this
bundle over $M_{\tau_0}$. Let
$(\epsilon_1,\epsilon_2)$ denote its first Chern class, so that
$(t_1,t_2) = (\e^{\epsilon_1},\e^{\epsilon_2})$ parameterize the holonomy of a flat connection on the
$\torus^2$-bundle, and let $\vec
a=(a_1,\dots,a_N)$ denote the Higgs vevs of the complex scalar field $\phi$
that is induced from the horizontal components of the six-dimensional gauge connection
on $M_{\tau_0}\to T^2$.

The definition of the $\Omega$-background can be formalised into the
setting of toric geometry, which we shall need in the following in
order to generalize it to other spaces. Let $L$ be a lattice of rank two, and let $L^*=\Hom_\zed(L,\zed)$ be
the dual lattice; the dual pairing between
lattices is denoted $\langle-,-\rangle: L^*\otimes_\zed
L\to\zed$. Let
$\torus^2=L\otimes_\zed\complex^\times$ be the associated complex torus of
dimension~two over $\complex$. Upon fixing a $\zed$-basis $\{e_1,e_2\}$ for
$L$, with corresponding dual basis $\{e_1^*,e_2^*\}$ for $L^*$,
one has $L\cong\zed\oplus \zed$, $L^*\cong \zed\oplus \zed$, and
$\torus^2 \cong\complex^\times\times \complex^\times$. The torus $\torus^2$ acts on $\complex^2$ in the
standard way by scaling $t\triangleright(z_1,z_2)=(t_1\, z_1,t_2\, z_2)$ for
$t=e_1\otimes t_1 +e_2\otimes t_2\in \torus^2$. Then the pair
$(\epsilon_1,\epsilon_2)$ are the equivariant parameters which
generate the coefficient ring $H_{\torus^2}^2({\rm pt},\zed)$ of
$\torus^2$-equivariant cohomology.

The instanton part of Nekrasov's partition function for pure $\cN=2$
$U(N)$ gauge theory on $\IC^2$ is the generating function for equivariant volumes of the moduli spaces $\frM_{N,n}(\IC^2)$ defined by 
\bea
\cZ_{\IC^2}^{\rm inst}(\epsilon_1,\epsilon_2, \vec a\,;\sfq) =
\sum_{n=0}^\infty\, \sfq^n \ \int_{\frM_{N,n}(\IC^2)} \, 1 \ .
\label{eq:NekinstC2} \eea
Here the integrals are understood as 
pushforward maps to a point in $\widetilde{\torus}$-equivariant
cohomology $H_{\widetilde{\torus}}^\bullet\big(\frM_{N,n}(\IC^2)\big)$; they can
be computed using the localization theorem in equivariant
cohomology which evaluates them as a sum over $\widetilde{\torus}$-fixed points of the moduli spaces, and gives (\ref{eq:NekinstC2}) as a combinatorial
expression in $\sfq$ and the equivariant parameters
$(\epsilon_1,\epsilon_2,\vec a\,)$.

The torus fixed points $\frM_{N,n}(\IC^2)^{\widetilde{\torus}}$ are isolated
and parameterized by $N$-vectors of Young diagrams $\vec
Y=(Y_1,\dots,Y_N)$ of total weight $|\vec Y|=\sum_l\, |Y_l|=n$ (see \S\ref{app:2DYoung}). 
The localization theorem in equivariant cohomology yields the combinatorial expansion
\bea
\int_{\frM_{N,n}(\IC^2)} \, 1 = \sum_{ |\vec
Y|=n}\
\frac{1}{\Euler\big(T_{\vec Y}\frM_{N,n}(\IC^2) \big)} \ ,
\nonumber \eea
and by explicitly computing the equivariant Euler class $\Euler\big(T_{\vec
  Y}\frM_{N,n}(\IC^2) \big) \in H^\bullet_{\widetilde{\torus}}\big(\frM_{N,n}(\IC^2) \big)$ of the tangent space at the fixed point $\vec Y$~\cite{Flume:2002az,Bruzzo:2002xf,NakYosh} one finds that the instanton partition function on $\IC^2$ is then given by
\bea
\cZ_{\IC^2}^{\rm inst}(\epsilon_1,\epsilon_2, \vec a\,;\sfq)= \sum_{\vec Y}\,
\sfq^{|\vec Y|} \
\prod_{l,l'=1}^N \, \frac{1}{\sfT^{\vec Y}_{l,l'}(\epsilon_1,\epsilon_2,
  a_l-a_{l'})} \ ,
\label{eq:NekinstC2comb} \eea
where
\bea
\sfT^{\vec Y}_{l,l'}(\epsilon_1,\epsilon_2,
  a) &=&\prod_{s\in Y_l}\, \Big(a -L_{Y_{l'}}(s)\, \epsilon_1+ \big(A_{Y_l}(s)+1\big)\, \epsilon_2\Big) \nonumber \\ && 
\times \ \prod_{s'\in Y_{l'}}\, \Big(a+\big(L_{Y_l}(s'\, )+1\big)\, \epsilon_1-A_{Y_{l'}}(s'\, )\, \epsilon_2 \Big) \ .
\nonumber \eea
These considerations can be extended to more general $\cN=2$ quiver gauge
theories on $\IC^2$, wherein the contributions of matter fields are
represented by Euler classes of universal vector bundles on
$\frM_{N,n}(\IC^2)$. They are also applicable to the more general $\cN=2$
gauge theories in four dimensions of class
$\cS$~\cite{Gaiotto:2009we,Gaiotto:2009hg}, which we discuss in \S\ref{sec:20theory}.

In the rank one case $N=1$, the combinatorial series \eqref{eq:NekinstC2comb} can be explicitly summed to give
the pure $\cn=2$ gauge theory partition function as a simple exponential~\cite[\S4]{NakYosh}
\bea
\cZ^{\rm inst}_{\IC^2}(\epsilon_1,\epsilon_2; \sfq) = \exp\Big(\,
\frac \sfq{\epsilon_1\, \epsilon_2}\, \Big) \ .
\label{eq:corN2C2U1} \eea
This formula can be derived from the Cauchy--Stanley formula for Jack symmetric functions, see \S\ref{sec:uglovfunctions}. See~\cite[\S4]{Pedrini:2014yoa} for a direct representation theoretic proof of this expression.

For later comparisons with the partition functions of the six-dimensional $(2,0)$
theory that we discuss in \S\ref{sec:20theory} and of the six-dimensional
$\cN=2$ gauge theory that we consider in \S\ref{sec:N26D},
we should further multiply the instanton partition function
(\ref{eq:NekinstC2comb}) by the classical contribution
\bea
\cZ_{\IC^2}^{\rm cl}(\epsilon_1,\epsilon_2,\vec a\, ;\sfq)=\prod_{l=1}^N\, \sfq^{-\frac{a_l^2}{2\,\epsilon_1\,\epsilon_2}} \ ,
\label{eq:4Dclass} \eea
and also by the purely perturbative contribution which is $\sfq$-independent and given by
\bea
\cZ_{\IC^2}^{\rm pert}(\epsilon_1,\epsilon_2,\vec a\,) = \exp \Big( -
\sum_{l,l'=1}^N \, \gamma_{\epsilon_1 , \epsilon_2 } (a_{l'} - a_l ; \Lambda)  \Big) \ .
\label{eq:4Dpert} \eea
Here we defined the Barnes double zeta-function~\cite[\S E.2]{nakajima2}
\bea
\gamma_{\epsilon_1 , \epsilon_2} (x ; \Lambda) :=
\lim_{s\to0}\, \frac{\dd}{\dd s}\, \frac{\Lambda^s}{\Gamma(s)} \,
\int_0^{\infty} \, \dd t \ t^{s-1} \, \frac{\e^{ - t \, x }}{\big(1-\e^{t\, \epsilon_1}\big)\,
  \big(1-\e^{t\, \epsilon_2}\big)} \ ,
\nonumber \eea
which is a regularization of the formal expression
\begin{equation}
  \gamma_{\epsilon_1 , \epsilon_2 } (x  ; \Lambda) = \sum_{m,n
    \ge 0} \, \log \Big( \, \frac{x - m \, \epsilon_1 - n \,
    \epsilon_2}{\Lambda} \, \Big) \ .
\label{eq:gammaformal} \end{equation}

\subsection{Toric geometry of ALE spaces}\label{se:cdal}

A natural class of complex surfaces $X$ on which these
considerations may be extended consists of orbifolds of $\IC^2$ and their
resolutions. We begin by describing ALE spaces of type $A_{k-1}$ regarded as toric
varieties, following~\cite{GL1,CA-KS,Bruzzo:2013daa}; in this paper all cones are understood to be strictly convex rational polyhedral cones in a real vector space.
For any non-negative integer $i$, define the lattice vector in $L$ by
\bea
v_i =i\, e_1- ( i-1) \, e_2 \ .
\nonumber \eea
Given an integer
$k\geq2$, let the cyclic group $\zed_k$ of order $k$ act on
$\complex^2$ by
\bea
\zeta\triangleright(z_1,z_2)=
(\omega \, z_1,\omega^{-1}\, z_2) \ ,
\label{eq:ZkactionC2}\eea
where $\zeta$ is the
generator of $\zed_k$ with $\zeta^k=1$ and $\omega=\e^{2\pi\ii/k}$ is
a primitive $k$-th root of unity. The $\zed_k$-action commutes
with the $\torus^2$-action on $\complex^2$, so $\complex^2/\zed_k$ is an
affine toric variety defined by the fan consisting of the single
two-dimensional simplicial cone $\IR_{\geq0}v_0+\IR_{\geq0}v_k$ spanned by the lattice
vectors $v_0$ and~$v_k$; it has a unique singular $\torus^2$-fixed point at the
origin of order $k$ and can be depicted schematically as
$$
\xymatrix{
& {} & \\
\IC^2/\IZ_k \ = \ 
 & {} \ar[dr]_{v_k} \ar[u]^{v_0} & \\
& {} & 
}
$$
The toric variety $\IC^2/\IZ_k= {\sf
  Spec}\big(\IC[z_1,z_2]^{\IZ_k} \big)$ corresponding to this fan is dual to the coordinate
subalgebra of invariant polynomials $\IC[z_1,z_2]^{\IZ_k}= \IC[z_1^k,z_2^k, z_1\,z_2]$; hence the
singular orbit space
$\IC^2/\IZ_k$ can also be regarded as the subvariety of $\IC^3$ cut
out by the equation
\bea
x\, y -z^k=0 \ .
\label{eq:C2ZkC3}\eea

Let $X_k$ be the toric resolution of $\complex^2/\zed_k$,
defined by a simplicial fan $\Sigma_k$ in the real vector space
$L_\real=L\otimes_\zed\real$. Let
$\Sigma_k(n)$, $n=0,1,2$, be the set of $n$-dimensional cones in
$\Sigma_k$, so that
\bea
\Sigma_k(0)&=& \big\{ \{0\}\big\} \ , \nonumber \\[4pt]
\Sigma_k(1) &=& \big\{\real_{\geq0}v_i\big\}_{i=0,1,\dots, k} \ ,
\nonumber \\[4pt]
\Sigma_k(2) &=& \big\{\real_{\geq0}v_{i-1} +
\real_{\geq0}v_i\big\}_{i=1,\dots, k} \ .
\nonumber \eea
This fan is a subdivision of the fan describing the orbit space
$\IC^2/\IZ_k$ which can be depicted schematically as
$$
\xymatrix{
 & {}  & \\
X_k \ = \ &
 {} \ar[dr]_{v_k} \ar[u]^{v_0} \ar[r]^{v_i} & \\
&  {} & 
}
$$
An \emph{ALE space} is a smooth Riemannian four-manifold which is
diffeomorphic to $X_k$ and carries a K\"ahler metric that is
asymptotically locally Euclidean (ALE), i.e. that approximates the
standard flat metric on the orbit space $\IC^2/\IZ_k$ ``at infinity''. It
can be realised by adding polynomials in $\IC[x,y,z]$ to
\eqref{eq:C2ZkC3} of degree $<k$.

A natural normal toric compactification of $\complex^2/\zed_k$ is the global quotient
$\PP^2/\zed_k$, where $\zed_k$ acts on the projective plane $\PP^2$ by
\bea
\zeta\triangleright[w_0,w_1,w_2]=
[w_0,\omega \, w_1,\omega^{-1} \, w_2] \ ,
\label{eq:zetaP2action}\eea
and the torus
action is given by
$t\triangleright[w_0,w_1,w_2]=
[w_0,t_1\, w_1,t_2\, w_2]$; it is a projective toric surface with finite quotient singularities. Let
$\ell_\infty=\big\{[0,w_1,w_2]\in\PP^2\big\}\cong\PP^1$ be a smooth
$\torus^2$-invariant divisor in
the projective plane with self-intersection number
$\ell_\infty\cdot\ell_\infty=1$. There is a disjoint union
\bea
\PP^2 /\zed_k = \big(\complex^2/\zed_k\big) \cup
\big(\ell_\infty/\zed_k\big) \ .
\nonumber\eea
Note that for $k=2$, the $\zed_2$-action on $\ell_\infty$ is trivial as $[-w_1,-w_2]=[w_1,w_2]$; in general the two fixed points of $\ell_\infty$ are orbifold points and hence it is a ``football''.
Then the orbit space $\PP^2/\zed_k$ is the coarse moduli space
underlying the global quotient stack $\big[\PP^2/\zed_k\big]$ which is the compact toric orbifold defined by the stacky fan~\cite{BCS} $\mbf\Sigma_k'=(L,\Sigma_k',\beta_k')$, where
$\Sigma_k'\subset L_\real$ is the simplicial fan with
\bea
\Sigma_k'(0)&=& \big\{ \{0\}\big\} \ , \nonumber \\[4pt]
\Sigma_k'(1) &=& \big\{\real_{\geq0}v_0 \,,\, \real_{\geq0}v_k
\,,\,\real_{\geq0}(-v_0- v_k)\big\} \ ,
\nonumber \\[4pt]
\Sigma_k'(2) &=& \big\{\real_{\geq0}v_{0} +
\real_{\geq0}v_k \,,\, \real_{\geq0}v_{0} +
\real_{\geq0}(-v_0-v_k) \,,\, \real_{\geq0}v_{k} +
\real_{\geq0}(-v_0-v_k) \big\} \ ,
\nonumber \eea
and $\beta_k':\zed b_1'\oplus \zed b_2'\oplus \zed b_3'\to L$ is the map
\beq
\beta_k'\, :\, \zed^3 \ \xrightarrow{\displaystyle{\begin{pmatrix} 0& k& -k\\ 1 &k-1 &-k \end{pmatrix}}} \ \zed^2
\nonumber \eeq
determined by the minimal lattice points $b'_1,b'_2,b'_3$ generating
the one-cones $\Sigma_k'(1)$; here and in the following we shall usually
implicitly assume that $k$ is odd for simplicity (see~\cite{Bruzzo:2013daa} for the general case), although many of our
conclusions hold more generally. Note that for $k=2$, the divisor $[\ell_\infty/\zed_2]\cong \PP^1\times B\zed_2$ corresponding to the vector $b_3'$ is a trivial $\zed_2$-gerbe (the quotient stack of $\PP^1$ by the trivial $\zed_2$-action), where $B\zed_2$ is the quotient groupoid $\big[{\sf Spec}(\complex)/\zed_2 \big]$.

Minimal resolution of the singularity at the origin $[1,0,0]$ gives a
stacky toric compactification
\bea
\big[\, \overline{X_k}\, \big] = \big[X_k\cup (\ell_\infty/\zed_k) \big]
\nonumber \eea
of $X_k$, which is defined by a stacky fan $\bar{\mbf\Sigma}_k=(L,\bar\Sigma_k,\bar\beta_k)$, where $\bar\Sigma_k\subset L_\real$ is the simplicial fan with
\bea
\bar \Sigma_k(0) &=& \big\{ \{0\}\big\} \ , \nonumber \\[4pt]
\bar \Sigma_k(1) &=& \big\{\real_{\geq0}v_i\big\}_{i=0,1,\dots, k}
\cup \big\{ \real_{\geq0}(-v_0- v_k)\big\} \ ,
\nonumber \\[4pt]
\bar \Sigma_k(2) &=& \big\{\real_{\geq0}v_{i-1} +
\real_{\geq0}v_i\big\}_{i=1,\dots, k} \nonumber \\ && \qquad \cup \ \big\{\real_{\geq0}v_{0} +
\real_{\geq0}(-v_0-v_k) \,,\, \real_{\geq0}v_{k} +
\real_{\geq0}(-v_0-v_k) \big\} \ ,
\nonumber \eea
and $\bar\beta_k:\zed \bar b_1\oplus\cdots\oplus\zed\bar b_{k+2} \to L$ is the map
\beq
\bar\beta_k\,:\, \zed^{k+2} \ \xrightarrow{\displaystyle{\begin{pmatrix} 0& 1& 2& \dots & k& -k \\ 1& 0& 1& \dots & k-1& -k \end{pmatrix}}} \ \zed^2
\nonumber \eeq
defined by the minimal lattice points $\bar b_1,\dots,\bar b_{k+2}$
generating the one-cones $\bar\Sigma_k(1)$. The coarse moduli space
$\overline{X_k}$ of this toric orbifold is the simplicial toric
variety defined by the fan $\bar\Sigma_k$; in particular for $k=2$, $\overline{X_2}$ is the second Hirzebruch surface $\mathbb{F}_2$.

Let $\Gamma$ be the toric graph with vertex set
$V(\Gamma)=\{{\rm v}_1,\dots,{\rm v}_k\}$ given by vertices of two-cones in $\Sigma_k(2)$, which is in bijective correspondence with the
set of 
smooth torus fixed points $p_i=p_{{\rm v}_i}$ in $X_k$, i.e. $X_k^{\torus^2}=\{p_1,\dots,p_k\}$. The tangent weights of the toric action
on the ${\torus^2}$-fixed point $p_i$ are
\bea
w_1^i= -(i-2)\, \epsilon_1- (i-1)\, \epsilon_2 \qquad \mbox{and} \qquad w_2^i=
(i-1) \, \epsilon_1-i\, \epsilon_2
\nonumber \eea
for $i=1,\dots,k$. Let $U[\sigma_i]$ be the affine toric variety generated by the smooth two-cones $\sigma_i:= \real_{\geq0}v_{i-1}+\real_{\geq0}v_i$ for $i=1,\dots,k$. Then
\bea
X_k=\bigcup_{i=1}^k\, U[\sigma_i] \ .
\nonumber \eea

Let $E(\Gamma)=\{{\rm e}_1,\dots,{\rm e}_{k-1}\}$ be the set of edges in the graph
$\Gamma$, where ${\rm e}_i$ is the edge connecting vertices ${\rm v}_i$ and
${\rm v}_{i+1}$, and let $\ell_i=\ell_{{\rm e}_i}$ be the corresponding
exceptional divisors of the minimal resolution
$X_k\to \complex^2/\zed_k$. Then $\ell_i$ is a ${\torus^2}$-invariant projective line connecting the
${\torus^2}$-fixed points $p_i$ and $p_{i+1}$. For each $i=1,\dots,k-1$ and $j=1,\dots,k$, there is a unique lattice vector $m_i{}^j\in L^*$ such that
\bea
\ell_i\big|_{U[\sigma_j]} = -\sum_{l=1}^{k-1}\, \big\langle m_i{}^j,v_l\big\rangle \ \ell_l\big|_{U[\sigma_j]} \ ;
\nonumber \eea
it is given explicitly by
\bea
m_i{}^j = \left\{ \begin{array}{ll}
(i-2)\, e^*_1+(i-1)\, e^*_2 \ , & j=i \ , \\[4pt]
-i\, e^*_1-(i+1)\, e^*_2 \ , & j=i+1 \ , \\[4pt]
0 \ , & \mbox{otherwise} \ .
\end{array} \right.
\nonumber \eea
Hence the stalk of the
associated line bundle $\cO_{X_k}(\ell_i)$ has weights
at $p_j$ given by
$$
w_{\ell_i}^{p_j}=\left\{ \begin{array}{ll}
-(i-2)\, \epsilon_1-(i-1)\, \epsilon_2 \ , & j=i \ , \\[4pt]
i\, \epsilon_1+(i+1)\, \epsilon_2 \ , & j=i+1 \ , \\[4pt]
0 \ , & \mbox{otherwise} 
\end{array} \right.
$$
for $i=1,\dots,k-1$ and $j=1,\dots,k$. The intersection form $C$ of
the exceptional divisors
coincides with minus the $(k-1)\times
(k-1)$ Cartan matrix
$$
C=\big(\ell_i\cdot \ell_j \big)=\begin{pmatrix}
-2& 1& 0 & \dots & 0 \\
1& -2& 1& \dots & 0 \\
0& 1 &-2 & \dots & 0 \\
\vdots & \vdots & \vdots & \ddots & \vdots \\
0 &0 &0 & \dots & -2
\end{pmatrix}
$$
of the Dynkin diagram for the $A_{k-1}$ Lie algebra. Since $X_k$ is
non-compact, the intersection form is not unimodular and the inverse
$C^{-1}$ has generically rational-valued matrix elements
given by~\cite{GSST}
$$
\big(C^{-1}\big)^{ij}= \frac{i\, j}k -{\min}(i,j)
$$
for $i,j=1,\dots,k-1$.

Any compactly supported divisor $D\in H_c^2(X_k,\zed)$ is a linear combination
\bea
D= \sum_{i=1}^{k-1}\, m_i\, \ell_i \qquad \mbox{with} \quad m_i\in
\zed \ ,
\nonumber \eea
so that the corresponding weights are
\bea
w_D^{p_j}= \sum_{i=1}^{k-1}\, m_i\, w_{\ell_i}^{p_j} = \big((j-1)\,
\epsilon_1+j\, \epsilon_2 \big)\, m_{j-1} - \big((j-2)\, \epsilon_1 +(j-1)\,
\epsilon_2\big)\, m_j
\nonumber \eea
for $j= 1,\dots, k$, where $m_0=m_k:=0$. However, to properly account for flat line bundles
on $X_k$ with non-trivial holonomy at
infinity, we define a dual generating set, with respect to the
intersection pairing linearly extended to non-compact divisors, by~\cite{CA-KS}
$$
\varepsilon^i=\sum_{j=1}^{k-1}\, \big(C^{-1}\big)^{ij}\,\ell_j \ .
$$
Then $\varepsilon^i$, $i=1,\dots,k-1$, extend the pair of non-compact torically
invariant divisors of $X_k$ to an integral generating set for the
Picard group of line bundles
\beq
\Pic(X_k) = H^2(X_k,\zed) \cong \zed^{k} \ ,
\nonumber \eeq
with intersection product $\varepsilon^i\cdot
\varepsilon^j=(C^{-1})^{ij}$; this set corresponds to the basis of tautological line bundles
$\Rcal_1,\dots,\Rcal_{k-1}$ constructed by Kronheimer and Nakajima
in~\cite{KN} with $\int_{X_k}\, c_1(\Rcal_i )\wedge c_1(\Rcal_j)=(C^{-1})^{ij}$.
We can then parameterize the class of a
divisor $D=D_{\vec u}$ as
$$
D_{\vec u} = \sum_{i=1}^{k-1}\, u_i\, \varepsilon^i
$$
with $\vec u=(u_1,\dots,u_{k-1})\in\zed^{k-1}$. The corresponding
(fractional) weights are
$$
w_{D_{\vec u}}^{p_i}=\sum_{j,l=1}^{k-1}\,u_j\, \big(C^{-1}\big)^{jl}\,
w_{\ell_l}^{p_i} = -\frac{(k-1) \,\epsilon_1+ k\, \epsilon_2}k\,
\sum_{j=1}^{i-1}\, j\, u_j- \frac{\epsilon_1}k\, \sum_{j=i}^{k-1}\,
(k-j)\, u_j
$$
for $i=1,\dots,k$.

\subsection{Quiver varieties\label{sec:quivers}}

We will now consider the extensions of the moduli spaces $\frM_{N,n}(\IC^2)$ to the more general family of cyclic Nakajima quiver varieties. As in the case of gauge theories on $\IR^4$, $U(N)$ instantons on $X_k$ are asymptotically flat, i.e. they approach
  flat connections at infinity. However, while flat connections at infinity on $\IR^4$ are necessarily trivial, on $X_k$ they are not: At infinity $X_k$ looks topologically like the lens space $S^3/\IZ_k$, so a flat connection can have non-trivial holonomy which is parameterized by a representation $\rho$ of the fundamental group $\pi_1(S^3/\IZ_k)\cong\IZ_k$. On the toric compactification $\overline{X_k}$ they correspond to flat bundles over the compactification
divisor $\ell_\infty/\zed_k$. For this, let $\rho:\zed_k\to GL(N,\complex)$ be a representation of the cyclic group $\zed_k$. Then the flat bundle $E_\infty\to\ell_\infty/\zed_k$ of rank $N$ associated to $\rho$ is
\bea
E_\infty:= \PP^1\times_{\zed_k} \complex^N \ \longrightarrow \ \PP^1/\zed_k \ ,
\nonumber \eea
where the generator $\zeta$ of $\zed_k$ acts on $\PP^1\times \complex^N$ as
\bea
\zeta\triangleright \big([w_1,w_2]\,,\, v\big)=\big([\omega \,
w_1,\omega^{-1} \, w_2]\,,\, \rho(\zeta)v\big) \ .
\nonumber \eea
Let $\rho_i:\IZ_k\to U(1)$ denote the irreducible one-dimensional representation of $\IZ_k$  with weight $i$ for $i=0,1, \ldots, k-1$. By the McKay correspondence, the corresponding flat line bundles $E_\infty$ are precisely the restrictions of the tautological line bundles $\Rcal_i$ at infinity, where $\Rcal_0:=\cO_{X_k}$. 

In~\cite{KN}, Kronheimer and Nakajima construct moduli spaces $\frM(\vec v,\vec w\,)$ of $U(N)$ instantons on $X_k$ with the vectors $\vec v,\vec w\in \mathbb{Z}_{\geq0}^k$ parameterizing the Chern classes ${\rm ch}(\cE)$ and the holonomies $\rho=\rho_{\vec w}:= \bigoplus_{i=0}^{k-1}\, w_i\, \rho_i$. For this, we recall the parameterization of the moduli spaces $\frM_{N,n}(\IC^2)$ by linear algebraic ADHM data which neatly captures the instanton quantum mechanics of $n$ D0-branes inside $N$ D4-branes on $\IC^2$: One has $
\frM_{N,n}(\IC^2)=\mu^{-1}(0)^\xi/GL(n,\IC)$, where
\bea
\mu=[b_1,b_2]+i\,j \ .
\label{eq:muADHM} \eea
Here $(b_1,b_2)\in\End_\IC(V)\otimes Q$, $i\in\Hom_\IC(W,V)$ and $j\in\Hom_\IC(V,W)\otimes\bigwedge^2Q$, where $V$ and $W$ are complex vector spaces of dimensions $n$ and $N$, respectively, while $Q\cong\IC^2$ is the fundamental representation of $\torus^2$ with weight $(1,1)$ and $W\cong\IC^N$ the fundamental representation of the torus $(\IC^\times)^N$ with weight $(1,\dots,1)$. The group $GL(n,\IC)$ acts by basis change of $V\cong\IC^n$,
and the superscript $\xi$ indicates that the GIT quotient is
restricted to \emph{stable} matrices $(b_1,b_2,i,j)$, i.e. the image
of $i$ generates $V$ under the action of $b_1,b_2$. This enables one
to interpret $\frM_{N,n}(\IC^2)$ as the moduli variety of stable linear representations of the ADHM quiver
\bea
\xymatrix{
V \ \bullet \ \ar@(ur,ul)|{\, b_1 \,} \ar@(dl,dr)|{\, b_2 \,}
\ar@/^/[rr]|{\, j \,} & & 
\ \circ \ W \ar@/^/[ll]|{\, i \,}}
\label{eq:ADHMquiver} \eea
with relations (\ref{eq:muADHM}), which can be regarded as originating in the following way: The Jordan
quiver
\medskip
\bea
\xymatrix{
{\bullet}\ar@(ur,ul) []}
\nonumber \eea
has as underlying graph the (formal) affine extended Dynkin diagram of
type $\widehat{A}_0$. The corresponding framed quiver (obtained by
adding a node and arrow to each node) is

\medskip
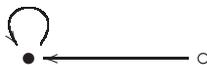
\bea
\xymatrix{
{\bullet}\ar@(ur,ul) [] & & \ar[ll] \circ} 
\nonumber \eea
and its double (obtained by adding an arrow in the opposite direction
to each arrow) is precisely the ADHM quiver \eqref{eq:ADHMquiver}. Alternatively, $\frM_{N,n}(\IC^2)$ can be regarded geometrically as a symplectic reduction of the cotangent bundle over the moduli space of representations of the framed Jordan quiver; passing to the cotangent bundle has the effect of doubling the quiver.

One can now take $V$ and $W$ to be $\IZ_k$-modules; then $\vec v$ and $\vec w$ are dimension vectors whose components give the multiplicities of their decompositions into irreducible $\IZ_k$-modules. By decomposing the linear maps $(b_1,b_2,i,j)$ of the ADHM construction as equivariant maps $(b_1,b_2)\in\End_{\IZ_k}(V)\otimes Q$, $i\in\Hom_{\IZ_k}(W,V)$ and $j\in\Hom_{\IZ_k}(V,W)\otimes\bigwedge^2Q$, one can define a moduli variety $\frM(\vec v,\vec w\,)$ of stable linear representations of the double of the framed quiver corresponding to the cyclic quiver
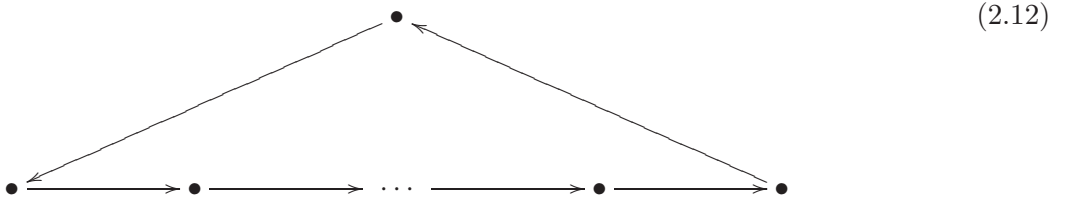
\bea
\xymatrix@C=20mm{
& & {\bullet}\ar[ddll] & & \\  & & & & \\
{\bullet}\ar[r] & {\bullet} \ar[r] & \ \cdots
\ \ar[r] & {\bullet}\ar[r] & {\bullet}\ar[uull]
}
\label{eq:cyclicquiver} \eea
whose underlying graph is the affine extended Dynkin diagram of type
$\widehat{A}_{k-1}$, together with relations provided by equivariant decomposition of (\ref{eq:muADHM}). More generally, one can define the Nakajima
quiver varieties $\frM_\xi(\vec v,\vec w\,)$ of
type~$\widehat{A}_{k-1}$ which depend on a suitable stability
parameter $\xi \in H^2(X_k, \IR)=\IR^k$~\cite{Nakajima1}. They are smooth
quasi-projective varieties of
dimension
\bea
\dim_\IC \frM_\xi(\vec v,\vec w\,) = 2\,\vec w\cdot\vec v-\vec
v\cdot\widehat{C} \vec v \ ,
\nonumber \eea
where 
\begin{equation} \nonumber 
\widehat{C} = \begin{pmatrix}
2& -1& 0 & \dots & -1 \\
-1& 2& -1& \dots & 0 \\
0& -1 &2 & \dots & 0 \\
\vdots & \vdots & \vdots & \ddots & \vdots \\
-1 &0 &0 & \dots & 2
\end{pmatrix}
\end{equation}
is the Cartan matrix of the extended Dynkin diagram of type $\widehat{A}_{k-1}.$
The space of parameters $\xi$ such that stable points coincide with semistable points (in the representation variety) has a subdivision into chambers, whereby the quiver varieties $\frM_\xi(\vec v,\vec w\,)$ with parameters in the same chamber are isomorphic (as complex algebraic varieties) while those with parameters in distinct chambers are only diffeomorphic. In this paper we will focus on two distinguished chambers.

There is a chamber $\sfC_0$ such that, for  $\xi_0\in \sfC_0$, the quiver
variety $\frM_{\xi_0}(\vec v,\vec w\,)$ parameterizes
$\IZ_k$-equivariant torsion-free sheaves $\cE$ on $\PP^2$ with a
$\IZ_k$-invariant framing isomorphism $\phi_\cE:\cE|_{\ell_\infty} \to
\cO_{\ell_\infty}\otimes\rho_{\vec w}$ and
$H^1\big(\PP^2,\cE\otimes\cO_{\PP^2}(\ell_\infty) \big)\cong V$. The $\IZ_k$-action on $\PP^2$ lifts to a natural $\IZ_k$-action on the moduli space $\frM_{N,n}(\IC^2)$ and one can compute
equivariant characters by taking the $\IZ_k$-invariant part of the
pertinent $\torus^2$-modules on $\frM_{N,n}(\IC^2)$; in fact there is
a decomposition of the $\IZ_k$-fixed point set
\bea
\frM_{N,n}(\IC^2)^{\IZ_k} = \bigsqcup_{|\vec v\,|=n} \, \frM_{\xi_0}(\vec v,\vec w\,)
\nonumber \eea
with $|\vec v\,|:= \sum_{i=0}^{k-1}\, v_i$, for fixed $\vec
w\in\IZ_{\geq0}^k$ with $| \vec w\,|=N$. This enables the computation
of partition functions, regarded as integrals over the moduli spaces~$\frM_{\xi_0}(\vec v,\vec w\,)$, for $\cN=2$ gauge theory on the resolution of the Kleinian singularity $\IC^2/\IZ_k$ provided by the quotient stack $[\IC^2/\IZ_k]$~\cite{art:fucitomoralespoghossian2004,art:fujii2005}. As an explicit example, in the following we treat the case of $U(1)$ gauge theory in detail. The $\zed_k$-action on $\IC^2$ given by
(\ref{eq:ZkactionC2}) endows $\IC[z_1,z_2]$ with a canonical $\IZ_k$-module structure which lifts to a $\IZ_k$-action on the Hilbert schemes $\Hilb^n(\IC^2)$. We set $\vec w=\vec w_0:=(1,0,\dots,0)$ and fix $\vec{v}=(v_0, v_1,\ldots, v_{k-1})\in\mathbb{Z}_{\geq 0}^{k}$. Then $\frM_{\xi_0}(\vec v,\vec w_0)\cong \Hilb^{|\vec v\,|}(\mathbb{C}^2)^{\mathbb{Z}_k, \vec{v}}$ is the moduli space parameterizing $\mathbb{Z}_k$-invariant ideals $I$ of codimension $|\vec v\,|$ in the polynomial ring $\IC[z_1,z_2]$ such that $\IC[z_1,z_2]/I \cong \bigoplus_{i=0}^{k-1}\, v_i\, \rho_i$. It is a smooth quasi-projective variety of dimension 
\begin{equation*}
\dim_\IC \Hilb^{|\vec v\,|}\big(\mathbb{C}^2 \big)^{\mathbb{Z}_k,
  \vec{v}}=2v_0- \vec{v}\cdot \widehat{C}\vec{v} \ .
\end{equation*}

There is another distinguished chamber $\sfC_\infty$ such that, for $\xi_\infty\in \sfC_\infty$ and for $\vec v=\vec{\delta}=(1, \ldots, 1)$, the moduli space $\frM_{\xi_\infty}(\vec\delta,\vec w_0)$ is the $\IZ_k$-Hilbert scheme of the
plane $\complex^2$ with $\widehat{C}\vec\delta=\vec 0$, $|\vec{\delta}\,|
=k$ and $\Hilb^{\IZ_k}(\IC^2):=
\Hilb^{|\vec{\delta}\,|}(\mathbb{C}^2)^{\mathbb{Z}_k,
  \vec{\delta}}\cong X_k$ under the Hilbert--Chow morphism
$\Hilb^{\IZ_k}(\IC^2)\to\IC^2/\IZ_k$
(cf.~\cite[\S3.1.2]{art:fujii2005}); in this case $\IC[z_1,z_2]/I$ is
isomorphic to the regular representation of the cyclic group
$\IZ_k$. This is the complex algebraic version of Kronheimer's
construction of ALE spaces~\cite{art:kronheimer1989}. Since the quiver
varieties $\frM_{\xi_0}(\vec v,\vec w\,)$ and $\frM_{\xi_\infty}(\vec
v,\vec w\,)$ are \emph{not} isomorphic, they have distinct universal
bundles. As it is difficult (for us) to work directly over the Nakajima quiver varieties, one needs to independently develop a means for studying gauge theories with instanton moduli associated to the chamber $\sfC_\infty$. Such a theory was developed by~\cite{Bruzzo:2013daa} in the framework of framed sheaves on a suitable orbifold compactification $\Xscr_k=X_k\cup\Dscr_\infty$, which is a smooth projective toric orbifold. The compactification divisor $\Dscr_\infty$ is a $\IZ_k$-gerbe over a football which as a toric Deligne--Mumford stack has a presentation as a global quotient stack
\bea
\Dscr_\infty \cong \left[\,
      \frac{\IC^2\setminus\{0\}}{\IC^\times\times \IZ_k}\,\right]
\nonumber \eea
with Deligne--Mumford torus $\IC^\times\times B\IZ_k$. Hence its Picard group is given by ${\sf Pic}(\Dscr_\infty)\cong \IZ\oplus
  \IZ_k$ and we denote the respective generators of the two factors by $\Lcal_1,\Lcal_2$. The fundamental group of the underlying topological stack is given by $\pi_1(\Dscr_\infty) \cong \IZ_k$, and each line bundle
  $\cO_{\Dscr_\infty}(i)= \Lcal_2^{\otimes i}$ can be endowed with a 
  unitary flat connection associated to the representation $\rho_i: \IZ_k \to U(1)$~\cite{EyssidieuxSala14}. The Picard group ${\sf Pic}(\Xscr_k)$ is generated by the line bundles $\cO_{\Xscr_k}(\Dscr_\infty)$ and $\mathscr{R}_1,\dots,\mathscr{R}_{k-1}$, where the restrictions of $\mathscr{R}_i$ to $X_k$ coincide with the tautological line
  bundles $\Rcal_i$ and to $\Dscr_\infty$ with $\cO_{\Dscr_\infty}(i)$.

For fixed $\vec w=
(w_0,w_1,\dots,w_{k-1}) \in \IZ_{\geq0}^k$, by using the general theory of framed sheaves on projective stacks developed by~\cite{BruzzoSala13} one can construct a fine moduli space $\frM_{\vec{u},\Delta,\vec{w}}(X_k)$ parameterizing torsion-free sheaves $\cE$ on $\Xscr_k$ with a framing isomorphism
$\phi_{\cE}:\cE\vert_{\Dscr_\infty} \to
\bigoplus_{i=0}^{k-1} \,
\cO_{\Dscr_\infty}(i)^{\oplus w_i}$ of rank $N=\sum_{i=0}^{k-1}\, w_i$, first Chern class $c_1(\cE) =\sum_{i=1}^{k-1} \, u_i \, c_1(\mathscr{R}_i)$, and discriminant $\Delta(\cE)=\Delta$ where
\bea
\Delta(\cE)=\int_{\Xscr_k}\, \Big(c_2(\cE)-\frac{N-1}N \,
c_1(\cE)^2\Big) \ .
\nonumber \eea
The framing condition restricts the Chern classes to $\cU_{\vec w}$ which is the set of integer vectors $\vec u \in \IZ^{k-1}$ that correspond to the sum of the weight $\sum_{i=1}^{k-1}\, w_i\,
  \omega_i$ and an element $\gamma_{\vec u} \in \cQ$ of the $A_{k-1}$ root lattice; here $\omega_1,\dots,\omega_{k-1}$ are the fundamental weights of type $A_{k-1}$. The moduli space $\frM_{\vec{u},\Delta,\vec{w}}(X_k)$ is a smooth quasi-projective variety of dimension
\bea
\dim_\IC \frM_{\vec{u},\Delta,\vec{w}}(X_k)=2\,r\, \Delta - \frac12\, \sum\limits_{j=1}^{k-1}\, \big(C^{-1}\big)^{jj}\, \vec{w}\cdot\vec{w}(j) \ ,
\nonumber \eea
where $\vec{w}(j):= (w_j, \ldots, w_{k-1}, w_0, w_1, \ldots,
w_{j-1})$, and it contains the moduli space of $U(N)$ instantons on
$X_k$ of first Chern class $\sum_i \, u_i\, c_1(\Rcal_i)$ and holonomy
at infinity $\rho=\bigoplus_j \, w_j\,
\rho_j$~\cite{EyssidieuxSala14}. One furthermore has~\cite{Bruzzo:2013daa,EyssidieuxSala14}
\begin{theorem}
There is a birational morphism
\bea
\frM_{\vec{u},\Delta,\vec{w}}(X_k) \ \longrightarrow \ \frM_{\xi_\infty}(\vec v,\vec w\,)
\nonumber \eea
for some $\vec v\in\IZ_{\geq0}^k$.
\end{theorem}
In the rank one case $N=1$, this morphism is an isomorphism: In this
instance $\frM_{\vec{u},n,\vec{w}}(X_k) \cong {\sf Hilb}^n(X_k)$ is the
Hilbert scheme of $n$ points on $X_k$ for all $\vec{u}$ and $\vec w$~\cite{Bruzzo:2013daa}, which is a Nakajima quiver variety by~\cite{art:kuznetsov2007}.

\subsection{$\cN=2$ gauge theory on $X_k$\label{sec:N2Xk}}

Let us first present the computation of the instanton partition function corresponding to the chamber $\sfC_\infty$. For pure $\cN=2$ gauge theory on $X_k$, it is defined by a weighted sum over topological sectors $\vec u$ of fractional instantons with chemical potentials $\vec\xi=(\xi_1,\dots,\xi_{k-1})$ as
\bea
\cZ^{\rm inst}_{X_k}\big(\epsilon_1, \epsilon_2,
\vec{a}\, ; \sfq, \vec{\xi}\ \big)_{\vec w}=
\sum_{\vec u\in \cU_{\vec w}} \, \vec{\xi}^{\ \vec{u}} \ \sum_{\Delta\in \frac{1}{2\,N\, k}\, \mathbb{Z}}\,
  \sfq^{\Delta+\frac{1}{2N}\, \vec{u}\cdot C^{-1} \vec{u}} \ 
  \int_{\frM_{\vec{u},\Delta,\vec w}(X_k)} \, 1 \ ,
\nonumber \eea
where $\vec\xi\ ^{\vec
  u}=\prod_{i=1}^{k-1}\, \xi_i^{u_i}$. 
The torus fixed points $\frM_{\vec{u},\Delta,\vec{w}}(X_k)^{\widetilde{\torus}}$ are parameterized by vectors of Young diagrams $\vec{\boldsymbol{Y}}=(\vec{Y}_1, \ldots, \vec{Y}_N)$, with $\vec{Y}_l =\{Y_l^i\}_{i=1, \ldots, k}$, and vectors of integers $\vec{\boldsymbol{u}} =(\vec{u}_1, \ldots, \vec{u}_N)$ such that $\vec u=\sum_{l=1}^N\, \vec
  u_l$ with $\vec u_l$ corresponding to $\gamma_{\vec u_l}+\omega_i$ for $i= 1, \ldots, k-1$ and $\sum_{j=0}^{i-1}\, w_j < l \leq \sum_{j=0}^i\, w_j$, and
\bea
\Delta =\sum_{l=1}^N \, |\vec{Y}_l| +\frac{1}{2}\,
\sum\limits_{l=1}^N\, \vec{u}_l\cdot C^{-1}\vec{u}_l-\frac{1}{2N}\,
\sum\limits_{l,l'=1}^N \, \vec{u}_l\cdot C^{-1}\vec{u}_{l'} \ .
\nonumber \eea

The localization theorem then gives the factorization formula
\begin{multline*}
\cZ^{\rm inst}_{X_k}\big(\epsilon_1, \epsilon_2, \vec{a}; \sfq, \vec{\xi}\ \big)_{\vec w} = \sum_{\vec u\in \cU_{\vec w}} \, \vec{\xi}^{\ \vec{u}} \
  \sum_{\vec{\boldsymbol{u}}} \, \sfq^{\frac{1}{2}\,
    \sum\limits_{l=1}^N \, \vec{u}_l\cdot C^{-1}\vec{u}_l} \\ \times \
  \prod_{l,l'=1}^N \ \prod_{n=1}^{k-1} \,
  \frac{1}{\ell^{(n)}_{\vec{u}_{l'}-\vec u_l}\big(\epsilon_1^{(n)},
    \epsilon_2^{(n)}, a_{l'}-a_l\big)} \ \prod_{i=1}^k\,
  \cZ_{\IC^2}^{\rm inst}\big(\epsilon_1^{(i)}, \epsilon_2^{(i)}, \vec{a}\,^{(i)};
  {\sfq} \big) \ ,
\end{multline*}
which determines the partition function in terms of the fan of $X_k$ as a product of $k$ copies
of the instanton partition function \eqref{eq:NekinstC2comb} on ${\IC^2}$; it depends on the equivariant parameters of the
torus action on the affine toric patches $U[\sigma_i]\cong\IC^2$ of
$X_k$ (which can be read off from the weights of \S\ref{se:cdal}), and
on the leg factors $\ell^{(n)}_{\vec{u}}$ which are determined by the
geometry of the exceptional divisors $\ell_1,\dots,\ell_{k-1}$ of the resolution $X_k\to
\IC^2/\IZ_k$ (whose explicit forms can be found
in~\cite{Bruzzo:2013daa,Bruzzo:2014jza}). This expression generalizes
the Nakajima--Yoshioka blowup formulas~\cite{NakYosh}; it was
originally conjectured
in~\cite{art:bonellimaruyoshitanzini2011,art:bonellimaruyoshitanzini2012,art:bonellimaruyoshitanziniyagi2012}
and then rigorously proven in~\cite{Bruzzo:2013daa}. This form of the partition function is expected to nicely capture physical features of $\cN=2$ gauge theories on the ALE space $X_k$ and their relations to two-dimensional conformal field theory that we discuss in \S\ref{sec:20theory}.

In the rank one case $N=1$, with $\vec w_j$ the $j$-th coordinate
vector of $\IZ^k$ for $j=0,1,\dots,k-1$, the leg factors are unity and the partition function simplifies to~\cite{Pedrini:2014yoa}
\bea
\cZ^{\rm inst}_{X_k}\big(\epsilon_1,\epsilon_2; \sfq, \vec{\xi} \ \big)_{\vec w_j}
= \widehat{\eta}(\sfq)^{k-1} \ \chi^{\widehat{\omega}_j}\big(\sfq,\vec\xi \ \big) \ \exp\Big(\, \frac \sfq{k\,\epsilon_1\, \epsilon_2}\, \Big) \ ,
\label{eq:cZXk} \eea
where
\begin{equation} 
\chi^{\widehat{\omega}_j}\big(q,\vec\xi \ \big)=\frac{1}{\widehat{\eta}(q)^{k-1}} \
\sum_{\vec{u}\in\cU_{\vec w_j}} \, q^{\frac{1}{2}\, \vec{u}\cdot
  C^{-1}\vec{u}}\ \vec{\xi}^{\ \vec{u}}
\nonumber \end{equation}
is the character
of the integrable highest weight representation of the affine Lie algebra $\widehat{\frsl}_{k}$
at level one, with highest weight the $j$-th fundamental weight
$\widehat{\omega}_j$ of type $\widehat{A}_{k-1}$ for $j=0,1, \ldots,
k-1$, and
\bea
\widehat{\eta}(q)^{-1} = \prod_{n=1}^\infty\, \frac1{1-q^n } 
\nonumber \eea
is the character of the Fock space representation of
the Heisenberg algebra $\frh$, i.e. the Euler function which is the generating function for Young diagrams. This formula generalizes the $k=1$ expression \eqref{eq:corN2C2U1}; its representation theoretic content will be elucidated further in \S\ref{sec:20theory}.

\subsection{$\cN=2$ gauge theory on $[\complex^2/\zed_k]$\label{sec:N2orbifold}}

Let us now compare the computation of \S\ref{sec:N2Xk} with that
corresponding to the chamber $\sfC_0$; see \S\ref{app:Maya} for the
relevant combinatorial definitions and properties used below. Consider again the rank one case to begin with. The instanton part of Nekrasov's partition function for the pure
$\cN=2$ gauge theory on the quotient stack $[\IC^2/\IZ_k]$ is defined by
$$
\cZ^{\rm inst}_{[\IC^2/\IZ_k]}(\epsilon_1,\epsilon_2; \tilde\sfq,\vec\sfq\,) :=
\sum_{\vec v\in\IZ_{\geq0}^k}\, \tilde\sfq^{|\vec v\,|} \,
\vec\sfq^{\,\vec v} \ \int_{\Hilb^{|\vec v\,|}(\mathbb{C}^2)^{\mathbb{Z}_k, \vec{v}}}\, 1 \ ,
$$
where $\vec\sfq=(\sfq_0,\sfq_1,\dots,\sfq_{k-1})$ and $\vec\sfq^{\,\vec
  v}:=\prod_{i=0}^{k-1} \, \sfq_i^{v_i}$. The action of the torus $\torus^2=\mathbb{C}^\times \times \mathbb{C}^\times$ on
$\Hilb^{|\vec v\,|}(\mathbb{C}^2)$ restricts to a torus action on
$\Hilb^{|\vec v\,|}(\mathbb{C}^2)^{\mathbb{Z}_k, \vec{v}}.$ As
described in~\cite[\S3.2]{art:fujii2005}, the torus fixed points
of $\Hilb^{|\vec v\,|}(\mathbb{C}^2)^{\mathbb{Z}_k, \vec{v}}$
correspond to $k$-coloured Young diagrams $Y$ with $\vert Y\vert=|\vec
v\,|$ such that $\nu_i(Y)=v_i$ for $i=0,1, \ldots, k-1.$ Since all nodes
on the line $b=a-\mu\,k-i$ for $\mu\in\IZ$ are coloured by the same
colour $i$, we have
$$
v_i=\sum_{\mu\in\IZ}\, N_{\mu\, k+i}(Y) \ .
$$
Since an $r$-hook for $r=n\,k$ has $n$ $i$-nodes for each
$i=0,1,\dots,k-1$ and $c_h(Y)=v_{h-\frac12}-v_{h+\frac12}$, all $k$-coloured Young diagrams $Y$ corresponding
to dimension vectors $\vec v=n\,\vec\delta$ have $k$-core $\vec c\, (Y)=\vec0$, i.e.
$\underline{Y}=\emptyset$.

By the localization theorem we have
\begin{eqnarray*}
\int_{\Hilb^{|\vec v\,|}(\mathbb{C}^2)^{\mathbb{Z}_k, \vec{v}}}\, 1 =
\sum_{\nu_i(Y)=v_i} \ \prod_{\stackrel{\scriptstyle s\in
    Y}{\scriptstyle h(s)\equiv0\, {\rm mod}\,
    k}}\, \frac{1}{\big((L(s)+1)\, \epsilon_1-A(s)\, \epsilon_2
  \big)\big(-L(s)\, \epsilon_1+
    (A(s)+1)\, \epsilon_2\big)} \ .
\end{eqnarray*}
By decomposing Young diagrams $Y$ into their $k$-cores and
$k$-quotients $(\vec c\, (Y),\vec q\, (Y))$, and repeating the same argument as in~\cite[\S4.2.1]{art:fujii2005},
we then find the factorization formula
$$
\cZ^{\rm inst}_{[\IC^2/\IZ_k]}(\epsilon_1,\epsilon_2;\tilde\sfq,\vec\sfq\,) = \cZ_{[\IC^2/\IZ_k]}(\tilde\sfq,\vec\sfq\,)^{\rm
  core} \ \cZ_{[\IC^2/\IZ_k]}(\epsilon_1,\epsilon_2;\tilde\sfq,\vec\sfq\,)^{\rm
  quot} \ ,
$$
where
$$
\cZ_{[\IC^2/\IZ_k]}(\tilde\sfq,\vec\sfq\,)^{\rm
  core} := \sum_{\vec q\, (Y)=\vec\emptyset}\, \tilde\sfq^{|Y|} \
\prod_{i=0}^{k-1}\, \sfq_i^{\sum\limits_{\mu\in\IZ}\, N_{\mu\,k+i}(Y)}
$$
is the contribution from $k$-cores (Young diagrams with no $k$-hooks),
while
\begin{multline*}
\cZ_{[\IC^2/\IZ_k]}(\epsilon_1,\epsilon_2;\tilde\sfq,\vec\sfq\,)^{\rm
  quot} \ := \  \sum_{v_0=v_1=\cdots=v_{k-1}}\, \tilde\sfq^{|\vec v\,|} \,
\vec\sfq^{\, \vec v} \\ \times \ \sum_{\nu_i(Y)=v_i} \ \prod_{\stackrel{\scriptstyle s\in
    Y}{\scriptstyle h(s)\equiv0\, {\rm mod}\,
    k}}\, \frac{1}{\big((L(s)+1)\, \epsilon_1-A(s)\, \epsilon_2
  \big)\big(-L(s)\, \epsilon_1+
    (+1)\, \epsilon_2\big)} \nonumber
\end{multline*}
is the contribution from $k$-quotients ($\vec c\, (Y)=\vec0\, $).

The minimal resolution $X_k\to\IC^2/\IZ_k$ induces a
natural $\torus^2$-equivariant morphism $\Hilb^{n}(X_k) \to
\Sym^n(\IC^2/\IZ_k)$. It is shown in~\cite[\S8.1]{Pedrini:2014yoa} that the
generating function for equivariant volumes of the Hilbert scheme can
be summed explicitly to give (cf. \eqref{eq:cZXk})
$$
\sum_{n=0}^\infty\, \sfq^{n} \ \int_{\Hilb^n(X_k)}\, 1 =
\exp\Big(\, \frac\sfq{k\,\epsilon_1\, \epsilon_2}\, \Big) \ .
$$
The coefficient of $\sfq^n$ is given by
$$
\int_{\Hilb^n(X_k)}\, 1 =
\frac1{n!\, k^n\, (\epsilon_1\, \epsilon_2)^n} \ ,
$$
which is the expected $\torus^2$-equivariant volume of the $n$-th symmetric product
$\Sym^n(\IC^2/\IZ_k)$. 
Using geometric arguments on quiver varieties, Fujii and Minabe show in~\cite{art:fujii2005}
that $\Hilb^{|\vec
  v\,|}(\IC^2)^{\IZ_k,\vec v}\to \Sym^n(\IC^2/\IZ_k)$ is a
$\torus^2$-equivariant resolution of singularities for $n=|\vec
  v\,|$, and thus $\int_{\Hilb^{|\vec
  v\,|}(\IC^2)^{\IZ_k,\vec v}}\, 1$ also gives
the equivariant volume of $\Sym^n(\IC^2/\IZ_k)$. Hence the series over $k$-hooks can be summed explicitly to give~\cite[Prop.~A.5]{art:fujii2005}
$$
\sum_{\nu_i(Y)=v_i} \ \prod_{s\in\vec q\, (Y)} \, \frac{1}{\big((L(s)+1)\, \epsilon_1-A(s)\, \epsilon_2
  \big)\big(-L(s)\, \epsilon_1+
    (A(s)+1)\, \epsilon_2\big)} =
\frac1{n!\, k^n \, (\epsilon_1\,\epsilon_2)^n}
$$
where $n=|\vec q\, (Y)|$. It follows that
\begin{eqnarray*}
\cZ_{[\IC^2/\IZ_k]}(\epsilon_1,\epsilon_2;
\tilde\sfq,\vec\sfq\,)^{\rm quot} = \sum_{n=0}^\infty\, \tilde\sfq^{n\, k} \
\vec\sfq^{\,n\, \vec\delta}\, \frac1{n!\, k^n \, (\epsilon_1\,\epsilon_2)^n}
= \exp\Big(\,
\frac \sfq{k\, \epsilon_1\, \epsilon_2}\, \Big)
\end{eqnarray*}
where 
$$
\sfq:= \tilde\sfq^k\ \prod_{i=0}^{k-1}\, \sfq_i \ .
$$

For the $k$-core contribution, we identify the set of $k$-cores
$\big(\IZ^{\tilde I}\,\big)_0$ with the $A_{k-1}$ root lattice $\cal Q$,
and proceed similarly to the proof of~\cite[Lem.~4.9]{art:fujii2005}
to compute
\bea
\cZ_{[\IC^2/\IZ_k]}(\tilde\sfq,\vec\sfq\,)^{\rm
  core} &=& \sum_{\rmc\in{\cal Q}}\, \sfq^{\frac12\, \sum\limits_{j=0}^{k-1}\,
  c^2_{j+\frac12}} \ \prod_{i=1}^{k-1}\, (\tilde\sfq \,
\sfq_i)^{\sum\limits_{j=i}^{k-1}\, c_{j+\frac12}} \nonumber \\[4pt] &=& \sum_{\rmc\in{\cal Q}}\, \sfq^{\frac12\, \sum\limits_{j=0}^{k-1}\,
  (v_j(\rmc)-v_{j+1}(\rmc))^2} \ \prod_{i=1}^{k-1}\, (\tilde\sfq \,
\sfq_i)^{\sum\limits_{j=i}^{k-1}\, (v_j(\rmc)-v_{j+1}(\rmc)) } \nonumber
\\[4pt] &=& \sum_{\rmc\in{\cal Q}}\, \sfq^{\frac12\, \vec v(\rmc)\cdot \widehat{C}\vec v(\rmc)} \ \prod_{i=1}^{k-1}\, (\tilde\sfq \,
\sfq_i)^{v_i(\rmc)-v_0(\rmc)} \ . \nonumber
\eea
Here we use the fact that, for any element
$\mathrm{c}:=(c_{\frac{1}{2}}, \ldots, c_{k-\frac{1}{2}})$ of the root
lattice $\mathcal{Q}$, there is a unique vector $\vec{v}=(v_0, v_1,
\ldots, v_{k-1})\in\mathbb{Z}_{\geq 0}^{k}$ such that
$c_{\frac{2i+1}{2}}=v_{i\ \mathrm{mod}\, k}-v_{i+1\ \mathrm{mod}\, k}$
for $i=0,1, \ldots, k-1$ and $\frac{1}{2}\,\vec{v}\cdot
\widehat{C}\vec{v}+v_0\in \mathbb{Z}_{>0}$. To make contact with the
partition function on the resolution from \S\ref{sec:N2Xk}, we introduce the first Chern numbers $\vec u(\rmc)=-\widehat{C} \vec v(\rmc)$ as before, and define $\vec
w(\rmc)=(w_1(\rmc),\dots,w_{k-1}(\rmc))$ by $w_i(\rmc)=
v_i(\rmc)-v_0(\rmc)$ for $i=1,\dots,k-1$. Then $\vec v(\rmc)\cdot \widehat{C} \vec
v(\rmc)=\vec w(\rmc)\cdot C\vec w(\rmc)$ and $u_i(\rmc)=-\sum_{j=1}^{k-1}\, C_{ij}\, w_j(\rmc)$
for $i=1,\dots,k-1$. It follows that
$$
\cZ_{[\IC^2/\IZ_k]}(\tilde\sfq,\vec\sfq\,)^{\rm
  core} = \sum_{\rmc\in{\cal Q}}\, \sfq^{\frac12\, \sum\limits_{i,j=1}^{k-1}\,
  u_i(\rmc)\, (C^{-1})^{ij}\, u_j(\rmc)} \ \prod_{i=1}^{k-1}\,
\xi_i^{u_i(\rmc)} \ ,
$$
where
$$
\xi_i:= \prod_{j=1}^{k-1}\, (\tilde\sfq\, \sfq_j)^{(C^{-1})^{ij}}
$$
for $i=1,\dots,k-1$.

Putting everything together, we have thus shown that the gauge theory
partition functions on the quotient stack and on the minimal resolution in this case
coincide:
$$
\cZ^{\rm inst}_{[\IC^2/\IZ_k]}(\epsilon_1,\epsilon_2;\tilde\sfq,\vec\sfq\,) = \cZ^{\rm inst}_{X_k}\big(\epsilon_1,\epsilon_2; \sfq,
\vec\xi\ \big)_{\vec w_0} \ .
$$

The Nekrasov partition functions in the generic case of $U(N)$ gauge theory on $[\IC^2/\IZ_k]$ can be similarly defined and regarded as $\widetilde{\torus}$-equivariant integrals over the moduli spaces $\frM_{\xi_0}(\vec v,\vec w\,)$, see e.g.~\cite{art:wyllard2011,Belavin:2011sw,Alfimov:2013cqa,art:itomaruyoshiokuda2013,Bruzzo:2014jza}; the $\widetilde{\torus}$-fixed points in this case are vectors of $\vec w$-coloured Young diagrams $\vec Y$ with $k$ colours. However, in this case the torus fixed point sets $\frM_{\xi_0}(\vec v,\vec w\,)^{\widetilde{\torus}}$ and $\frM_{\xi_\infty}(\vec v,\vec w\,)^{\widetilde{\torus}}$ are not in bijective correspondence, and the explicit matching with the partition function on the minimal resolution $X_k$ is not presently understood in generality.

\subsection{$\cN=1$ gauge theory in five dimensions\label{sec:5D}}

Let us now sketch how to derive our $\cN=2$ gauge theory partition
functions directly from the six-dimensional $\cN=1$ supersymmetric
gauge theory defining the $\Omega$-background, as discussed in
\S\ref{sec:N2C2}. We consider the limit in which the torus $T^2$
collapses to a circle $S^1$ of radius $\varrho$, which yields five-dimensional $\cN=1$ gauge theory on the flat affine bundle $M_\varrho \to S^1$ where $M_\varrho :=(\IC^2\times\IR)/\IZ$ with the $\IZ$-action given by
\bea
n\triangleright(z_1,z_2,x)=(q^n\, z_1,t^n\, z_2,x+2\pi\, n\, \varrho) \ .
\nonumber \eea
Here $n\in\IZ$ is the instanton charge of the four-dimensional description, $(z_1,z_2)\in\IC^2$, $x\in\IR$, and $(q,t)$ is an
element of the maximal torus of the complex rotation group $GL(2,\IC)$
which parameterizes the holonomy of a connection on the flat
$\torus^2$-bundle discussed in \S\ref{sec:N2C2}.
Then we take the collapsing limit $\varrho\to0$ to get four-dimensional $\cn=2$ gauge theory on the $\Omega$-background. We shall study the Nekrasov partition function for $\cn=2$
gauge theory on $[\IC^2/\IZ_k]$ by studying the Uglov limits of the
corresponding gauge theories on $M_\varrho$ (see \S\ref{sec:uglovfunctions}). Partition functions of gauge
theories on $M_\varrho$ also compute the K-theory versions of Nekrasov partition
functions on~$\IC^2$.

Let $\varrho \in\IC$ be a parameter. For a vector bundle $V$
over $\Hilb^n(\IC^2)$, let $\widehat{A}_\varrho(V)$ be the
deformation of the $\widehat{A}$-genus of $V$ defined in~\cite[\S4.6]{GL1}. The instanton partition function of pure
$\cN=1$ 
five-dimensional gauge theory on
$M_\varrho$ is defined by
$$
\cZ_{M_\varrho}^{\rm inst}(q,t; \sfq):=
\sum_{n=0}^\infty\, \sfq^{n} \ \int_{\Hilb^n(\IC^2)}\, 
\widehat{A}_\varrho\big({T}\Hilb^n(\IC^2) \big) \ .
$$
By the localization theorem one has
\begin{eqnarray}
\int_{\Hilb^n(\IC^2)}\, 
\widehat{A}_\varrho\big({T}\Hilb^n(\IC^2) \big) =\sum_{\vert\lambda\vert=n} \
\prod_{s\in Y_\lambda}\,
\frac{\varrho^2}{\big(1-q^{-A(s)-1}\,
  t^{-L(s)} \big)\, \big(1-q^{-A(s)}\,
  t^{-L(s)-1}\big)} \ ,
\label{eq:5dbilinear}\end{eqnarray}
where we introduced the parameters
$$
q=t_2^\varrho=\e^{\varrho\,\epsilon_2} \qquad \mbox{and} \qquad 
t=t_1^{-\varrho}=\e^{-\varrho\,\epsilon_1}=q^\beta
$$
with $\beta=-\epsilon_1/\epsilon_2$. This partition
function is a $q$-deformation of the partition function \eqref{eq:NekinstC2comb} for
four-dimensional pure $\cn=2$ gauge theory on $\IC^2$, to which it reduces in the collapsing limit $\varrho\to0$,
$$
\lim_{\varrho\to0}\, \cZ^{\rm inst}_{M_\varrho}\big(q=t_2^\varrho,t=q^\beta; \sfq\big) =
\cZ^{\rm inst}_{\IC^2}(\epsilon_1,\epsilon_2; \sfq) \ .
$$
In~\cite[\S4]{NakYosh} the same expression is
derived as the K-theory version of the Nekrasov partition function on $\IC^2$,
wherein we replace integration in equivariant cohomology by
integration in equivariant K-theory. The series over partitions can be explicitly
summed (see e.g.~\cite[\S4]{NakYosh}) to get
$$
\cZ^{\rm inst}_{M_\varrho}(q,t; \sfq) = \prod_{i,j=0}^\infty\
\frac1{1-t^i\,q^{-j}\,\sfq\, \varrho^2} = \exp\Big(\,
\sum_{r=1}^\infty\, \frac{\sfq^r\,\varrho^{2r}}{r\, \big(1-q^{-r}\big)\, 
  \big(1-t^r\big)} \, \Big)\ .
$$
This formula can be derived from the Cauchy--Stanley identity for
Macdonald symmetric functions, see \S\ref{app:Macdonald}; it
reduces to (\ref{eq:corN2C2U1}) in the limit $\varrho\to0$. In the following we
regard $\cZ^{\rm inst}_{M_\varrho}(q,t; \sfq)$ as a
function of arbitrary $q,t\in\IC$.

We will now consider this partition function in the \emph{Uglov limit} by instead
setting
$$
q=\omega\, \e^{\varrho\,\epsilon_2} = \omega\, p \qquad \mbox{and} \qquad t=
\omega\, \e^{-\varrho\,\epsilon_1} = \omega \, p^\beta \qquad \mbox{with}
\quad p=t_2^\varrho \ ,
$$
and defining the {orbifold partition function} as
$$
\cZ^{\rm inst}_{[\IC^2/\IZ_k]}(\epsilon_1,\epsilon_2; \sfq):=
\lim_{\varrho\to0}\, \cZ^{\rm inst}_{M_\varrho}(q=\omega\,t_2^\varrho,t=\omega\, t_1^{-\varrho}; \sfq) \
.
$$
Multiplication with the primitive $k$-th root of unity
$\omega:=\e^{2\pi\ii/k}$ implements the
orbifold projection onto $\IZ_k$-invariant sectors~\cite{Kimura:2011zf}. For each partition
$\lambda$ occuring in the bilinear form product (\ref{eq:5dbilinear}), one easily sees that
the product gives $0$ in the limit $\varrho\to0$ unless $\omega^{h(s)}=1$ for all nodes
$s\in Y_\lambda$. Thus the series truncates to a sum over $k$-coloured
Young diagrams and one has
\begin{eqnarray*}
\cZ^{\rm inst}_{[\IC^2/\IZ_k]}(\epsilon_1,\epsilon_2; \sfq) =
\sum_\lambda\, \sfq^{|\lambda|} \ \prod_{\stackrel{\scriptstyle s\in
    Y_\lambda}{\scriptstyle h(s)\equiv0\, {\rm mod}\, k}} \ \prod_{s\in
  Y_\lambda}\frac{1}{\big((L(s)+1)\, \epsilon_1-A(s)\, \epsilon_2
  \big)\, \big(-L(s)\, \epsilon_1+
    (A(s)+1)\, \epsilon_2\big)} \ ,
\end{eqnarray*}
which agrees with the partition function of \S\ref{sec:N2orbifold}
at
$\tilde\sfq=\sfq$ and $\sfq_i=1$ for $i=0,1,\dots,k-1$.

\subsection{Quantization of coarse moduli spaces}

In the remainder of this section we shall initiate some of the geometric constructions underlying a putative reformulation of the partition functions of $\cN=2$ gauge theories on ALE spaces in terms of contributions from noncommutative instantons. The Pontrjagin dual group
$\widehat{{\torus^2}}=\Hom_\complex({\torus^2},\complex^\times)\cong\zed\oplus \zed$ is the group
of characters $\{\chi_p\}_{p\in L^*}$ parameterized by elements of the
dual lattice $p=p_1\,e^*_1+ p_2\,e^*_2\in L^*$ as
\beq
\chi_p(t)=t^p := t_1^{p_1}\, t_2^{p_2} \ . 
\label{Tchar}\eeq
The unital algebra $\hil=\alg({\torus^2})$ of coordinate functions on the torus ${\torus^2}$ is the Laurent
polynomial algebra
$$
\hil:=\complex(t_1,t_2)
$$
generated by elements $t_1,t_2$. It is equipped
with the Hopf algebra structure
$$
\Delta(t^p)=t^p\otimes t^p \ , \qquad \varepsilon(t^p)=1 \qquad \mbox{and} \qquad
S(t^p)=t^{-p}
$$
for $p\in L^*$, with the coproduct and the counit respectively extended as algebra
morphisms $\Delta:\hil\to \hil\otimes \hil$ and $\varepsilon:\hil\to\complex$, and the
antipode as an anti-algebra morphism $S:\hil\to \hil$. 

The canonical right
action of ${\torus^2}$ on itself by group multiplication dualizes to give a
left {$\hil$-}coaction
\beq
\Delta_L\,:\, \alg({\torus^2})~\longrightarrow~ \hil\otimes \alg({\torus^2}) \ , \qquad
\Delta_L(u_i)= t_i\otimes u_i \ , \quad \Delta_L\big(u_i^{-1} \big)=
t_i^{-1}\otimes u_i^{-1} \ ,
\label{DeltaLT}\eeq
where we write $u_i$, $u_i^{-1}$, $i=1,2$, for the generators of $\alg({\torus^2})$
viewed as a left comodule algebra over itself, 
when distinguishing the coordinate algebra $\alg({\torus^2})$ from the Hopf algebra $\hil$. 
This coaction is
equivalent to a grading of the algebra $\alg({\torus^2})$ by the dual lattice
$L^*$, for which the homogeneous elements are the characters
(\ref{Tchar}).

The functorial quantization constructed in~\cite[\S1.2]{CLSII}
twists the {algebra} multiplication in $\alg=\alg({\torus^2})$ into a new
product. Let $\alg(\torus^2_\theta)$ be the Laurent polynomial algebra
generated by $u_1$ and $u_2$ with
this product, where $\theta\in\complex$ is the deformation parameter. It has relations
\bea
u_1^{p_1}\, u_2^{p_2} =q^{2p_1\, p_2}~ u_2^{p_2}\, u_1^{p_1} \qquad \mbox{for} \quad
p_1,p_2\in\zed \ ,
\label{Tthetarels}\eea
where $q=\exp(\frac\ii 2\, \theta)$. This quantizes the torus ${\torus^2}$ into the
noncommutative complex torus
$\torus^2_\theta$ {dual to the algebra
  $\alg(\torus^2_\theta)$}. We will sometimes write
$\alg(\torus^2_\theta)=\complex_\theta(u_1,u_2)$ for this coordinate algebra.

Noncommutative affine toric varieties
correspond to finitely-generated
$\hil_\theta$-comodule subalgebras of the algebra $\alg(\torus^2_\theta)$ of the
noncommutative torus~\cite[\S3]{CLSI}, where $\hil_\theta$ is the Hopf algebra $\hil$
with twisted coquasitriangular structure
$\Rcal_\theta:\hil_\theta\otimes\hil_\theta\to\complex$ defined on
generators by
$\Rcal_\theta(t_i\otimes t_i)=1$ for $i=1,2$ and
$\Rcal_\theta(t_1\otimes t_2)=q^{-2} =\Rcal_\theta(t_2\otimes t_1)^{-1}$. To each cone
$\sigma\subset L_\real$, we define the coordinate algebra
$\complex_\theta[\sigma]$ dual to a noncommutative affine variety
$U_\theta[\sigma]$ to be the subalgebra of $\alg(\torus^2_\theta)$ spanned by
the Laurent monomials $u^{m_a}$, $a=1,\dots,r\geq2$, subject to
relations derived from (\ref{Tthetarels}), where $m_a$ are the
generators of the semigroup $\check\sigma\cap L^*$ and
\bea
\check\sigma=\big\{m\in L_\real^* \ \big| \ \langle m,v\rangle\geq0
\quad \forall v\in\sigma\big\}
\nonumber \eea
is the dual cone to $\sigma$. We will sometimes denote this
noncommutative algebra by $\complex_\theta[\sigma]=
\complex_\theta[u^{m_1},\dots,u^{m_r}]$.

Let us begin by spelling out the noncommutative toric geometry of the
compactified orbit space; we write
$\alg'=\alg\big((\PP^2/\zed_k)_\theta\big)$ for the coordinate
algebra. For the two-cones $\sigma_{0,k}':= \real_{\geq0}v_{0} +
\real_{\geq0}v_k$, $\sigma_{0,\infty}':=\real_{\geq0}v_{0} +
\real_{\geq0}(-v_0-v_k)$, and $\sigma_{k,\infty}' :=\real_{\geq0}v_{k} +
\real_{\geq0}(-v_0-v_k)$, the relations among the generators of the
subalgebras $\complex_\theta[\sigma_{i,j}']\subset \alg(\torus^2_\theta)$ are
as follows:

\begin{itemize}

\item[(O1)] 
The semigroup $\check\sigma_{0,k}'\cap L^*$ is generated over
$\zed_{\geq0}$ by the
lattice vectors $m_1'=(k-1) \,e_1^*+k\, e_2^*$, $m_2'=e_1^*$, and $m_3'=e_1^*+e_2^*$,
with the relation $m_1'+m_2'=k\, m_3'$. The coordinate algebra
$\complex_\theta[\sigma_{0,k}']$ of the noncommutative affine variety
$U_\theta[\sigma_{0,k}']$ is thus the polynomial algebra
$\complex[x,y,z]$ generated over $\complex$ by $x=u^{m_1'}= u_1^{k-1} \, u_2^{k}$,
$y=u^{m_2'}= u_1$, and $z=u^{m_3'}= u_1\, u_2$ subject to the commutation
relations
\bea
x\, y= q^{-2k}\ y\, x \ , \qquad x\, z= q^{-2}\ z\, x \qquad \mbox{and}
\qquad y\, z= q^{2}\ z\, y
\label{xyzorbrels}\eea
together with
\bea
x\, y- q^{k\,(k-3)}\ z^k=0 \ .
\label{xyzkorbrel} \eea
These relations demonstrate that $U_\theta[\sigma_{0,k}']$ is a toric
noncommutative deformation of the $A_{k-1}$ orbifold singularity $\complex^2/\zed_k$; this
generalizes the $k=2$ construction of~\cite[\S3.4]{CLSI}.

\item[(O2)]
The semigroup $\check\sigma_{0,\infty}'\cap L^*$ is generated by
$m_1'=(k-2) \, e_1^*+k\, e_2^*$, $m_2'=e_1^*$, and $m_3'=(k-1)\, e_1^*+k\, e_2^*$ with the
relation $m_1'+m_2'=m_3'$. The coordinate algebra
$\complex_\theta[\sigma_{0,\infty}']$ of
$U_\theta[\sigma_{0,\infty}']$ is generated by $x=u_1^{k-2}\, u_2^{-2}$,
$y=u_1$, and $z=u_1^{k-1}\, u_2^k$ with the commutation relations
\bea
x\, y= q^{-2k}\ y\, x \ , \qquad x\, z= q^{-2k}\ z\, x \qquad \mbox{and}
\qquad y\, z= q^{2k}\ z\, y
\label{O2commrels} \eea
together with
\bea
x\, y-q^{-2k}\ z=0 \ .
\nonumber \eea

\item[(O3)]
The semigroup $\check\sigma'_{k,\infty}\cap L^*$ is generated by exactly the same lattice vectors as in item~(O2), and hence the coordinate algebra $\complex_\theta[\sigma'_{k,\infty}]$ of $U_\theta[\sigma'_{k,\infty}]$ also has the same generators and relations.

\end{itemize}

Away from the orbifold points, we can glue these subalgebras of
$\alg(\torus^2_\theta)$ together via algebra automorphisms in the
braided monoidal category ${}^{\hil_\theta}\Module$ of left
$\hil_\theta$-comodules to form the global noncommutative toric orbit space $(\PP^2/\zed_k)_\theta$~\cite[\S3]{CLSI}. We denote the orbit one-cones by $\tau_0'=\real_{\geq0}v_0$, $\tau_k'=\real_{\geq0}v_k$, and $\tau_\infty'=\real_{\geq0}(-v_0-v_k)$. Their quantization is described as follows:

\begin{itemize}

\item
For the face $\tau_0'=\sigma_{0,k}'\cap \sigma_{0,\infty}'$, the semigroup $\check\tau_0'\cap L^*$ is generated by $m_1'=e_2^*$, $m_2'=e_1^*$, and $m_3'=-e_1^*=-m_2'$. The generators over $\complex$ of the subalgebra $\complex_\theta[\tau_0'\, ]=\complex_\theta[s,t,t^{-1}]$ are $s=u^{m_1'}=u_2$ and $t=u^{m_2'}=u_1$ with the relations
\beq
s\, t= q^{-2}\ t\, s \ , \qquad s\, t^{-1}= q^{2}\ t^{-1}\, s \qquad \mbox{and} \qquad t\, t^{-1}= 1= t^{-1}\, t \ .
\label{stqcommrels}\eeq
These relations show that the noncommutative affine variety $U_\theta[\tau_0'\,]\cong \PP_\theta^1$ is that of a toric noncommutative projective line~\cite{CLSI}.
Recalling the generators of $\complex_\theta[\sigma_{0,k}']$ and $\complex_\theta[\sigma'_{0,\infty}]$, the inclusions $\tau_0'\hookrightarrow \sigma_{0,k}'$ and $\tau_0'\hookrightarrow \sigma_{0,\infty}'$ induce canonical morphisms in ${}^{\hil_\theta}\Module$ of noncommutative algebras $\complex_\theta[\sigma_{0,k}']\to \complex_\theta[\tau_0'\, ]$ and $\complex_\theta[\sigma_{0,\infty}']\to \complex_\theta[\tau_0'\, ]$ which are both natural inclusions of subalgebras.

\item
For the face $\tau_k'=\sigma'_{0,k}\cap \sigma_{k,\infty}'$, the semigroup $\check\tau_k'\cap L^*$ is generated by $m_1'=e_2^*$, $m_2'=e_1^*-k\, e_2^*$, and $m_3'=k\, e_2^*-e_1^*=-m_2'$. The generators of $\complex_\theta[\tau_k'\,]=\complex_\theta[s,t,t^{-1}]$ are $s=u_2$ and $t=u_1\, u_2^{-k}$, again with the commutation relations (\ref{stqcommrels}). There are natural subalgebra inclusion morphisms $\complex_\theta[\sigma_{0,k}']\to \complex_\theta[\tau_k'\,]$ and $\complex_\theta[\sigma_{k,\infty}']\to \complex_\theta[\tau_k'\,]$.

\item
The face $\tau_\infty'=\sigma'_{0,\infty}\cap \sigma_{k,\infty}$ is the orbit ``line at infinity''.
The semigroup $\check\tau'_\infty\cap L^*$ is generated by $m_1'=e_2^*$, $m_2'=(k-2) \, e_1^*+k\, e_2^*$, $m_3'=-(k-2)\, e_1^*-k\,e_2^*=-m_2'$, and $m_4'=-e_2^*=-m_1'$. The generators of the subalgebra $\complex_\theta[\tau'_\infty]=\complex_\theta(s,t)$ are $s=u_2$ and $t=u_1^{k-2}\, u_2^{k}$ with the relations
\beq
s\, t= q^{-2(k-2)}\ t\, s \qquad \mbox{and} \qquad s\, t^{-1} = q^{2(k-2)} \ t^{-1}\, s
\nonumber \eeq
together with
\beq
s\, s^{-1}= 1 = s^{-1}\, s \qquad \mbox{and} \qquad t\, t^{-1}= 1= t^{-1}\, t \ .
\nonumber \eeq
Again there are natural subalgebra inclusion morphisms
$\complex_\theta[\sigma_{0,\infty}']\to \complex_\theta[\tau_\infty']$
and $\complex_\theta[\sigma_{k,\infty}']\to
\complex_\theta[\tau_\infty']$. Note that for $k=2$ the algebra
$\complex_\theta(s,t)$ describes a \emph{commutative} orbit line.

\end{itemize}

Let us now turn to the noncommutative toric resolution; we write
$\bar\alg=\alg\big((\,\overline{X_k}\,)_\theta\big)$ for the dual
coordinate algebra. The generators of the subalgebras
$\complex_\theta[\sigma'_{i,\infty}]$, $i=0,k$, associated to the
two-cones of the orbit compactification divisor $[\ell_\infty/\zed_k]$
are described by items~(O2) and~(O3) above, while those of the
subalgebras $\complex_\theta[\sigma_i]\subset\alg(\torus^2_\theta)$
associated to the two-cones
$\sigma_i=\real_{\geq0}v_{i-1}+\real_{\geq0}v_i$, $i=1,\dots,k$, of
$\overline{X_k}$ are described as follows. The generators of the
semigroup $\check\sigma_i\cap L^*$ over $\zed_{\geq0}$ are $\bar
m_1=-(i-2)\, e_1^*- (i-1)\, e_2^*$ and $\bar m_2=(i-1) \,e_1^*+i\, e_2^*$ for each $i=1,\dots,k$. In this case the algebra $\complex_\theta[\sigma_i]=\complex_\theta[\bar x,\bar y]$ is generated over $\complex$ by $\bar x=u^{\bar m_1}=u_1^{i-2}\, u_2^{i-1}$ and $\bar y=u^{\bar m_2}=u_1^{-(i-1)}\, u_2^{-i}$ with the quadratic relations
\beq
\bar x\, \bar y = q^2\ \bar y \, \bar x \ .
\nonumber \eeq
It follows that the noncommutative affine varieties $U_\theta[\sigma_i]\cong\complex_\theta^2$ for $i=1,\dots,k$ are each copies of the two-dimensional algebraic Moyal plane~\cite{CLSI}.

The ``inner'' one-cones of the toric resolution are denoted $\tau_i:=
\real_{\geq0}v_i=\sigma_i\cap\sigma_{i+1}$ for $i=1,\dots,k-1$. The
semigroup $\check\tau_i\cap L^*$ is generated by $\bar m_1=e_2^*$,
$\bar m_2=e_1^*-i\, e_2^*$, and $\bar m_3= i\, e_2^*-e_1^*=-\bar
m_2$. The generators of the noncommutative coordinate algebra
$\complex_\theta[\tau_i]=\complex_\theta[s,t,t^{-1}]$ are $s=u_2$ and
$t=u_1\, u_2^{-i}$ with the relations (\ref{stqcommrels}) of the
noncommutative projective line $\PP_\theta^1$. Again there are natural
subalgebra inclusion morphisms $\complex_\theta[\sigma_i]\to
\complex_\theta[\tau_i]$ and $\complex_\theta[\sigma_{i+1}]\to
\complex_\theta[\tau_i]$, and in $\complex_\theta[\tau_i]$ a natural
algebra automorphism $\complex_\theta[\sigma_i]\to
\complex_\theta[\sigma_{i+1}]$ is given on generators by
$(u_1,u_2)\mapsto (u_1\, u_2^{-1} ,u_2)$ for each $i=1,\dots,k-1$; this is similar to the automorphisms defining the toric noncommutative projective plane $\PP_\theta^2$~\cite[\S3.3]{CLSI}. The ``outer'' one-cones $\tau_0'=\sigma_1\cap\sigma_{0,\infty}'$ and $\tau_k'=\sigma_k\cap \sigma'_{k,\infty}$, as well as the orbit ``line at infinity'' $\tau'_\infty$, are treated as above.

\subsection{Noncommutative instantons on quotient stacks}

To proceed now with a construction of moduli spaces of noncommutative instantons, we
need a better ``global'' description of these toric deformations as
in~\cite{CLSI,CLSII} which is moreover a quantization of the overarching quotient groupoids. An extension of the
Nekrasov--Schwarz noncommutative ADHM construction~\cite{NS} to the Kleinian
singularities $\complex^2/\zed_k$ was given by Lazariou~\cite{Lazaroiu}. In particular, he
obtains a noncommutative geometry interpretation of their minimal
resolutions, which is natural from the point of view of Yang--Mills matrix models and
the resolution of orbifold singularities by D-branes, and provides
another interpretation of Nakajima's quiver varieties of type~$\widehat{A}_{k-1}$~\cite{Nakajima1}
as moduli spaces of $\zed_k$-equivariant noncommutative instantons. We seek an analogous interpretation for our deformations.

For this, we follow the construction of deformations of Kleinian singularities (and of Hilbert
schemes of points in $\complex^2$) in terms of
multi-homogeneous coordinate algebras which was considered
in~\cite{GS1,GS2,Boyarchenko,KR}; their toric geometry aspects are
studied in~\cite{Musson,BK}. Classically, the quotient stack
$[\complex^2/\zed_k]$, or the corresponding dual coordinate algebra
$\complex[z_1,z_2]\rtimes \zed_k$, is a noncommutative crepant
resolution of the $\complex^2/\zed_k$ orbifold singularity. By the
McKay correspondence, there is an equivalence between
the derived category of coherent sheaves on the minimal crepant resolution
$\Hilb^{\zed_k}(\complex^2)\cong X_k$ of the quotient singularity $\complex^2/\zed_k$ and the derived category of finitely
generated modules over $\complex[z_1,z_2]\rtimes \zed_k$. This equivalence generalizes to the noncommutative setting: One can
construct an algebra which simultaneously gives a noncommutative deformation of
$\complex^2/\zed_k$ and of its minimal resolution of singularities
$\Hilb^{\zed_k}(\complex^2)\to \complex^2/\zed_k$; in contrast to the classical McKay
correspondence, which involves only derived categories, this is given by an equivalence of abelian categories
of modules. 

We shall apply this equivalence to the noncommutative toric deformation of the quotient stack $[\PP^2/\zed_k]$. For this, we recall the construction of the homogeneous coordinate algebra $\alg=\alg(\PP^2_\theta)$ of the noncommutative projective plane $\PP_\theta^2$ from~\cite{CLSI,CLSII,CLSIII}. It is the graded polynomial algebra in three generators $w_i$, $i=0,1,2$ of degree~one with the
quadratic relations
\bea
w_{0}\,w_i= w_i\,w_{0} \qquad \mbox{for} \quad i=1,2 \qquad \mbox{and} \qquad 
w_1\,w_2=q^2~w_2\,w_1 \ .
\label{eq:wirels}\eea
The algebra $\alg$ is an Artin--Schelter regular algebra of global homological dimension three which is naturally an $\hil_\theta$-comodule algebra. Each monomial $w_i$ generates a
left denominator set in $\alg$, and the degree~zero
subalgebra of the left Ore localization of $\alg$ with respect to
$w_i$ is naturally isomorphic to the noncommutative coordinate algebra
of the $i$-th maximal cone in the fan of $\PP_\theta^2$, for each
$i=0,1,2$. The category of coherent sheaves on the noncommutative
projective plane $\PP^2_\theta$ is equivalent to the category of
finitely-generated graded $\alg$-modules.

The cyclic group $\zed_k$ acts by automorphisms of the graded algebra $\alg$
via (\ref{eq:zetaP2action}). The subalgebra of invariants
$\alg^{\zed_k}\subset\alg$ is dual to the quantization of the orbit space
$\PP^2/\zed_k$; it is generated by the central element $w_0$ together with $x:=w_1^k$, $y:=w_2^k$ and $z:=w_1\, w_2$ subject to the commutation relations
\bea
x\, y= q^{2k^2}\ y\, x \ , \qquad x\, z= q^{2k}\ z\, x \qquad \mbox{and} \qquad y\, z=q^{-2k}\ z\, y
\nonumber \eea
along with
\bea
x\, y-q^{k\, (k-1)}\ z^k= 0 \ .
\nonumber \eea
Just as in the commutative case, we can construct the McKay quiver $\sfQ$ associated to this $\zed_k$-action on $\alg$ by assigning vertices to the character group $\{0,1,\dots,k-1\}$ of the cyclic group and arrows $w_i^{(l)}$, $i=0,1,2$, $l=0,1,\dots,k-1$, whose target vertex shifts the source vertex $l$ by the character of the $\zed_k$-eigenvector $w_i$ minimally generating $\alg$; hence $w_0^{(l)}:l\to l$ are vertex loops, while $w_1^{(l)}:l\to l+1$, $w_2^{(l)}:l\to l-1$ and the quiver $\sfQ$ is given by

\medskip
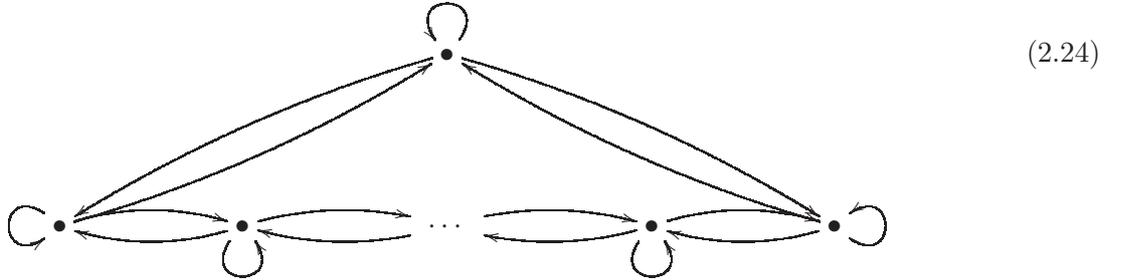
\bea
\xymatrix@C=20mm{
& & {\bullet}\ar@/_/[ddll]\ar@{>}@(ur,ul)[]{} \ar@/^/[ddrr]{} & & \\ & & & & \\
{\bullet}\ar@{>}@(ul,dl)[]\ar@/^/[r]{}\ar@/_/[uurr]{} & {\bullet} \ar@{>}@(dl,dr)[] \ar@/^/[l]{}\ar@/^/[r]{} & \ \cdots
\ \ar@/^/[l]{}\ar@/^/[r]{} & {\bullet}\ar@{>}@(dl,dr)[] \ar@/^/[l]{}\ar@/^/[r]{} & {\bullet}\ar@{>}@(dr,ur)[]\ar@/^/[l]{}\ar@/^/[uull]{}
}
\label{eq:NCquiver} \eea

\bigskip
\noindent
The relations $\sfR$ of the quiver are induced by the relations (\ref{eq:wirels}) of the algebra $\alg$, where composition of arrows makes sense, giving
\bea
w_0^{(l+1)}\, w_1^{(l)} = w_1^{(l)}\, w_0^{(l)} \ , \qquad w_0^{(l-1)}\, w_2^{(l)} = w_2^{(l)}\, w_0^{(l)} \ , \qquad w_1^{(l-1)}\, w_2^{(l)} = q^2\ w_2^{(l+1)}\, w_1^{(l)} \ .
\label{eq:wilrels}\eea
Then the category of linear representations of the McKay quiver with relations $(\sfQ,\sfR)$ is equivalent to the category of $\zed_k$-equivariant $\alg$-modules~\cite{Smith}.

Alternatively, the toric noncommutative deformation of the quotient stack
$[\PP^2/\zed_k]$ is dual to the skew-group algebra
$\alg\rtimes\zed_k$ over the algebra $\alg$, which replaces the algebra
$\alg^{\zed_k}$. It is the free $\alg$-module $
\alg\otimes\complex[\zed_k]$ over the group algebra of the cyclic
group $\zed_k$, with the twisted noncommutative product $(a\otimes
\xi)\cdot(a'\otimes\xi'\,):= \big(a\, (\xi\triangleright a'\, )\big)\otimes(\xi\,\xi'\,)$ for $a,a'\in\alg$ and $\xi,\xi'\in\zed_k$; it has generators $w_i^{(l)}$ with the relations (\ref{eq:wilrels}), and hence the representation category of the quiver $(\sfQ,\sfR)$ is equivalent to the category of modules over $\alg\rtimes\zed_k$. This toric noncommutative deformation is analogous to the deformations of Kleinian singularities considered
in~\cite{HW,MM} in terms of crossed product $C^*$-algebras and the
twisted group algebra of $\zed_k$ associated with a projective regular
representation, which may be characterised in terms of noncommutative gauge
theories and D-branes.

In fact, there is a stronger statement which generalises the
well-known situation in the commutative case~\cite{Smith}: If
$\sfA=\sfA(\sfQ,\sfR)$ denotes the path algebra of the quiver
$(\sfQ,\sfR)$, i.e. the algebra of $\complex$-linear combinations of
paths of the quiver with product defined by path concatenation
whenever paths compose and $0$ otherwise, modulo the ideal
$\langle\sfR\rangle$ generated by the relations (\ref{eq:wilrels}),
then
\bea
\sfA\cong\alg\rtimes\zed_k \ .
\nonumber \eea
In this sense the algebra $\alg\rtimes\zed_k$ plays the role of a
``noncommutative crepant resolution'' of the algebra of invariants
$\alg^{\zed_k}$. Note that localising with respect to the central
element $w_0$ reduces the quiver $\sfQ$ to the standard (unframed)
McKay quiver associated to the Kleinian singularity
$\complex^2/\zed_k$, i.e. the double of the cyclic quiver \eqref{eq:cyclicquiver} of type~$\widehat{A}_{k-1}$, and
the relations $\sfR$ to a $q$-deformation of the moment map \eqref{eq:muADHM} (after framing) defining a Nakajima quiver variety of
type~$\widehat{A}_{k-1}$ which is obtained by replacing the commutator with the \emph{braided commutator} $[b_1,b_2]_\theta:=b_1\,b_2-q^2\, b_2\, b_1$. In this way, a construction of instanton
moduli spaces over the toric noncommutative deformation of
$[\PP^2/\zed_k]$ via a braided version of the ADHM construction should go through exactly as in~\cite{CLSII}, and as
in~\cite{CLSII,CLSIII} the corresponding equivariant noncommutative
instanton partition functions for $\cN=2$ gauge theory should coincide with their classical
counterparts. It would be interesting to study further the properties
of these new moduli spaces which are (commutative) deformations of the
usual quiver varieties, as well as their relations with the
noncommutative quiver varieties obtained via deformation quantization
of their standard symplectic structure.

It is much more difficult to construct an analogous quantization of
the resolution $\big[\, \overline{X_k}\, \big] $ in terms of
homogeneous coordinate algebras. The stacky fan $\bar{\mbf\Sigma}_k=(L,\bar\Sigma_k,\bar\beta_k)$ yields a description of $\big[\, \overline{X_k}\, \big] $ as a global quotient stack in the following way~\cite{BCS} (see also~\cite{Bruzzo:2013daa}). There is a short exact sequence
\bea
0 \ \longrightarrow \ L^* \ \xrightarrow{\ \bar\beta_k^* \ } \ \big(\zed \bar b_1\oplus\cdots\oplus\zed\bar b_{k+2} \big)^* \ \xrightarrow{ \ \bar\beta_k^\vee \ } \ \Pic\big(\, \overline{X_k}\, \big) \ \longrightarrow \ 0
\nonumber \eea
from which one can represent the Gale dual $\bar\beta_k^\vee:(\IZ^{k+2})^*\to {\rm coker}(\bar\beta_k^*)=\IZ^k \cong\Pic\big(\, \overline{X_k}\, \big)$ of the map $\bar\beta_k$. Applying the exact functor $\Hom_{\IZ}(-,\IC^\times)$ yields the exact sequence
\bea
1 \ \longrightarrow \ G_k \ \longrightarrow \ (\IC^\times)^{k+2} \ \longrightarrow \ \torus^2 \ \longrightarrow 1 \ ,
\nonumber \eea
where $G_k=\Hom_{\IZ}\big({\rm coker}(\bar\beta_k^*),\IC^\times\big)$. Then the toric orbifold $\big[\, \overline{X_k}\, \big] $ is isomorphic to the stack quotient
\bea
\big[\, \overline{X_k}\, \big] \cong \big[Z_k\,\big/\, G_k\big] \ ,
\nonumber \eea
where $Z_k\subset\IC^{k+2}$ is the union over $(z_1,\dots,z_{k+2})\in\IC^{k+2}$ with pairwise neighbouring entries $(z_i,z_{i+1})\in\torus^2$ for $i=1,\dots,k+2$ (${\rm mod}\ k+2$), and the abelian affine algebraic group $G_k\subset(\IC^\times)^{k+2}$ consists of torus points $(t_1,\dots,t_{k+2})$ satisfying the relations
\bea
t_2\, t_3^2\cdots t_{k+1}^k=t_{k+2}^k= t_1\, t_3\cdots t_{k+1}^{k-1}
\nonumber \eea
and acting on $Z_k$ via the standard scaling action of $(\IC^\times)^{k+2}$ on $\IC^{k+2}$; the coarse moduli space $\overline{X_k}$ is isomorphic to the GIT quotient $Z_k/G_k$. It is unclear whether or not the corresponding noncommutative coordinate algebras will have nice smoothness properties, or even how to deal with their higher homological dimension.
For $k=2$, isospectral deformations of the minimal resolution of the $\complex^2/\zed_2$ orbifold singularity are
studied in~\cite{Yang} by regarding it as an Eguchi--Hanson space,
i.e. the total space $\cO_{\PP^1}(-2)$ of the canonical line bundle
over $\PP^1$.

\bigskip

\section{$\cN=(2,0)$ superconformal theories in six dimensions\label{sec:20theory}}

We now move on to the celebrated conjectural relations between certain
two-dimensional conformal field theories and four-dimensional $\cN=2$
gauge theories; from a mathematical perspective these dualities generally
assert the existence of geometric highest weight representations of
vertex algebras on the cohomology of moduli spaces of torsion-free sheaves. At
present there are explicit results available only for the examples
discussed in \S\ref{sec:braidedsym}, so we focus our discussion to an
overview of some of the results that have been obtained in those
instances. We begin with the physical background to these conjectures
as it is worthy of a brief account in order to understand properly the set-ups and origins of some of the statements.

\subsection{$\cN=2$ field theories of class $\cS$}

Let $\Gamma$ be a finite subgroup of $SU(2)$. By the McKay
correspondence, $\Gamma$ can be identified with the Dynkin diagram of
ADE type of a simply-laced Lie algebra $\frg$; let $G$ be the
corresponding connected Lie group. The group $\Gamma$ acts naturally on $\IC^2$
(viewed as the fundamental representation of $SU(2)$), and
isometrically on the standard round three-sphere $S^3\subset
\IC^2$. We consider Type IIB string theory on the ten-manifold
$M\times \IC^2/\Gamma$ which is the product of a Riemannian
six-manifold $M$ and the ADE singularity $\IC^2/\Gamma$. Its
reduction in the limit where $S^3/\Gamma$ is collapsed to a point
defines an $\cN=(2,0)$ supersymmetric theory on $M$ which is
conformally invariant~\cite{Witten:1995zh}. The six-dimensional $(2,0)$ theory is not a
gauge theory but rather a quantum theory of gerbes (instead of
principal bundles) with connection, and it has no known Lagrangian description. For the
A-series, which we shall primarily focus on in this section, it provides the worldvolume
description of M5-branes in M-theory~\cite{Strominger:1995ac}. Its
partition function is a section of a vector bundle of rank $r>1$ over the
moduli space of field configurations, i.e. the theory has a vector
space of possible partition functions (instead of a unique one) called
the space of \emph{conformal blocks}. This is completely analogous to
the well-known situation in two-dimensional conformal field theory, where partition functions are sections of Friedan--Shenker bundles over moduli spaces of punctured Riemann surfaces.

Consider now the topologically twisted $(2,0)$ theory on a six-manifold which is a product
\bea
M = X\times C \ ,
\nonumber \eea
where as before $X$ is a complex toric surface, and $C$ is a Riemann
surface of genus $g$ with $n$ marked points carrying labels
$(\rho_i,\mu_i)$ for $i=1,\dots,n$; here $\rho_i:\frsu_2\to\frg$ are
Lie algebra homomorphisms and $\mu_i\in\IC$ are ``mass''
parameters. Using conformal invariance one can consider two equivalent
reductions of this theory. On the one hand, the limit where $C$ collapses to a point defines a four-dimensional quantum field theory on $X$ which, in addition to the labels $(\rho_i,\mu_i)$, depends only on the complex structure of the punctured Riemann surface $C$ and its area. When $\mu_i=0$ for $i=1,\dots,n$, this is an $\cN=2$ supersymmetric gauge theory with structure group $\prod_{i=1}^n\, G^{\rho_i}$ where $G^{\rho_i}$ is the centralizer of $\rho_i$ in $G$; these are the $\cN=2$ gauge theories in four dimensions of class $\cS$~\cite{Gaiotto:2009we,Gaiotto:2009hg}. On the other hand, the limit where $X$ collapses to a point defines a two-dimensional theory on $C$, and therefore we see an equality between a four-dimensional and a two-dimensional quantum field theory. 

While this relation seems mysterious from a purely two-dimensional or
four-dimensional perspective, it is not surprising from the
six-dimensional point of view. The four-dimensional field theory of class~$\cS$ is completely determined by the Riemann surface $C$ and its
punctures, and when it has $\cN=2$ superconformal invariance the
two-dimensional theory on $C$ is a conformal field theory; the chiral
fields on $C$ arise as the zero modes of the $(2,0)$ tensor multiplet
in six dimensions. In this instance the Seiberg--Witten curve
$\Sigma$, which captures the low-energy dynamics
of the $\cN=2$ gauge theory~\cite{art:seibergwitten1994-I}, is a branched cover of $C$; for example,
when $X=\IC^2$ and $\frg=\frsl_2$, $\Sigma\subset T^*C$ is a double
cover of $C$ of genus $h=3g-3+n$. The geometry of $\Sigma$ is completely encoded by the Seiberg--Witten prepotential $\cF_{X}(\vec a\, ; \sfq)$ which determines its periods
$$
\tau_{ll'} = -\frac1{2\pi\ii}\
\frac{\partial^2\cF_{X}}{\partial a_l\, \partial a_{l'}}(\vec
a\,;\sfq) \ ,
$$
and it describes the low-energy
effective action of the $\cN=2$ gauge theory. Different canonical homology bases for $H_1(\Sigma,\IZ) / H_1(C,\IZ)$ are related by $Sp(2h,\IZ)$ modular transformations which describe S-dualities in the low-energy physics. The four-dimensional perspective
lends a multitude of new tricks and insights into two-dimensional
conformal field theories, and conversely the two-dimensional
perspective can naturally account for many gauge theoretic phenomena
such as S-duality which is manifested as modular invariance of
conformal field theory correlation functions.

More generally, Nekrasov's partition function can be defined and
studied for \emph{any} theory of class~$\cS$, and it can be regarded
as defining a two-dimensional holomorphic quantum field theory on the
Riemann surface $C$. On general grounds, any $\cN=2$ gauge theory in
four dimensions can be naturally associated to an algebraic integrable system,
called the Donagi--Witten integrable system~\cite{Martinec:1995by,Donagi:1995cf}, such that the
Seiberg--Witten curve $\Sigma$ coincides with the spectral curve of an
associated Hitchin system on $C$. The $\Omega$-deformed theories of
class $\cS$ are quantizations of these Hitchin systems which
are obtained in the limit $\epsilon_1,\epsilon_2\to0$. For $X=\IC^2$
this perspective was used in~\cite{Nekrasov:2009rc} to conjecturally
capture the spectrum of a quantum integrable system in the limit
$\epsilon_2\to0$ of the $\Omega$-deformation; we return to this
description in~\S\ref{sec:N26D}.

This construction is at the heart of recent 2d/4d dualities, and in
particular the AGT and BPS/CFT correspondences. In this setting the
(localized) $\widetilde{\torus}$-equivariant cohomology $\Hscr_N(X)$ of the instanton
moduli space $\frM_N(X)=\bigsqcup_{n\geq0}\, \frM_{N,n}(X)$
conjecturally carries a geometric highest weight representation of the vertex
algebra $\alg_N(X)$ underlying the conformal field theory on $C$. The
Nekrasov partition functions for pure $\cN=2$ gauge theory on $X$ are
conjecturally given as geometric inner products involving a Whittaker vector (or irregular
conformal block) for this representation, while the Nekrasov functions
for $\cN=2$ quiver gauge theories on $X$ should coincide with
conformal blocks on $C$. In the remainder of this section we summarize
the current state of affairs for some of these results in the classes of examples considered in~\S\ref{sec:braidedsym}.

\subsection{AGT duality on $\IC^2$}

Let us look first at the case $X=\IC^2$, where the holomorphic field
theory on $C$ is believed to be the theory of $\cW({\frgl}_N)$
conformal blocks (see \S\ref{app:Walgebras}). Consider the infinite-dimensional $\IZ_{\geq0}$-graded vector space
\bea
\Hscr_N(\IC^2) =\bigoplus_{n=0}^\infty\,\Hscr_{N,n}(\IC^2) := \bigoplus_{n=0}^\infty \, H_{\widetilde{\torus}}^\bullet\big(\frM_{N,n}(\IC^2)\big)_{\rm loc} \ ,
\nonumber \eea
where the subscript designates the \emph{localized} equivariant
cohomology which is the corresponding $\IF$-vector space, with $\IF$
the quotient field of the coefficient ring
$H_{\widetilde{\torus}}^\bullet({\rm pt})$. Let $\iota_{\vec Y}$
denote the inclusion in $\frM_{N,n}(\IC^2)$ of the fixed point parameterized by the
$N$-vector of Young diagrams $\vec Y$, and
define distinguished classes $[\vec Y]:=\iota_{\vec Y*}(1)$ in
$H_{\widetilde{\torus}}^\bullet\big(\frM_{N,n}(\IC^2)\big)$. Using the
projection formula 
\bea
[\vec Y]\cup [\vec Y{}'\,] = \delta_{\vec Y,\vec Y{}'} \ \Euler\big(T_{\vec Y}\frM_{N,n}(\IC^2)\big)^{-1} \ [\vec Y]
\nonumber \eea
for the cup product in equivariant cohomology, one
defines a non-degenerate symmetric bilinear form $\langle-,-\rangle:\Hscr_N(\IC^2)
\times\Hscr_N(\IC^2) \to\IF$ such that
\bea
\big\langle[\vec Y]\,,\,[\vec Y{}'\,]\big\rangle = \delta_{\vec Y,\vec Y{}'} \ \Euler\big(T_{\vec Y}\frM_{N,n}(\IC^2)\big)^{-1}
\label{eq:Yorthogonal} \eea
on $\Hscr_{N,n}(\IC^2)\times \Hscr_{N,n}(\IC^2)$.
By the localization theorem, it thus follows that the collection of vectors $[\vec Y]$ for all $n\geq0$ forms an orthogonal $\IF$-basis for $\Hscr_N(\IC^2)$. In particular, the equivariant fundamental class of the moduli space $\frM_{N,n}(\IC^2)$ in $H_{\widetilde{\torus}}^\bullet\big(\frM_{N,n}(\IC^2)\big)$ is given by $[\frM_{N,n}(\IC^2)]=\sum_{\vec Y}\, [\vec Y]$.
\begin{theorem} \ (a) The vector space $\Hscr_N(\IC^2)$ is the Verma
  module for the $\cW({\frgl}_N)$-algebra of central charge
\bea
c=N+(N-1)\, N\, (N+1)\, \varepsilon^2 \qquad \mbox{with} \quad \varepsilon^2:=
\frac{(\epsilon_1+ \epsilon_2)^2}{\epsilon_1\, \epsilon_2}
\nonumber \eea
and highest weight $\vec\lambda= \frac{\vec a}{\sqrt{\epsilon_1\,
    \epsilon_2}}+\varepsilon \, \vec\rho$ where $\vec\rho$ is the Weyl vector of
$\frgl_N$.
\\[1mm]
(b) The vector
\bea
\psi_{\IC^2}=\sum_{n=0}^\infty\, \big[\frM_{N,n}(\IC^2)\big]
\nonumber \eea
in the completion $\prod_{n\geq0}\,\Hscr_{N,n}(\IC^2)$ is a Whittaker vector for this highest weight representation.
\end{theorem}
Part (a) of this theorem generalizes the constructions of
Nakajima~\cite{art:nakajima1997,book:nakajima1999} and
Vasserot~\cite{art:vasserot2001} of Fock space
representations, using geometric Hecke correspondences on the (equivariant) cohomology of Hilbert schemes
$\Hilb^n(\IC^2)$, of the Heisenberg algebra
$\alg_1(\IC^2)=\frh=\widehat{\frgl}_1$ in the rank one case $N=1$;
this is the vertex algebra underlying the $c=1$ free boson conformal
field theory. For $N>1$ the theorem was proved independently by
Schiffmann--Vasserot~\cite{art:schiffmannvasserot2013} and by
Maulik--Okounkov~\cite{art:maulikokounkov2012}. The vertex algebra
$\alg_N(\IC^2)=\frh\oplus \cW(\frsl_N)$ underlies the $A_{N-1}$ Toda
conformal field theories; for $N=2$ the $\cW(\frsl_2)$-algebra is the
Virasoro algebra underlying Liouville theory. The technical
difficulties involved in the proofs is the lack of presentation of
generic $\cW(\frsl_N)$-algebras in terms of generators and relations
(see \S\ref{app:Walgebras}), which are overcome by realising the induced action
of the vertex algebra $\cW(\frgl_N)$ via its embedding into a larger
infinite-dimensional Hopf algebra. The
Whittaker vector $\psi_{\IC^2}$ is sometimes also called a \emph{Gaiotto
  state}~\cite{art:gaiotto2009}; it plays the role of a kind of
``coherent state'' for the geometric highest weight representation.

This theorem is regarded as providing a rigorous proof of the AGT
conjectures~\cite{art:aldaygaiottotachikawa2010,art:wyllard2009,art:aldaytachikawa2010}
for pure $\cN=2$ gauge theory on $\IC^2$ in the following
sense. Define an operator $\sfq^{L_0}$ on $\Hscr_N(\IC^2)$ by letting
it act as scalar multiplication by $\sfq^n$ on each graded component
$\Hscr_{N,n}(\IC^2)$ for $n\geq0$. It then follows immediately from
the projection formula \eqref{eq:Yorthogonal} that Nekrasov's
partition function (\ref{eq:NekinstC2comb}) for $\cN=2$ gauge theory
on $\IC^2$ takes the form of a conformal block
\bea
\cZ_{\IC^2}^{\rm inst}(\epsilon_1,\epsilon_2, \vec a\,;\sfq) =
\big\langle \psi_{\IC^2} \,,\, \sfq^{L_0}\psi_{\IC^2} \big\rangle \ .
\nonumber \eea

Via term by term matching of explicit expressions, the partition
functions of $\cN=2$ quiver gauge theories on $\IC^2$ have been found
to agree with the conformal blocks of $A_{N-1}$ Toda field
theories~\cite{art:aldaygaiottotachikawa2010,art:wyllard2009}. A
rigorous proof of this correspondence requires a geometric
construction of chiral vertex operators whose suitable matrix elements
on $\Hscr_N(\IC^2)$ coincide with the Nekrasov partition functions. It
is always possible to define such operators in terms of the Euler
classes of Carlsson--Okounkov type
Ext-bundles~\cite{art:carlssonokounkov2012} which are defined using
the universal sheaf on $\frM_{N,n}(\IC^2)\times\PP^2$; they have
completely factorised matrix elements in the fixed point basis $[\vec
Y]$ of $\Hscr_N(\IC^2)$. However, a characterization of these
operators as primary fields of the $\cW(\frgl_N)$-algebra is currently
not known in generality. For $N=1$, this construction produces vertex
operators which are primary fields of the Heisenberg algebra $\frh$ in
terms of bosonic exponentials of Nakajima
operators~\cite{art:carlssonokounkov2012}. Hence the AGT conjecture is
completely proven in the rank one case. For $N=2$ it was shown
by~\cite{Fateev:2009aw,Hadasz:2010xp} that these matrix elements
correspond to primary field insertions in conformal blocks. One of the
main interests in the AGT correspondence from the perspective of
conformal field theory is in fact the existence of the
distinguished basis $[\vec
Y]$ for the Verma module which has no natural origin from the purely two-dimensional
point of view.

For the pure $\cN=2$
gauge theory on $X=\IC^2$, the instanton part of the Seiberg--Witten prepotential is
given by~\cite{art:nekrasov2003,NakYosh,art:nekrasovokounkov2006}
\bea
\cF_{\IC^2}^{\rm inst}(\vec a\, ; \sfq) = \lim_{\epsilon_1, \epsilon_2\to0}\, -\epsilon_1\, \epsilon_2\, \log
\cZ_{\IC^2}^{\rm inst}(\epsilon_1,\epsilon_2, \vec a\,;\sfq) \ ,
\nonumber \eea
where the normalization is the equivariant volume of $X=\IC^2$,
i.e. $\int_{X}\, 1 = \frac1{\epsilon_1\, \epsilon_2}$. The
Donagi--Witten integrable system in this case is the twisted Toda
integrable system of type $A_{N-1}$, which is a Hitchin system on the
Riemann sphere $C=\PP^1$ with two irregular punctures at
$z=0,\infty$. The Whittaker vector $\psi_{\IC^2}$ is a state in the
representation corresponding to an irregular puncture; ``irregular''
here refers to wild singularities of the Seiberg--Witten spectral
curve $\Sigma$ corresponding to non-conformal behaviour of the
conformal blocks in this case.

In the $\cN=2$ gauge theory, the vector multiplet scalars $\boldsymbol I_p= \Tr\phi^p$ for all $p\geq 1$ give an infinite of commuting Hamiltonians underlying the Donagi--Witten integrable system. The quantization of this integrable system provided by the $\Omega$-deformation is realised by the action of an infinite-dimensional commutative algebra of integrals of motion acting in the representation $\Hscr_N(\IC^2)$, which are diagonalised in the fixed point basis $[\vec Y]$. They are realised \emph{geometrically} on $\Hscr_N(\IC^2)$ as cup multiplication by the Chern classes $\boldsymbol{I}_p= c_{p-1}(\boldsymbol V^N)$ of the \emph{natural vector bundle} $\mbf V^N$ on $\frM_N(\IC^2)$~\cite{Smirnov:2013hh}; they can be read off from the $\widetilde{\torus}$-equivariant Chern character
\bea
\Ch\big(\mbf V^N_{\vec Y}\big) = \sum_{l=1}^r \ \sum_{s\in Y_l}\, \e^{-L^t(s)\, \epsilon_1- A^t(s)\, \epsilon_2}
\nonumber \eea
at the fixed point $[\vec Y]$. In the rank one case $N=1$, they can be
identified with Nakajima operators which span a commutative subalgebra
of the universal enveloping algebra of the Heisenberg algebra $\frh$;
in particular, this identifies $\mbf I_1$ with the energy operator
$L_0$ and $\mbf I_2$ with the bosonised Hamiltonian of a quantum trigonometric Calogero--Sutherland model with infinitely many particles and coupling constant $\beta^{-1}$. In fact, a stronger statement is true for any $N\geq1$: With $\Lambda_{\IF}:= \Lambda\otimes_{\IC}\IF$ the algebra of symmetric functions over the field $\IF$ (see \S\ref{sec:symmetric}), there is an isomorphism $\Hscr_N(\IC^2)\cong \Lambda^{\otimes N}_{\IF}$, as representations of the $\cW(\frgl_N)$-algebra, which sends
\bea
[\vec Y] \ \longmapsto \ J_{\vec Y} \ .
\nonumber \eea
Here $J_{\vec Y}$ are the generalized Jack symmetric functions which are the eigenfunctions of a quantum deformation of the Calogero--Moser--Sutherland integrable system~\cite{Alba:2010qc,Morozov:2013rma,Smirnov:2014npa}; for $N=1$ they coincide with the usual Jack functions $J_Y(x;\beta)$, see \S\ref{sec:uglovfunctions}.

\subsection{AGT duality on $[\IC^2/\IZ_k]$\label{sec:AGTonALE}}

Next we consider the quotient stack $X=[\IC^2/\IZ_k]$. Then the $\widetilde{\torus}$-equivariant cohomology $\Hscr_{\vec w}\big([\IC^2/\IZ_k]\big)$ of the corresponding quiver variety $\frM_{\xi_0}(\vec w\,):= \bigsqcup_{\vec v\in\IZ_{\geq0}^k} \, \frM_{\xi_0}(\vec v,\vec w\,)$ for $\xi_0\in \sfC_0$ conjecturally carries a representation of the coset construction \cite{art:belavinfeigin2011,art:nishiokatachikawa2011,art:belavinbelavinbershtein2011,art:wyllard2011} 
\begin{eqnarray*}
\mathcal{A}_N\big([\IC^2/\IZ_k]\big) \ := \ \frac{\big(\,
  \widehat{\frgl}_{\kappa}\big)_N}{\big(\, \widehat{\frgl}_{{\kappa}-k}\big)_N} \ \cong \ \mathfrak{h}\oplus\big(\,\widehat{\mathfrak{sl}}_k\big)_N\oplus \frac{\big(\, \widehat{\mathfrak{sl}}_N\big)_k\oplus\big(\, \widehat{\mathfrak{sl}}_N\big)_{{\kappa}-k}}{\big(\,\widehat{\mathfrak{sl}}_N\big)_{\kappa}} \ ,
\end{eqnarray*}
where ${\kappa}\in\IC$ is given in terms of the equivariant parameters $(\epsilon_1,\epsilon_2,\vec a\,)$ and the isomorphism is a consequence of level-rank duality.
This conjecture generalises Nakajima's seminal
construction~\cite{Nakajima1} of level $N$ representations of the affine Lie
algebra $\widehat{\mathfrak{gl}}_k$ on the cohomology of
$\widehat{A}_{k-1}$ quiver varieties $\frM_\xi(\vec v,\vec w\,)$ via
geometric Hecke correspondences. A gauge theory realisation of
Nakajima's construction was provided by Vafa and
Witten~\cite{art:vafawitten1994} in the framework of a topological twisting of $\cN=4$
gauge
theory in four dimensions, which is defined as the dimensional reduction of
ten-dimensional $\cN=1$ supersymmetric Yang--Mills theory on the trivial complex vector bundle of rank three in the
limit where the fiber collapses to a point. They showed that the corresponding
partition functions, which are generating
functions for the Euler characteristics of instanton moduli spaces,
compute characters of highest weight representations of
$\widehat{\mathfrak{gl}}_k$ at level $N$ (see also~\cite{Dijkgraaf:2007fe}); a rigorous derivation, based
on realising the instanton moduli space as the quiver variety $\frM_{\xi_0}(\vec w \,)$, is given in~\cite{art:fujii2005}.

For $N=1$, a vertex algebra realization of the $j$-th fundamental representation of
$\mathcal{A}_1\big([\IC^2/\IZ_k]\big)= \mathfrak{h}\oplus\big(\,
\widehat{\mathfrak{sl}}_k\big)_1$ on the equivariant cohomology of the
quiver variety $\frM_{\xi_0}(\vec w_j)$ is given
in~\cite{art:nagao2009}; this is the chiral algebra underlying the
$U(k)$ WZW conformal field theory at level one. For $N=2$, the algebra
$\mathcal{A}_2\big([\IC^2/\IZ_k]\big)$ is the sum of
$(\,\widehat{\mathfrak{gl}}_k)_2$ and the $\cN=1$ supersymmetric
extension of the Virasoro algebra; for $k=2$ this is the vertex
algebra underlying $\cN=1$ supersymmetric Liouville theory. For
general $N$ and $k$, the pertinent two-dimensional conformal field
theory is the $\IZ_k$-parafermionic Toda field theory of type
$A_{N-1}$. Beyond these results and some explicit checks (see e.g.~\cite{Ito:2011mw,Alfimov:2011ju,Tan:2013tq}), not much is
rigorously proven at present for the AGT conjecture on the quotient
stack $[\IC^2/\IZ_k]$. One line of attack has been to take the Uglov limit of the related five-dimensional gauge theory discussed in \S\ref{sec:5D}, wherein a $q$-deformed vertex algebra conjecturally acts on the $\widetilde{\torus}$-equivariant
K-theory of the instanton moduli space
$\frM_{N}(\IC^2)$~\cite{Awata:2011dc}; this algebra is called the
elliptic Hall algebra (among various other names). It was conjectured by~\cite{art:belavinbershteintarnopolsky2013} that the Uglov limit of the level $N$ representation of the elliptic Hall algebra tends to the conformal
algebra $\mathcal{A}_N\big([\IC^2/\IZ_k]\big)$ preserving the special
fixed point bases where the vertex operators have completely factorised
forms; checks of this proposal can be found in
e.g.~\cite{Itoyama:2013mca,Spodyneiko:2014qsa}. In the rank one case
$N=1$, the geometric action of the elliptic Hall algebra is
constructed using Hecke correspondences in~\cite{SVHall} and vertex operators in~\cite{Feigin}, while the
K-theory versions of the vertex operators geometrically defined using
Ext-bundles that compute matrix elements in $q$-deformed conformal
field theory are  constructed in~\cite{Carlsson:2013jka} and shown to
have a bosonic exponential form in terms of deformed Heisenberg
operators as in~\cite{Awata:2011dc}.

We can also connect this duality to quantum integrability. For $N=1$, there are isomorphisms $\Hscr_{\vec w_j}\big([\IC^2/\IZ_k]\big)\cong\Lambda_{\IF}$ for each $j=0,1,\dots,k-1$, as level one representations of the affine algebra $\widehat{\mathfrak{gl}}_k$, which sends
\bea
[Y] \ \longmapsto \ U_Y(x;\beta,k) \ .
\nonumber \eea
Here $U_Y(x;\beta,k)$ are the Uglov symmetric functions of rank $k$ associated with the coloured Young diagram $Y$ (see \S\ref{sec:uglovfunctions}), which are the eigenfunctions of a spin generalization of the Calogero--Sutherland model~\cite{art:uglov1998}. The first few integrals of motion are given in~\cite{art:belavinbershteintarnopolsky2013}.

\subsection{AGT duality on $X_k$\label{sec:AGTXk}}

We finally turn to the minimal resolution $X=X_k$. As a first check, let us compute the Vafa--Witten partition function in the chamber $\sfC_\infty$ which is the generating function
\bea
\cZ^{\rm VW}_{X_k}\big(\sfq, \vec{\xi}\ \big)_{\vec w}:=
\sum_{\vec u\in \cU_{\vec w}} \, \vec{\xi}^{\ \vec{u}} \ \sum_{\Delta\in \frac{1}{2\,N\, k}\, \mathbb{Z}}\,
  \sfq^{\Delta+\frac{1}{2N}\, \vec{u}\cdot C^{-1} \vec{u}} \ 
  \int_{\frM_{\vec{u},\Delta,\vec w}(X_k)} \, \Euler\big(T\frM_{\vec{u},\Delta,\vec w}(X_k) \big)
\nonumber \eea
for the Euler characteristic of the moduli space $\frM_{\vec w}(X_k):= \bigsqcup_{\vec u,\Delta}\, \frM_{\vec{u},\Delta,\vec w}(X_k)$. Applying the localization theorem, the Euler classes cancel between the denominator and numerator, and the resulting expression simply enumerates fixed points $(\vec{\boldsymbol{Y}}, \vec{\boldsymbol{u}} )$. Analogously to the rank one computation of \S\ref{sec:N2Xk}, the weighted combinatorial sum gives~\cite{Bruzzo:2013daa}
\bea
\cZ^{\mathrm{VW}}_{X_k}\big(\sfq, \vec{\xi}\
\big)_{\vec w} = 
\prod\limits_{j=0}^{k-1}\, \Big(\, \frac{\chi^{\widehat{\omega}_{j}}(\sfq, \vec{\xi}\
\big)}{\widehat{\eta}(\sfq)}\, \Big)^{w_j}
\label{eq:VWXk} \eea
which is the character of the affine Lie algebra $\widehat{\frgl}_k$
at level $N$ expected from Nakajima's highest weight representations
on the cohomology of quiver varieties $\frM_\xi(\vec v,\vec w\,)$. The
formula (\ref{eq:VWXk}) formally includes the case $k=1$, where it reduces
to the Vafa--Witten partition function
\bea
\cZ^{\mathrm{VW}}_{\IC^2}(\sfq)_{N} = \widehat{\eta}(\sfq)^{-N}
\label{eq:VWC2} \eea
for $\cN=4$ $U(N)$ gauge theory on $\IC^2$.

This result also confirms the $SL(2,\IZ)$ modularity (S-duality) of the $\cN=4$ gauge theory partition function, which has a natural explanation in the $(2,0)$ theory on the product Riemannian six-manifold
\bea
M= X_k\times T^2 \ ,
\nonumber \eea
where the real torus $T^2$ has complex structure modulus $\tau$, interpreted as the complexified gauge coupling constant, and
local coordinates $(x,y)\in \IR^2$. For instance, in the abelian case
$N=1$ the $(2,0)$ theory governs a self-dual closed real
three-form $H\in \Omega_{\IZ}^3(M)$ with integral periods which is the curvature of an abelian gerbe with connection. It can be parameterized as
\bea
H=F\wedge \dd x+*F\wedge \dd y \ ,
\nonumber \eea
where $F\in\Omega_{\IZ}^2(X_k)$ is regarded as the curvature two-form of a
complex line bundle over $X_k$ with connection. The limit in which the torus $T^2$ collapses to a point thereby sends the six-dimensional abelian gerbe theory on $M$ to four-dimensional $U(1)$ gauge theory on $X_k$. The S-duality symmetry of the quantum gauge theory is induced geometrically in the quantum gerbe theory by the automorphism group of $T^2$ which acts as $SL(2,\IZ)$ modular transformations of $\tau$. For the higher rank case $N>1$, we first take the limit where the torus $T^2$ collapses to a circle to decribe the non-abelian gerbe theory as five-dimensional $\cN=1$ $U(N)$ gauge theory on the product $X_k\times S^1$, and then further collapse the remaining circle $S^1$ to get the $U(N)$ Vafa--Witten theory on $X_k$; see~\cite{Witten:2009at} for further details and explanations.

Using blowup formulas, for $N= 2$ the partition functions of $\cN=2$
gauge theories on $X_2$ have been matched with conformal blocks of
$\cN=1$ supersymmetric Liouville
theory~\cite{art:bonellimaruyoshitanzini2011,art:bonellimaruyoshitanzini2012}, while a representation theoretic interpretation of the blowup equations in this case has been provided in terms of vertex operator algebras and conformal blocks by~\cite{Bershtein:2013oka,Bershtein:2014yia}. But beyond this very little is known about the AGT correspondence in generality for Nakajima quiver varieties with stability parameters from the chamber $\sfC_\infty$. For $N=1$ we can say a bit more. We note first of all that the localized equivariant cohomology of the moduli spaces $\frM_{\vec w_j}(X_k)$ can be decomposed as
\bea
\Hscr_{\vec w_j}(X_k) = \bigoplus_{\vec u\in\cU_{\vec w_j}}\, \Hscr_{\vec u, \vec w_j}(X_k) \qquad \mbox{with} \quad \Hscr_{\vec u, \vec w_j}(X_k) \cong \bigoplus_{n=0}^\infty \, H_{\widetilde{\torus}}^\bullet\big({\sf Hilb}^n(X_k) \big)_{\rm loc} \, \otimes \, \IC(\epsilon_1,\epsilon_2)[\cQ+\omega_j] \ .
\nonumber \eea
\begin{theorem} \ (a) The vector space $\Hscr_{\vec w_j}(X_k)$ is the $j$-th fundamental representation of $\widehat{\mathfrak{gl}}_k$ at level one with weight spaces $\Hscr_{\vec u, \vec w_j}(X_k)$ of highest weight $\gamma_{\vec u}+\omega_j$.
\\[1mm]
(b) The vector space $\Hscr_{\vec u, \vec w_j}(X_k)$ is the Verma module for the Virasoro algebra with central charge $c=k$ and conformal dimension $\Delta_{\vec u}=\frac12\, \vec u\cdot C^{-1}\vec u$.
\\[1mm]
(c) The vector
\bea
\psi_{X_k}=\sum_{n=0}^\infty \ \sum_{\vec u\in\cU_{\vec w_j}} \, \big[\frM_{\vec u,n,\vec w_j}(X_k)\big]
\nonumber \eea
in the completion $\prod_{n,\vec u}\,\Hscr_{\vec u,n,\vec w_j}(X_k)$ is a Whittaker vector for this highest weight representation.
\end{theorem}
This geometric action of $\alg_1\big([\IC^2/\IZ_k] \big)=\big(\,\widehat{\frgl}_k\big)_1$ is constructed in~\cite{Pedrini:2014yoa} by using Nakajima correspondences for the affine algebra $\mathfrak{h}\oplus\mathfrak{h}_\cQ$ on the equivariant cohomology of the Hilbert schemes ${\sf Hilb}^n(X_k)$, and then applying the Frenkel--Kac construction to the Heisenberg algebra $ \mathfrak{h}_\cQ$ associated to the $A_{k-1}$ root lattice $\cQ$. At present, however, the geometric relationship between the level one $\widehat{\frgl}_k$-modules $\Hscr_{\vec w_j}\big([\IC^2/\IZ_k]\big)$ and $\Hscr_{\vec w_j}(X_k)$ is not understood.

The underlying Seiberg--Witten geometry is found by computing the limit~\cite{Bruzzo:2013daa}
\bea
\lim_{\epsilon_1,\epsilon_2\to 0}\, - {k}\, \epsilon_1\, \epsilon_2\, \log
\cZ_{X_k}^{\rm inst}(\epsilon_1,\epsilon_2,\vec{a}\, ; \sfq, \vec{\xi} \ )_{\vec w} =
\frac{1}{k} \, \cF_{\IC^2}^{\rm inst}(\vec{a}\, ; \sfq) \ .
\nonumber \eea
This result shows that the $\Omega$-deformed $\cN=2$ gauge theory on $X_k$ is a quantization of the same $U(N)$-Hitchin system on a two-punctured sphere $C=\PP^1$
(the $A_{N-1}$-Toda system) as for the pure $\cN=2$ gauge theory on
$\IC^2$. For $N=1$ we can again elucidate further the structure of
this quantum integrable system. With $\boldsymbol V_{\vec w_j}$
denoting the natural vector bundle on the moduli space $\frM_{\vec
  w_j}(X_k)$, one can again show that the equivariant Chern classes
$\mbf I_{p}= c_{p-1}(\mbf V_{\vec w_j})$ with $p\geq1$ form an
infinite system of commuting multiplication operators which are
diagonalized in fixed point basis $[\vec Y,\vec u\, ]$ of $\Hscr_{\vec
  w_j}(X_k)$~\cite{Pedrini:2014yoa}; in particular, $\mbf I_1$ can be
identified with the Virasoro operator $L_0$ for
$\widehat{\mathfrak{gl}}_k$ and $\mbf I_2$ as the sum of $k$
non-interacting quantum Calogero--Sutherland Hamiltonians with
prescribed couplings $\beta^{(i)-1}=-\epsilon_2^{(i)}/\epsilon_1^{(i)}$
for $i=1,\dots,k$. In fact, in this case there is an isomorphism
\bea
\Hscr_{\vec w_j}(X_k) \cong \Lambda_{\IF}^{\otimes k}\otimes \IC(\epsilon_1,\epsilon_2)[\cQ+\omega_j] \ ,
\nonumber \eea
as representations of the affine algebra $\widehat{\mathfrak{gl}}_k$, which sends
\bea
[\vec Y,\vec u\,] \ \longmapsto \ J_{Y^1}\big(x;\beta^{(1)}\big)\otimes\cdots\otimes J_{Y^k}\big(x;\beta^{(k)}\big)\otimes(\gamma_{\vec u}+\omega_j) \ .
\nonumber \eea
Understanding the relations between the two level one representations
of $\widehat{\mathfrak{gl}}_k$ provided by the chambers $\sfC_0$ and
$\sfC_\infty$ at the level of symmetric functions and their
combinatorics would serve as a good step towards a clearer
understanding of the relationships between the AGT dualities, as well
as a closer step towards their proofs. This would also entail unveiling the relations between the different fixed point bases of $\Hscr_{\vec w_j}\big([\IC^2/\IZ_k]\big)$ and $\Hscr_{\vec w_j}(X_k)$, as well as the corresponding quantum integrable systems.

\bigskip

\section{$\cN=2$ gauge theories in six dimensions\label{sec:N26D}}

The purpose of this section is to compute and understand the equivariant partition
functions of the topologically twisted maximally supersymmetric Yang--Mills theory in six
dimensions in parallel to our four-dimensional $\cN=2$ gauge theories. In the Coulomb branch of the $U(N)$ gauge theory on
$\complex^3$, this has been computed in~\cite{Cirafici:2008sn} using
equivariant localization techniques in the case when the complex equivariant
parameters
$\vec\epsilon=(\epsilon_1,\epsilon_2,\epsilon_3)$ of the
natural scaling action of the three-dimensional torus $\torus^3$ on the affine
space $\complex^3$ satisfy
the Calabi--Yau constraint $\epsilon=0$ where
\bea
\epsilon:= \epsilon_1+\epsilon_2+\epsilon_3 \ ,
\nonumber \eea
i.e. the action of $\torus^3\subset SL(3,\complex)$ is reduced to an
action of a two-dimensional torus. This defines the
``Calabi--Yau specialization'' of the $\Omega$-deformation, as it is the
condition that the $\torus^3$-action preserve the holomorphic
three-form of $\complex^3$, and it is the six-dimensional analog
of the anti-diagonal torus action which is commonly used in four
dimensions; there it has various physical interpretations, e.g. as the
condition for self-duality of the graviphoton background in
four-dimensional $\cN=2$ supergravity, or in the context of the AGT
duality as the condition for vanishing background $U(1)$ charge $\varepsilon=0$ in
the dual conformal field theory. Here we extend this computation to arbitrary torus
actions, and then describe some intriguing four-dimensional reductions of the six-dimensional gauge theory.

\subsection{Instanton partition function\label{sec:6Dinst}}

The $\cN=2$ gauge theory of interest can be defined on any complex K\"ahler threefold $(X,\omega)$ by dimensional reduction of ten-dimensional $\cN=1$ supersymmetric Yang--Mills theory on the trivial complex vector bundle $X\times \IC^2$ of rank two over $X$ in the limit where the fiber collapses to a point. Restricting for simplicity to the case that $X$ is Calabi--Yau with no compact divisors, supersymmetric localization identifies the relevant moduli problem as that associated to solutions of the $\delta$-fixed point equations~\cite{Iqbal:2003ds,Cirafici:2008sn}
\bea
F^{1,1}\wedge \omega\wedge\omega=0 \qquad \mbox{and} \qquad \nabla_A\phi=0
\nonumber \eea
for an integrable connection $A$ on a holomorphic vector bundle over
$X$ and a local section $\phi$ of its adjoint bundle. In analogy with
the four-dimensional case we shall refer to solutions of these
equations as (generalised) instantons. In some instances this moduli problem is associated to instantons on a noncommutative deformation of $X$, for example in Type~IIA string theory when it describes bound states of D0-branes and D2-branes in D6-branes~\cite{Witten:2000mf}. Depending on the choice of stability parameters, the instanton moduli space can also be parameterized as a moduli space of torsion-free sheaves on $X$~\cite{Iqbal:2003ds,Cirafici:2008sn,Cirafici:2010bd}; in the rank one case this gauge theory is identified by~\cite{Nekrasov:2014nea} with the K-theoretic Donaldson--Thomas theory of $X$.

In this section we will work primarily in the Coulomb branch of $U(N)$ gauge theory on $X=\IC^3$. The $\Omega$-deformed gauge theory is defined by realising it as the worldvolume theory of $N$ separated D6-branes in Type~IIA string theory and lifting it to its dual M-theory description on a ${\sf TN}_N\times \IC^3$-bundle $\cM_\varrho$ over a circle $S^1$ whose radius $\varrho$ is identified with the string coupling constant, where ${\sf TN}_N$ is the hyper-K\"ahler $N$-centered Taub-NUT space which is a local $S^1$-fibration over $\IR^3$ (for the present discussion we could also replace the generic $S^1$ fibre with $\IR$, which gives an ALE space $X_N$). The total space of $\cM_\varrho$ is the quotient of ${\sf TN}_N\times \IC^3\times \IR$ by the $\IZ$-action
\bea
n\triangleright (p,z_1,z_2,z_3,x)= \big(g^n(p) \,,\, t^n_1\,z_1
\,,\,t_2^n\,z_2 \,,\,t^n_3\, z_3\,,\,x+2\pi\,n\,\varrho \big) \ ,
\nonumber \eea
where $p\in{\sf TN}_N$, $(z_1,z_2,z_3)\in\IC^3$, $x\in\IR$, $g$ is an
isometry of ${\sf TN}_N$ and $(t_1,t_2,t_3)\in \torus^3$ is an element
of the maximal torus of the complex rotation group $GL(3,\IC)$ with
$t_i:=\e^{\epsilon_i}$, while $n\in\IZ$ is the D0-brane (instanton)
charge of the dual Type~IIA string theory description in ten dimensions. Taking the collapsing limit $\varrho\to0$ then gives the desired $\Omega$-deformation.

The instanton partition function for $\cN=2$ gauge theory on $\IC^3$ is then given by
\beq
\cZ_{\IC^3}^{\rm inst}(\vec\epsilon,\vec a\,;\sfq) = \sum_{n=0}^\infty\, \sfq^n
\ \int_{[\frM_{N,n}(\IC^3)]^{\rm vir}} \, 1 \ , 
\label{ZNdef}\eeq
where in this case $\sfq=\e^{-\varrho}$. Here $[\frM_{N,n}(\IC^3)]^{\rm vir}$ is the virtual fundamental class of the
moduli scheme $\frM_{N,n}(\IC^3)$ of $U(1)^N$ (generalised) $n$-instanton
solutions of the six-dimensional noncommutative $\cN=2$ gauge theory,
or equivalently a particular
moduli space of torsion-free sheaves $\cE$ on $\mathbb{P}^3=\IC^3\cup \wp_{\infty}$ of rank
$N$ with $\ch_3(\cE)=-n$ which are framed on a plane $\wp_{\infty} \cong\PP^2$ at
infinity~\cite{Cirafici:2010bd}, and $\vec a=(a_1,\dots,a_N)$ is the
vector of Higgs vevs which are the
parameters of the induced $(\IC^\times)^N$-action on the moduli space. The
moduli spaces $\frM_{N,n}(\IC^3)$ have generic complex dimension $3\, N\, n$, and here
we will treat them in the framework of perfect obstruction
theory, as they are generally not smooth varieties. We
will understand the integral in (\ref{ZNdef}) as a $\widetilde\torus$-equivariant integral, where $\widetilde\torus=
\torus^3\times (\IC^\times)^N$, and define it properly via the localization theorem; it arises formally from localizing the path
integral to $\int_{\frM_{N,n}(\IC^3)}\, \Euler(N\frM_{N,n}(\IC^3))$, where
$\Euler(N\frM_{N,n}(\IC^3))$ is the equivariant Euler class of the obstruction bundle on the
instanton moduli space which arises from integration of the quartic
fermion terms in the action of the supersymmetric gauge theory. Since $\Euler(N\frM_{N,n}(\IC^3))\in
CH^\bullet_{\widetilde\torus}(\frM_{N,n}(\IC^3))$, it takes values in the coefficient ring
$CH^\bullet_{\widetilde\torus}({\rm pt})=
\complex[\epsilon_1,\epsilon_2,\epsilon_3,a_1,\dots,
a_N]_{(\vec\epsilon,\vec a\,)}$
for the equivariant Chow theory of the moduli space, which is the
localization of the ring $\complex[\epsilon_1,\epsilon_2,\epsilon_3,a_1,\dots,
a_N]$ at the maximal ideal $(\vec\epsilon,\vec a\,)$ generated by $\epsilon_1,\epsilon_2,\epsilon_3,a_1,\dots,
a_N$. 

The $\widetilde\torus$-fixed points of the instanton moduli
space $\mathfrak{M}_{N,n}(\IC^3)$ are isolated and
parameterized by $N$-vectors of three-dimensional Young diagrams (plane
partitions) $\vec\pi=(\pi_1,\dots,\pi_N)$ of total size $|\vec\pi\, |=\sum_l\, |\pi_l|= n$ (see~\S\ref{app:planepart}). Each component partition
$\pi_l$ specifies a $\torus^3$-invariant $U(1)$ noncommutative instanton of topological
charge $|\pi_l|$, or equivalently
a monomial ideal $I\subset\IC[z_1,z_2,z_3]=\cO_{\IC^3}$ of codimension $|\pi_l|$.  Applying the virtual localization theorem in
equivariant Chow theory to the integral in (\ref{ZNdef}) yields the
combinatorial expansion
\beq
\int_{[\frM_{N,n}(\IC^3)]^{\rm vir}} \, 1 = \sum_{ |\vec
\pi\,|=n}\
\frac{\Euler\big(N_{\vec\pi}\frM_{N,n}(\IC^3)\big)}{\Euler\big(T_{\vec\pi}\frM_{N,n}(\IC^3)\big)} \ .
\label{ZNcombgen}\eeq
In contrast to the maximally supersymmetric gauge theory in four
dimensions (the Vafa--Witten theory), here the ratios of Euler classes do not cancel in general,
as the tangent and obstruction bundles do not necessarily
coincide (though they have the same rank). Hence the partition function is generally a complicated
function of the equivariant parameters and the Higgs field vevs. In
fact, the situation closely resembles the case of pure $\cN=2$ gauge
theory in four dimensions, except that we work with the virtual
tangent bundle
$T^{\rm vir}\frM_{N,n}(\IC^3)= T\frM_{N,n}(\IC^3)\ominus N\frM_{N,n}(\IC^3)$ rather than the stable tangent bundle
$T\frM_{N,n}(\IC^3)$ (which is generally not well-defined here); in particular, the equivariant characteristic classes
in our case involve rational functions of the equivariant parameters rather
than polynomials. Further details concerning such
equivariant integrals can be found in~\cite[\S3.5]{Szabo:2009vw}.

To compute the virtual equivariant characteristic classes in
(\ref{ZNcombgen}), we use the local model of the instanton moduli
space developed in~\cite{Cirafici:2008sn} from the instanton quantum
mechanics for $n$ D0-branes inside $N$ D6-branes on $\complex^3$. There it was shown that
these classes can be computed from the character of the equivariant instanton
deformation complex
\begin{equation} \label{adhmdefcomplex}
\xymatrix@1{
  \End_\IC(V_{\vec\pi})
   \quad\ar[r] &\quad
   {\begin{matrix} \End_\IC(V_{\vec\pi}) \otimes Q 
   \\ \oplus \\
   \Hom_\IC(W_{\vec\pi},V_{\vec\pi}) \\ \oplus \\ \End_\IC(V_{\vec\pi}) \otimes \bigwedge^3 
   Q \end{matrix}}\quad \ar[r] & \quad
   {\begin{matrix} \End_\IC(V_{\vec\pi} ) \otimes \bigwedge^2
       Q \\ \oplus \\ 
       \Hom_\IC(V_{\vec\pi},W_{\vec\pi}) \otimes \bigwedge^3 Q
   \end{matrix}}
}
\end{equation}
where $Q\cong\complex^3$ is the fundamental representation of
$\torus^3$ with weight $(1,1,1)$, while $V_{\vec\pi}$ and
$W_{\vec\pi}$ are complex vector spaces of dimensions $n$
and $N$ parameterizing the D0-branes and D6-brane gauge degrees of freedom, respectively. They admit the decompositions
\begin{eqnarray}
V_{\vec\pi} = \sum_{l=1}^N \,e_l~ \sum_{(i,j,k)\in \pi_l}\,
t_1^{i-1} \,t_2^{j-1}\,t_3^{k-1} \qquad \mbox{and} \qquad W_{\vec\pi} =
\sum_{l=1}^N\,e_l
\label{decompos}\end{eqnarray}
as $\widetilde\torus$ representations regarded as polynomials in
$t_i$, $i=1,2,3$, and $e_l:=\e^{a_l}$,
$l=1,\dots,N$, i.e. as elements in the representation ring of $\widetilde\torus$. The first cohomology of this complex is a local model
of the tangent space to the moduli space at the fixed point $\vec\pi$, while the second cohomology
parameterizes obstructions. Its equivariant Chern character is easily computed as
\begin{eqnarray} 
\Ch\big(T_{\vec\pi}^{\rm vir}\frM_{N,n}(\IC^3)\big) &=& W^*_{\vec\pi}
\otimes V_{\vec\pi} - (t_1\, t_2\, t_3)^{-1}\, 
 V^*_{\vec\pi} \otimes W_{\vec\pi} \nonumber \\  && +\, \big(t_1^{-1}-1\big)\, \big(t_2^{-1}-1\big)\, \big(t_3^{-1}-1\big)\, 
 V^*_{\vec\pi} 
\otimes V_{\vec\pi} \ ,
\label{character}\end{eqnarray}
and the inverse of the corresponding top Chern polynomial yields the desired
ratio in (\ref{ZNcombgen}); here the dual involution acts on the weights as
$t_i^*=t_i^{-1}$ and $e_l^*=e_l^{-1}$.

In~\cite{Cirafici:2008sn} it is shown that at the Calabi--Yau specialization
$\epsilon=0$ of
the $\Omega$-deformation, the Euler classes in (\ref{ZNcombgen}) coincide up to a
sign,
\beq
\Euler\big(N_{\vec\pi}\frM_{N,n}(\IC^3)\big)
\big|_{\epsilon=0} = (-1)^{N\,n} \
\Euler\big(T_{\vec\pi}\frM_{N,n}(\IC^3)\big)\big|_{\epsilon=0} \ , 
\nonumber \eeq
and the partition function in this case is the generating function for
higher rank Coulomb branch Donaldson--Thomas invariants of
$\complex^3$ (D6--D0 states) given by the MacMahon function
\beq
\cZ_{\IC^3}^{\rm inst}(\vec\epsilon,\vec a\,;\sfq) \, \big|_{\epsilon=0} =
M\big((-1)^N\, \sfq\big)^N \qquad \mbox{with} 
\quad M(q)=\prod_{n=1}^\infty\, \frac1{\big(1-q^n \big)^n} \ ,
\label{ZNCY}\eeq
independently of the equivariant parameters and the Higgs vevs. This
is the six-dimensional version of the Vafa--Witten partition function
\eqref{eq:VWC2} on $\IC^2$. The
simplicity of the Coulomb branch invariants in this case may be
understood by rewriting them using the Joyce--Song formalism of generalized
Donaldson--Thomas invariants $\DT(n)$, as explained
in~\cite{Cirafici:2010bd}. In the present case they are defined
through
\beq
\cZ_{\IC^3}^{\rm inst}(\vec\epsilon,\vec a\,;\sfq) \, \big|_{\epsilon=0}  =:
\exp\Big(-\sum_{n=1}^\infty\, (-1)^{n\,N}\, n\, N \ \DT(n)\,\sfq^n \Big)
\ ,
\nonumber \eeq
and they lead to the generalized Gopakumar--Vafa BPS invariants
$\BPS(n)$ defined by
\beq
\DT(n)=: \sum_{m|n}\, \frac1{m^2}\, \BPS(n/m) \ .
\nonumber \eeq
By taking the logarithm of the MacMahon function and resumming we can express it as
\beq
M(q) = \exp\Big(\, \sum_{n=1}^\infty\, \frac1{\big(1-q^n\big)^2}\,
\frac{q^n}n\, \Big) \ ,
\nonumber \eeq
from which we find explicitly
\beq
\DT(n) = \sum_{m|n}\, \frac1{m^2} \qquad \mbox{and} \qquad \BPS(n)=1 \
.
\nonumber \eeq

We
will now generalize the result (\ref{ZNCY}) to arbitrary points
$\vec\epsilon$ in
the $\Omega$-deformation; see~\S\ref{app:planepart} for the set up of the relevant combinatorial notation.

\begin{proposition}
The instanton partition function of $\cN=2$ gauge theory on $\IC^3$ at
a generic point $\vec\epsilon$ of
the $\Omega$-deformation is given by
\beq
\cZ_{\IC^3}^{\rm inst}(\vec\epsilon,\vec a\,;\sfq)= \sum_{\vec\pi}\,
\big((-1)^N\, \sfq\big)^{|\vec\pi\, |} \
\prod_{l,l'=1}^N \, \frac{\sfN^{\vec\pi}_{l,l'}(\vec\epsilon,
  a_l-a_{l'})}{\sfT^{\vec\pi}_{l,l'}(\vec\epsilon,
  a_l-a_{l'})} \ ,
\label{ZNgencomb}\eeq
where
\bea
\sfN^{\vec\pi}_{l,l'}(\vec\epsilon,
  a) &=&\prod_{(i,j,k)\in\pi_l}\, \Big( \big(a +i\,
  \epsilon_1+j\, \epsilon_2+(k-(\pi_{l'})_{1,1})\,
  \epsilon_3\big) \nonumber \\ && \qquad \times\,
  \prod_{k'=1}^{(\pi_{l'})_{1,1}}\, \big(a+(i-(\pi_{l'\,
    (k'\,)}^{tt})_j)\, \epsilon_1+((\pi^t_{l\, (k)})_i-j+1)\,
  \epsilon_2+ (k'-k+1)\, \epsilon_3\big) \nonumber \\ &&
  \qquad \qquad \times\,
  \big(a+(i-(\pi^{tt}_{l'\, (k'\,)})_j-1)\,
  \epsilon_1+((\pi^t_{l\, (k)})_i-j)\, \epsilon_2+(k'-k-1)\,
  \epsilon_3\big) \Big) \nonumber
\eea
and
\bea
\sfT^{\vec\pi}_{l,l'}(\vec\epsilon,
  a) &=&\prod_{(i,j,k)\in\pi_l}\, \Big( \big( a+(i-1) \,
  \epsilon_1+(j-1)\, \epsilon_2+(k-(\pi_{l'})_{1,1}-1)\,
  \epsilon_3\big) \nonumber \\ && \qquad \times\,
  \prod_{k'=1}^{(\pi_{l'})_{1,1}}\, \big(a+(i-(\pi_{l'\,
    (k'\,)}^{tt})_j-1)\, \epsilon_1+((\pi^t_{l\, (k)})_i-j)\,
  \epsilon_2+ (k'-k)\, \epsilon_3\big) \nonumber \\ &&
  \qquad \qquad \times\, 
  \big(a+(i-(\pi^{tt}_{l'\, (k'\,)})_j)\,
  \epsilon_1+((\pi^t_{l\, (k)})_i-j+1)\, \epsilon_2+(k'-k)\,
  \epsilon_3\big) \Big) \ . \nonumber
\eea
\end{proposition}
\Proof{
We write the decomposition of the $\widetilde\torus$-module
$V_{\vec\pi}$ in (\ref{decompos}) as
\beq
V_{\vec\pi}= \sum_{l=1}^N\, e_l \ \sum_{k=1}^{(\pi_l)_{1,1}}\,
R_{\pi_{l\, (k)}^t}\, t_3^{k-1} \qquad \mbox{with} \quad R_{\pi_{l\,
    (k)}^t}= \sum_{(i,j)\in\pi_{l\,(k)}^t}\, t_1^{i-1}\, t_2^{j-1} \ .
\nonumber \eeq
Then the last term of the character (\ref{character}) may be expressed
as
\bea
&& \big(t_1^{-1}-1\big)\, \big(t_2^{-1}-1\big)\, \big(t_3^{-1}-1\big)\, 
 V^*_{\vec\pi} 
\otimes V_{\vec\pi} \nonumber \\ && \qquad \
= \ \big(t_3^{-1}-1\big)\, \sum_{l,l'=1}^N\, e_l^{-1}\, e_{l'} \ \sum_{k=1}^{(\pi_l)_{1,1}} \
\sum_{k'=1}^{(\pi_{l'})_{1,1}}\, t_3^{k'-k} \, \Big(\big(t_1^{-1}-1\big)\,
\big(t_2^{-1}-1\big)\, R_{\pi_{l\,(k)}^t}^*\otimes
R_{\pi^t_{l'\,(k'\,)}} \Big) \ . \nonumber
\eea
Using the calculation of~\cite[Thm.~2.11]{NakYosh} we have
\beq
\big(t_1^{-1}-1\big)\,
\big(t_2^{-1}-1\big)\, R_{\pi_{l\,(k)}^t}^*\otimes
R_{\pi^t_{l'\,(k'\,)}} = R_{\pi^t_{l'\,(k'\,)}}+ (t_1\, t_2)^{-1}\,
R_{\pi_{l\,(k)}^t}^*- \cT_{\pi_{l\,(k)}^t, \pi^t_{l'\,(k'\,)}}
\nonumber \eeq
where
\bea
\cT_{\pi_{l\,(k)}^t, \pi^t_{l'\,(k'\,)}} = \sum_{(i,j)\in\pi^t_{l\,(k)}}
\, t_1^{(\pi_{l'\,
    (k'\,)}^{tt})_j-i}\, t_2^{j-(\pi^t_{l\, (k)})_i-1} +
\sum_{(i,j)\in\pi^t_{l'\,(k'\,)}} \, t_1^{i-(\pi^{tt}_{l\,
      (k)})_j-1}\, t_2^{(\pi^t_{l'\, (k'\,)})_i-j} \ .
\nonumber \eea
Whence the full character (\ref{character}) is given by
\bea
\Ch\big(T_{\vec\pi}^{\rm vir}\frM_{N,k}\big) &=& V_{\vec\pi}\,
\sum_{l=1}^N\, e_l^{-1}\, \Big(1+\big(t_3^{-1}-1\big)\,
\sum_{k=1}^{(\pi_l)_{1,1}}\, t_3^{-k+1}\Big) \nonumber \\ && -\, (t_1\, t_2\,
t_3)^{-1}\, V_{\vec\pi}^*\, \sum_{l=1}^N\, e_l\, \Big(1+(t_3-1)\,
\sum_{k=1}^{(\pi_l)_{1,1}}\, t_3^{k-1} \Big) \nonumber \\ && -\,
\big(t_3^{-1}-1\big)\, \sum_{l,l'=1}^N\, e_l^{-1}\, e_{l'} \ \sum_{k=1}^{(\pi_l)_{1,1}} \
\sum_{k'=1}^{(\pi_{l'})_{1,1}}\, t_3^{k'-k}\, \cT_{\pi_{l\,(k)}^t,
  \pi^t_{l'\,(k'\,)}} \nonumber \\[4pt] &=& V_{\vec\pi}\,
\sum_{l=1}^N\, e_l^{-1}\,t_3^{-(\pi_l)_{1,1}} - (t_1\, t_2\,
t_3)^{-1}\, V_{\vec\pi}^*\, \sum_{l=1}^N\, e_l\, t_3^{(\pi_l)_{1,1}}
\nonumber \\ && -\,
\big(t_3^{-1}-1\big)\, \sum_{l,l'=1}^N\, e_l^{-1}\, e_{l'} \ \sum_{k=1}^{(\pi_l)_{1,1}} \
\sum_{k'=1}^{(\pi_{l'})_{1,1}}\, t_3^{k'-k}\, \cT_{\pi_{l\,(k)}^t,
  \pi^t_{l'\,(k'\,)}} \nonumber \\[4pt] &=& \sum_{l,l'=1}^N\,
e_l^{-1}\, e_{l'} \, \Big( \, \sum_{(i,j,k)\in\pi_{l'}}\, t_1^{i-1}\,
t_2^{j-1}\, t_3^{k-(\pi_l)_{1,1}-1}- \sum_{(i,j,k)\in\pi_{l}}\, t_1^{-i}\,
t_2^{-j}\, t_3^{-k+(\pi_{l'})_{1,1}} \nonumber \\ && -\,
\sum_{k'=1}^{(\pi_{l'})_{1,1}} \ \sum_{(i,j,k)\in\pi_{l}}\,t_1^{(\pi_{l'\,
    (k'\,)}^{tt})_j-i}\, t_2^{j-(\pi^t_{l\, (k)})_i-1}\,
\big(t_3^{k-k'-1}-t_3^{k-k'}\, \big) \nonumber \\ && -\,
\sum_{k'=1}^{(\pi_{l})_{1,1}} \ \sum_{(i,j,k)\in\pi_{l'}}\, t_1^{i-(\pi^{tt}_{l\,
      (k'\,)})_j-1}\, t_2^{(\pi^t_{l'\, (k)})_i-j}\,
  \big(t_3^{k'-k-1}-t_3^{k'-k} \big) \, \Big) \ .
\nonumber \eea
Collecting and relabelling terms, this finally yields the expression
for the ratio of fluctuation determinants in (\ref{ZNgencomb}).
}

It is easy to see that at the Calabi--Yau specialization one has
\beq
\sfN^{\vec\pi}_{l,l'}(\vec\epsilon,
  a)\big|_{\epsilon=0} = \sfT^{\vec\pi}_{l,l'}(\vec\epsilon,
  a)\big|_{\epsilon=0} \ ,
\nonumber \eeq
and hence that the partition function (\ref{ZNgencomb})
reduces to (\ref{ZNCY}) in this case. From this result it also follows
that there are no non-trivial instanton contributions at
the locus $\epsilon_1+\epsilon_2=0$ of the $\Omega$-deformation,
as then
\beq
\cZ_{\IC^3}^{\rm inst}(\sigma,-\sigma,\epsilon_3, \vec a \, ;\sfq)=1 \ ;
\nonumber \eeq
to see this, note that $(i-j)\,\sigma+(k-(\pi_l)_{1,1} )\,
\epsilon_3$ vanishes at $(i,j,k)=(1,1,(\pi_l)_{1,1}) \in \pi_l$,
hence $\sfN^{\vec\pi}_{l,l}(\sigma,-\sigma,\epsilon_3,
  0)=0$ whenever $\pi_l\neq\emptyset$. 

\subsection{$U(1)$ gauge theory}

Remarkably, as in four dimensions one can sum the combinatorial series (\ref{ZNgencomb})
explicitly in the rank one case. For $N=1$, the moduli space
$\frM_{1,n}(\IC^3) \cong \Hilb^n(\IC^3)$ is the Hilbert scheme of $n$ points on $\IC^3$ and the computation of the equivariant
instanton partition function is the content of~\cite[Thm.~1]{MNOPII}
which gives
\beq
\cZ_{\IC^3}^{\rm inst}(\vec\epsilon\, ;\sfq)=
M(-\sfq)^{- \chi_{\torus^3}(\IC^3)} \ ,
\label{Z1instexpl}\eeq
where
\beq
\chi_{\torus^3}(X)=\int_X\, \ch^{\torus^3}_3(TX) =
\frac{(\epsilon_1+\epsilon_2)\,
  (\epsilon_1+\epsilon_3)\,
  (\epsilon_2+\epsilon_3)}{\epsilon_1\, \epsilon_2\,
  \epsilon_3}
\nonumber \eeq
is the $\torus^3$-equivariant Euler characteristic of $X=\IC^3$,
evaluated by the Bott residue formula.
This formula is proven using geometric arguments from relative
Donaldson--Thomas theory. 

Our general result (\ref{ZNgencomb}) then computes equivariant Coulomb
branch invariants which may be regarded as a higher rank
generalization of the degree zero equivariant Donaldson--Thomas
invariants computed by (\ref{Z1instexpl}). In fact, the conjectures
of~\cite{NekJapan,Awata:2009dd} can be stated in our case as

\begin{conjecture}
The instanton partition function of $\cN=2$ gauge theory on $\IC^3$ at
a generic point $\vec\epsilon$ of
the $\Omega$-deformation is given by
\beq
\cZ_{\IC^3}^{\rm inst}(\vec\epsilon,\vec a\,;\sfq)= M\big((-1)^N\,
\sfq\big)^{-N\,\chi_{\torus^3}(\IC^3)} \ .
\nonumber \eeq
\label{conj:6D}\end{conjecture}

\subsection{Reductions to four dimensions}

In the remainder of this section we will present two applications of
the ``full'' $\Omega$-deformation which relates the six-dimensional
$\cN=2$ gauge theory to the four-dimensional $\cN=2$ gauge
theories. Consider first the action of an additional complex torus $\torus=\IC^\times$ whose generator $t$  acts on the homogeneous coordinates of the projective space $\PP^3$ as
\bea
t\triangleright[w_0,w_1,w_2,w_3] := [w_0,w_1,w_2,t\, w_3] \ .
\nonumber \eea
The fixed point set for this $\torus$-action is just the linearly embedded projective plane $(\PP^3)^\torus=\PP^2$. This $\torus$-action commutes with the $\torus^3$-action on $\PP^3$, and, as a toric variety, in this reduction we restrict to the subtorus $\torus^2\subset\torus^3$ defined by $t_3=t_2^*$, i.e. $\epsilon_2+\epsilon_3=0$, which acts on the fixed $\PP^2$ in the standard way. In this limit the instanton partition function on $\IC^3$ is trivial, but we need to look more closely at its definition when defining a suitable reduction of it.

The linear algebraic form of the instanton deformation complex \eqref{adhmdefcomplex} is based on the ADHM-type parameterization of the moduli space as the GIT quotient~\cite{Cirafici:2008sn}
\bea
\frM_{N,n}(\IC^3)= \mu_1^{-1}(0)\cap \mu_2^{-1}(0)\cap\mu_3^{-1}(0) \, \big/ \, GL(n,\IC)
\label{eq:ADHM6D} \eea
with
\bea
\mu_1=[b_1,b_2]+i_1\, i_2 \ , \qquad \mu_2=[b_1,b_3]+i_1\, i_3 \qquad
\mbox{and} \qquad \mu_3= [b_2,b_3] \ .
\label{eq:muP3} \eea
Here $(b_1,b_2,b_3)\in\End_\IC(V_{\vec \pi})\otimes Q$,
$i_1\in\Hom_\IC(W_{\vec \pi},V_{\vec \pi})$ and $(i_2,i_3,0) \in
\Hom_\IC(V_{\vec \pi},W_{\vec \pi})\otimes\bigwedge^2Q$, together with
a stability condition analogous to that of the usual ADHM
construction, i.e. the image of $i_1$ generates $V_{\vec\pi}$ under
the action of $b_1,b_2,b_3$, and as before the group $GL(n,\IC)$ acts by basis change automorphisms of the vector space $V_{\vec \pi}\cong\IC^n$. The first map of \eqref{adhmdefcomplex} is a linearised $GL(n,\IC)$ transformation, while the second map is the differential of the maps (\ref{eq:muP3}) that define the moduli space.

We shall now study the moduli of sheaves on $\PP^2$ which are pullbacks of framed torsion-free sheaves on $\PP^3$ by the inclusion $\PP^2\subset\PP^3$. The $\torus$-action on $\PP^3$ has a natural lift to the instanton moduli space $\frM_{N,n}(\IC^3)$ which is defined on the linear algebraic data as
\bea
t\triangleright(b_1,b_2,b_3,i_1,i_2,i_3) = (b_1,b_2,t\, b_3,i_1,i_2,t\, i_3) \ .
\label{eq:TactionADHM} \eea
The subvariety of invariants consists of sextuples
$(b_1,b_2,0,i_1,i_2,0)$ which are identified with matrices
$(b_1,b_2)\in \End_{\IC}(V_{\vec\pi})\otimes Q|_{\torus^2}$,
$i_1\in\Hom_\IC(W_{\vec \pi},V_{\vec \pi})$ and $i_2\in
\Hom_\IC(V_{\vec \pi},W_{\vec
  \pi})\otimes\bigwedge^2Q|_{\torus^2}$. The $\torus$-action
(\ref{eq:TactionADHM}) preserves the zero sets of the maps
\eqref{eq:muP3} and the action of the group $GL(n,\IC)$. The pullbacks
of the maps $\mu_2$ and $\mu_3$ to the subvariety of invariants
vanish, so their zero sets are trivial, while the pullback of $\mu_1$
coincides with the ADHM moment map \eqref{eq:muADHM} after relabelling
$i:=i_1$ and $j:=i_2$. One can also easily check that the stability
condition here pulls back to the one used in the ADHM construction of
\S\ref{sec:quivers}. Altogether we have shown that the moduli spaces of instantons in four dimensions can be described as smooth holomorphic submanifolds of the moduli spaces of generalized instantons in six dimensions,
\bea
\frM_{N,n}\big(\IC^2\big) = \frM_{N,n}\big(\IC^3\big)^\torus \ .
\nonumber \eea
We stress that this identification holds within the full $U(N)$ gauge theory and not just on its Coulomb branch, insofar as the matrix equations (\ref{eq:muP3}) describe moduli of $U(N)$ noncommutative instantons on $\IC^3$~\cite{Cirafici:2008sn}. Note how the full $\Omega$-deformation with $\epsilon\neq0$ is necessary here in order to reduce to a full $\torus^2$-action in four dimensions.

One can extend these considerations to some of the other Calabi--Yau threefolds for which the generalised instanton moduli can be described in terms of linear algebraic data. This is the case, for example, for the $A_{k-1}$-surface singularity $\IC^2/\IZ_{k} \times\IC$ with the $\IZ_{k}$-action given again by \eqref{eq:ZkactionC2}; its crepant resolution $\Hilb^{\IZ_k}(\IC^3)$ is a fibration of the minimal resolution $X_k$ over the affine line $\IC$. Its Coulomb branch partition functions at the Calabi--Yau specialisation $\epsilon =0$ are computed in~\cite[\S7]{Cirafici:2010bd}. By the three-dimensional version of the McKay correspondence, the instanton moduli space is the moduli variety of representations of the associated McKay quiver, which in this instance coincides exactly with \eqref{eq:NCquiver}; the matrix equations are given by the $\IZ_k$-equivariant decomposition of \eqref{eq:muP3}, analogously to the construction of \S\ref{sec:quivers}. For $\epsilon\neq0$ one can now use the above construction to realise the Nakajima quiver varieties $\frM_\xi(\vec v,\vec w\,)$ as $\torus$-invariant submanifolds of the moduli scheme of instantons on the (small radius) quotient stack $\big[\IC^2/\IZ_{k} \times\IC\big]$, respectively the (large radius) Calabi--Yau resolution $\Hilb^{\IZ_k}(\IC^3)$. It would be interesting to see if this six-dimensional perspective, and in particular the wall-crossing behaviour described in~\cite[\S7.4]{Cirafici:2010bd}, offers any insight into the chamber differences between the $\cN=2$ gauge theory partition functions discussed in \S\ref{sec:braidedsym} and hence to their associated AGT dualities. More generally, if $X\subset Y$ is a surface embedded in a threefold $Y$ as the fixed points of a $\torus$-action, it would be interesting to develop the conditions required for the generalised instanton moduli on $Y$ to correctly pullback to instanton moduli on $X$.

Our considerations are in harmony with the considerations of~\cite{Bonelli:2007hb} which show that the partition function of the six-dimensional $\cN=2$ gauge theory on a local surface can be reduced to the partition function of Vafa--Witten theory on the base surface via  equivariant localization with respect to the scaling action of $\torus$ on the fibers; it would be interesting to generalise this construction to the $\Omega$-deformed gauge theory.
The construction also fits nicely with the embedding of $U(N)$ Vafa--Witten theory on the ALE space $X_k$ in Type~IIA string theory as a D4--D6-brane intersection over a torus $T^2$~\cite{Dijkgraaf:2007sw}, which also has a lift to M-theory on the compactification ${\sf TN}_N\times T^2\times \IR^5$. The reductions described here are completely analogous to those of~\cite{Nishinaka:2013mba}, who show that the BPS moduli space of the supersymmetric gauge theory on D4--D2--D0-branes on a noncompact toric divisor $D$ of a toric Calabi--Yau threefold $X$ can be embedded in the moduli space of BPS D6--D2--D0-brane states on $X$; here the inclusion map is characterized combinatorially by a perfect matching in a brane tiling construction of the gauge theory on the D-branes (dimer model), and its image can be regarded as a $\torus$-invariant subspace. It would be interesting to understand whether the $\Omega$-deformed $\cN=2$ gauge theories in four dimensions can be similarly understood via the twisted M-theory lifts discussed in \S\ref{sec:6Dinst}, along the lines of the more general intersecting brane constructions considered in~\cite{Dijkgraaf:2007sw}.

Another way to effectively achieve a dimensional reduction of the theory is via the analog of the Nekrasov--Shatashvili limit of
four-dimensional $\cN=2$ supersymmetric gauge
theory~\cite{Nekrasov:2009rc}; in our case this corresponds to the
locus $\epsilon_3=0$ of the $\Omega$-deformation. In this case the M-theory fibration $\cM_\varrho$ from \S\ref{sec:6Dinst} contains an invariant line $\IC$ so we expect that the six-dimensional
gauge theory in this limit reduces to some effective exactly solvable four-dimensional
field theory, which may be thought of as arising through some sort of geometric engineering type
construction via a level-rank duality analogously to the constructions of~\cite{Dijkgraaf:2007sw}. Unfortunately, the
combinatorial expansion (\ref{ZNgencomb}) is not useful for studying
this limit, as it diverges at $\epsilon_3=0$. In analogy with the
four-dimensional case, we can conjecture that the free energy $\log
\cZ_{\IC^3}^{\rm inst}(\vec\epsilon,\vec a\,;\sfq)$ has only simple monomial poles in
the equivariant parameters, as in~(\ref{Z1instexpl}); they originate
from the equivariant volume $\int_X\, 1= \frac1{\epsilon_1\,
  \epsilon_2\, \epsilon_3}$ of $X=\IC^3$. 

One way to extract the $\epsilon_3=0$
limit analogously to the computations of~\cite{Nekrasov:2009rc} is to
again exploit the parameterization \eqref{eq:ADHM6D} of the instanton
moduli space and
use the matrix model representation of the integral in (\ref{ZNdef})
which was derived in~\cite[eq.~(5.23)]{Cirafici:2008sn}; it is given by
\bea
\int_{[\frM_{N,n}(\IC^3)]^{\rm vir}} \, 1 = \frac{(-1)^n}{n!} \, 
\int_{\real^n} \ \prod_{i=1}^n\, \frac{\dd\phi_i}{2\pi\ii} \
\frac{P_N(\phi_i+\epsilon)}{P_N(\phi_i)} \ 
\prod_{i\neq j} \, \Dcal(\phi_{i}-\phi_{j}) \ ,
\nonumber \eea
where 
\bea
P_N(x)&=& \prod_{l=1}^N\, (x-a_l) \ , \nonumber \\[4pt]
\Dcal(x)&=& \frac{x\, (x+\epsilon_1+\epsilon_2)\,
  (x+\epsilon_1+\epsilon_3)\,
  (x+\epsilon_2+\epsilon_3)}{(x+\epsilon)\,
  (x+\epsilon_1)\, (x+\epsilon_2)\, (x+\epsilon_3)} \ .
\nonumber \eea
The integrand here has simple poles at the fixed points of the
$\widetilde{\torus}$-action which are also parameterized by
three-dimensional Young diagrams $\vec\pi$ of size $|\vec\pi\,|= n$~\cite{Cirafici:2008sn}. A suitable evaluation of this integral could be useful for extracting the
$\epsilon_3=0$ limit via an integral equation for the free energy
analogously to that sketched in~\cite[\S6.2.6]{Nekrasov:2009rc}; note that $\Dcal(x)=1$ in the limit $\epsilon_3=0$.

\subsection{Classical and perturbative partition functions\label{sec:6Dclpertfns}}

Let us now see how the effective four-dimensional theory comes about in the limit $\epsilon_3=0$. The full partition function of the six-dimensional cohomological gauge
theory in the weak coupling regime receives classical, one-loop and
instanton contributions
\beq
\cZ_{\IC^3}(\vec\epsilon,\vec a\,;\sfq\, ) = \cZ_{\IC^3}^{\rm cl}(\vec\epsilon,\vec a\,;\sfq )\,
\cZ_{\IC^3}^{\rm pert}(\vec\epsilon,\vec a\,)\,
\cZ_{\IC^3}^{\rm inst}(\vec\epsilon,\vec a\,;\sfq) \ .
\nonumber \eeq
The easiest way to extract the classical and perturbative partition
functions is via a noncommutative deformation of the gauge theory. As concerns
the instanton contributions, the noncommutative gauge theory exhibits the
same properties as above
and seems to offer no new information. Setting $\epsilon_3=0$
implies that the noncommutative Higgs field $\Phi$ vanishes on a one-particle
subspace of the Fock space, hence the operator
$\exp(t\,\Phi)$ is not trace-class. Moreover, it is straightforward to
modify the fluctuation integrals of~\cite[eq.~(3.51)]{Cirafici:2008sn}
at a generic point $\vec\epsilon$ of the $\Omega$-deformation; the instanton integrals there then
reproduce our previous expression for the instanton partition function~(\ref{ZNgencomb}).

The classical part of the partition function is given
by~\cite[\S3.5]{Cirafici:2008sn}
\beq
 \cZ_{\IC^3}^{\rm cl}(\vec\epsilon,\vec a\,;\sfq) = \prod_{l=1}^N\,
 \sfq^{-\frac{a_l^3}{6\, \epsilon_1\, \epsilon_2\, \epsilon_3} } \ .
\nonumber \eeq

The expression for the vacuum contribution
in~\cite[eq.~(3.51)]{Cirafici:2008sn} holds at all points in parameter
space and gives the
perturbative partition function
\beq
\cZ_{\IC^3}^{\rm pert}(\vec\epsilon,\vec a\,) = \prod_{l,l'=1}^N\, 
\exp\Big(-\int_0^\infty\, \frac{\dd t}t \
\frac{\e^{t\,(a_l-a_{l'})}}{\big(1-\e^{t\, \epsilon_1}\big)\,
  \big(1-\e^{t\, \epsilon_2}\big)\, \big(1-\e^{t\,
    \epsilon_3}\big)} \Big) \ .
\nonumber \eeq
We can regularize the argument of the exponential function analogously
to~\S\ref{sec:N2C2} using the function
\begin{equation}
\label{gammadef}
\gamma_{\epsilon_1 , \epsilon_2 , \epsilon_3} (x ; \Lambda) =
\lim_{s\to0}\, \frac{\dd}{\dd s}\, \frac{\Lambda^s}{\Gamma(s)} \,
\int_0^{\infty} \, \dd t \ t^{s-1} \, \frac{\e^{ - t \, x }}{\big(1-\e^{t\, \epsilon_1}\big)\,
  \big(1-\e^{t\, \epsilon_2}\big)\, \big(1-\e^{t\,
    \epsilon_3}\big)}
\end{equation}
with $x=a_{l'}-a_l$. We expand
\begin{equation}
\label{cndef}
\frac{1}{\big(1-\e^{t\, \epsilon_1}\big)\,
  \big(1-\e^{t\, \epsilon_2}\big)\, \big(1-\e^{t\,
    \epsilon_3}\big)} = \sum_{n=0}^{\infty} \, \frac{c_n}{n!} \
t^{n-3} \ ,
\end{equation}
where the first few terms are given by
\begin{eqnarray}
c_0 &=& -\frac{1}{\epsilon_1 \, \epsilon_2 \, \epsilon_3} \ ,
\nonumber \\[4pt]
c_1 &=& \frac{\epsilon_1+\epsilon_2+\epsilon_3}{2 \,
  \epsilon_1\, \epsilon_2\, \epsilon_3} \ , \nonumber \\[4pt]
c_2 &=& -\frac{\epsilon_1^2+3 \, (\epsilon_2+\epsilon_3) \,
  \epsilon_1+\epsilon_2^2+\epsilon_3^2+3\, \epsilon_2 \,
  \epsilon_3}{6 \, \epsilon_1 \, \epsilon_2 \, \epsilon_3}
\ , \nonumber \\[4pt]
c_3 &=& \frac{(\epsilon_1+\epsilon_2+\epsilon_3) \,
  \big(\epsilon_2\, \epsilon_3+\epsilon_1 \,
  (\epsilon_2+\epsilon_3) \big)}{4 \, \epsilon_1 \,
  \epsilon_2 \, \epsilon_3} \ .
\nonumber \end{eqnarray}
It is easy to check that the coefficients $c_n$ all have denominator
$\epsilon_1 \, \epsilon_2 \, \epsilon_3$.
We can rescale $t \rightarrow \frac{t}{x}$ to generate gamma-functions and write
\begin{eqnarray}
\gamma_{\epsilon_1 , \epsilon_2 , \epsilon_3} (x ; \Lambda) =
\lim_{s\to0}\, \frac{\dd}{\dd s} \, \frac{\Lambda^s}{x^s} \
\sum_{n=0}^{\infty} \, \frac{c_n}{n!}  \, x^{3-n} \ \frac{\Gamma
  (n+s-3)}{\Gamma (s)} \ ,
\nonumber \end{eqnarray}
and taking the limit explicitly finally gives
\begin{eqnarray}
\gamma_{\epsilon_1 , \epsilon_2 , \epsilon_3} (x ; \Lambda)
&=& c_0 \, x^3 \, \Big( -\frac{11}{36} + \frac{1}{6} \, \log
\frac{x}{\Lambda} \, \Big) +c_1 \, x^2\, \Big( \, \frac{3}{4} -
\frac{1}{2} \log \frac{x}{\Lambda} \, \Big) 
 + c_2 \, \frac{x}{2} \, \Big(  -1 + \log \frac{x}{\Lambda} \, \Big)
 \nonumber \\ && - \, \frac{c_3}{6} \, \log \frac{x}{\Lambda} +
 \sum_{n=4}^{\infty} \, \frac{1}{n \, (n-1) \, (n-2) \, (n-3)} \ c_n
 \, x^{3-n} \ .
\nonumber \end{eqnarray}
Finally the perturbative part of the partition function is given by
\begin{equation}
\cZ_{\IC^3}^{\rm pert}(\vec\epsilon,\vec a\,) = \exp \Big( -
\sum_{l,l'=1}^N \, \gamma_{\epsilon_1 , \epsilon_2 ,
  \epsilon_3} (a_{l'} - a_l ; \Lambda)  \Big) \ ,
\nonumber \end{equation}
which is analogous to the one-loop partition function \eqref{eq:4Dpert} in four dimensions. However, here the appearance of logarithmic functions in this construction is
somewhat puzzling; in the four-dimensional case they arise because the
gauge theory is asymptotically free, however here the gauge theory is
topological (hence automatically infrared free). Similarly the meaning
of the cutoff $\Lambda$ is unclear, since it appears as a Yang--Mills
scale while it should be an ultraviolet cutoff. 

The function $\gamma_{\epsilon_1 , \epsilon_2 , \epsilon_3}
(x ; \Lambda) $ has interesting properties in the limit $\epsilon_3\to0$. Let us define
\begin{equation}
\Pi_{\epsilon_1 , \epsilon_2} (x ; \Lambda) :=
\lim_{\epsilon_3 \rightarrow 0} \, \epsilon_3 \,
\gamma_{\epsilon_1 , \epsilon_2 , \epsilon_3} (x ; \Lambda) \ .
\nonumber \end{equation}
 The analogous
quantity $\gamma_{\epsilon_1 , \epsilon_2} (x ; \Lambda)$ in the four-dimensional case replaces (\ref{cndef}) with
\begin{equation}
\frac{1}{\big(1-\e^{t\, \epsilon_1}\big)\,
  \big(1-\e^{t\, \epsilon_2}\big)} = \sum_{n=0}^{\infty} \,
\frac{d_n}{n!} \ t^{n-2} \ .
\nonumber \end{equation}
Multiplying (\ref{cndef}) by $\epsilon_3$ and taking the limit
$\epsilon_3 \to0$, by using l'H\^{o}pital's rule
\begin{equation}
 \lim_{\epsilon_3\rightarrow 0}  \ \frac{\epsilon_3}{1-\e^{t\, \epsilon_3}}=-\frac{1}{t}
\nonumber \end{equation}
we derive the relationship
\begin{equation} \label{strangeproperty}
\lim_{\epsilon_3 \rightarrow 0} \ \epsilon_3 \, c_n
(\epsilon_1 , \epsilon_2 , \epsilon_3) = - d_n(\epsilon_1
, \epsilon_2) \ .
\end{equation}
This also shows that
\begin{equation}
\frac{\dd}{\dd x} \, \Pi_{\epsilon_1 , \epsilon_2} (x ; \Lambda)
=  \gamma_{\epsilon_1 , \epsilon_2 } (x ; \Lambda) \ ,
\nonumber \end{equation}
where according to~\cite[eq.~(E.3)]{nakajima2} one has
\begin{eqnarray}
  \gamma_{\epsilon_1 , \epsilon_2 } (x ; \Lambda) &=&
  \frac{1}{\epsilon_1 \, \epsilon_2} \, \Big( - \frac12 \, x^2
  \, \log \frac{x}{\Lambda} + \frac34 \, x^2 \Big) +
  \frac{\epsilon_1 + \epsilon_2}{\epsilon_1 \, \epsilon_2}
  \, \Big( -x \, \log \frac{x}{\Lambda} + x \Big) \nonumber \\
  && - \, \frac{\epsilon_1^2 + \epsilon_2^2 + 3 \, \epsilon_1
    \, \epsilon_2}{12 \, \epsilon_1 \, \epsilon_2} \, \log
  \frac{x}{\Lambda} + \sum_{n=3}^{\infty} \, \frac{d_n \, x^{2-n}}{n
    \, (n-1) \, (n-2)} \ .
\nonumber \end{eqnarray}

Mimicking the approach of \cite{Nekrasov:2009rc}, we define the quantity
\beq
\cW_{\IC^3}(\epsilon_1,\epsilon_2,\vec a\,;\sfq) := \lim_{\epsilon_3\to0}\, \epsilon_3\, \log
\cZ_{\IC^3}(\vec\epsilon,\vec a\, ;\sfq) \ ,
\nonumber \eeq
which receives a sum of classical, one-loop and instanton
contributions. We have
\begin{equation}
\cW_{\IC^3}^{\rm cl}(\epsilon_1,\epsilon_2,\vec a\, ;\sfq) = -
\sum_{l=1}^N \, \frac{a_l^3}{6 \, \epsilon_1 \, \epsilon_2} \, \log \sfq
\nonumber \end{equation}
and
\begin{eqnarray}
\cW_{\IC^3}^{\rm pert}(\epsilon_1,\epsilon_2,\vec a\,) = -
\lim_{\epsilon_3 \rightarrow 0} \, \sum_{l,l'=1}^N  \,
\epsilon_3 \, \gamma_{\epsilon_1 , \epsilon_2 ,
  \epsilon_3} (a_{l'} - a_l ; \Lambda) = -  \sum_{l,l'=1}^N  \,
\Pi_{\epsilon_1 , \epsilon_2} (a_{l'} - a_l ; \Lambda) \ .
\nonumber \end{eqnarray}
Thus dropping signs the full semiclassical superpotential is then
\begin{equation}
\cW^{\rm sc}_{\IC^3}(\epsilon_1,\epsilon_2,\vec a\,;\sfq) =  \frac{1}{6 \,
  \epsilon_1 \, \epsilon_2} \, \log \sfq \,  \sum_{l=1}^N \, a_l^3
+ \sum_{l,l'=1}^N\, \Pi_{\epsilon_1 , \epsilon_2} (a_{l'} - a_l
; \Lambda) \ .
\nonumber \end{equation}
The relevant Bethe-type equations are obtained by minimizing this
superpotential with respect to each of the variables $a_l$ for
$l=1,\dots,N$, which gives
\begin{eqnarray}
\frac{1}{2 \, \epsilon_1 \, \epsilon_2} \, \log \sfq \, a_l^2 +
\sum_{l'\neq l}\, \gamma_{\epsilon_1 , \epsilon_2 } (a_l -
a_{l'}  ; \Lambda) -  \sum_{l' \neq l}\, \gamma_{\epsilon_1 ,
  \epsilon_2 } (a_{l'} - a_l  ; \Lambda)= 0 \ ,
\nonumber \end{eqnarray}
and by exponentiation our final equations are
\begin{equation}
\sfq^{\frac{a_l^2}{2 \, \epsilon_1 \, \epsilon_2}} = \prod_{l'\neq
  l}\, \frac{\e^{\gamma_{\epsilon_1 , \epsilon_2 } (a_{l'} - a_l
    ; \Lambda)}}{\e^{  \gamma_{\epsilon_1 , \epsilon_2 } (a_l -
    a_{l'}  ; \Lambda) }}
\nonumber \end{equation}
for $l=1,\dots,N$. 
These Bethe-type equations yield an alternative regularized version of
the formal expression (\ref{eq:gammaformal}) in terms of the contributions to the classical four-dimensional partition function \eqref{eq:4Dclass}.

In contrast to the context of four-dimensional gauge theory, the minimization
of the superpotential $\cW_{\IC^3}(\epsilon_1,\epsilon_2,\vec a\, ;\sfq)$
here does not appear to have a clear physical meaning; for this, one
should probably look directly at the dimensional reduction of the
physical untwisted gauge theory in six dimensions, which may have a more transparent meaning through its twisted M-theory lift discussed in \S\ref{sec:6Dinst}.
Now one can add instanton corrections and define similarly a nonperturbative
superpotential, which by \emph{assuming} Conj.~\ref{conj:6D} is given explicitly by
\beq
\cW_{\IC^3}^{\rm inst}(\epsilon_1,\epsilon_2,\vec a\,;\sfq) = N\,
(\epsilon_1+\epsilon_2)\, \sum_{n=1}^\infty\,
\frac1{\big(1-\big((-1)^N\, \sfq\big)^n\big)^2}\,
\frac{\big((-1)^N\, \sfq\big)^n}n \ .
\nonumber \eeq

\bigskip

\section*{Acknowledgments}

We thank Alastair Craw, Elizabeth Gasparim, Iain Gordon, Melissa Liu and Paul Smith for helpful
discussions, correspondence, and technical assistance on some of the
issues discussed in this paper, and Ugo Bruzzo, Michele Cirafici, Lucio Cirio, Zolt\'an K\"ok\'enyesi, Giovanni Landi, Mattia~Pedrini, Francesco~Sala and
Annamaria Sinkovics for collaboration on various parts of the material
presented here. We also thank Ugo~Bruzzo, Dimitri~Markushevich,
Vladimir~Rubtsov, Francesco~Sala and Sergey~Shadrin
for the invitation to contribute this article to the special issue, which is based on themes from the workshop ``Instanton Counting: Moduli
  Spaces, Representation Theory and Integrable Systems'' which was
  held at the Lorentz Center in June 2014. 
This work was supported in part by the Consolidated Grant ST/L000334/1 from the UK Science and Technology
Facilities Council (STFC) and by the Action MP1405 QSPACE from the European Cooperation in Science and Technology
(COST).

\appendix

\bigskip

\section{Combinatorics}\label{sec:Young}

\subsection{Young diagrams\label{app:2DYoung}}

A \emph{partition} of a positive integer $n$ is a nonincreasing
sequence of positive numbers $\lambda=(\lambda_1\geq \lambda_2\geq
\cdots\geq \lambda_\ell> 0)$ such that
$\vert\lambda\vert:=\sum_{i=1}^\ell\, \lambda_i=n.$ We call
$\ell=\ell(\lambda)$ the \emph{length} of the partition $\lambda.$ A
partition $\lambda$ of $n$ may also be described as a list
$\lambda=(1^{m_1}\ 2^{m_2}\ \ldots)$, where
$m_i=\#\{l\in\mathbb{Z}_{>0} \,\vert \lambda_l=i\}.$ Then $\sum_i\, i\, m_i=n$ and $\sum_i\, m_i=\ell.$ We also write $\|\lambda\|^2:=\sum_{i=1}^\ell\, \lambda_i^2.$

One can associate to a partition $\lambda$ a \emph{Young
  diagram}, which is the set $Y_\lambda=\{(a,b)\in
\mathbb{Z}_{>0}^2\,\vert\, 1\leq a \leq \ell,\, 1\leq b \leq \lambda_a\}.$
Thus $\lambda_a$ is the length of the $a$-th column of $Y_\lambda;$ we
write $|Y_\lambda|=|\lambda|$ for the \emph{weight} of the Young diagram $Y_\lambda$. We identify a partition $\lambda$ with its Young
diagram $Y_\lambda.$ For a partition $\lambda$, the \emph{transpose
  partition} $\lambda^t$ is the partition with Young diagram
$Y_{\lambda^t}:=\{(j,i)\in \mathbb{Z}_{>0}^2\,\vert\,(i,j)\in
\lambda\}.$ On the set $\Pi$ of all Young diagrams there is a
natural partial ordering called \emph{dominance ordering}: For two
partitions $\mu$ and $\lambda$, we write $\mu\leq\lambda$ if $|\mu|=|\lambda|$ and
$\mu_1+\cdots+\mu_i\leq\lambda_1+\cdots+\lambda_i$ for all $i\geq 1.$
We write $\mu<\lambda$ if $\mu\leq\lambda$ and
$\mu\neq\lambda.$

We call the elements of a Young diagram $Y$ the \emph{nodes} of $Y.$ For a
node $s=(a,b)\in Y$, we call the \emph{arm-length} of $s$ the quantity
$A(s):=A_Y(s)=\lambda_a-b$ and the \emph{leg-length} of $s$ the quantity
$L(s):= L_Y(s)=\lambda_b^t-a.$ The \emph{hook-length} of $s$ is
$h(s):= h_Y(s)=A_Y(s)+L_Y(s)+1.$ The \emph{content} of a node $s=(a,b)$ is
the number $a-b.$ The \emph{arm-colength} and \emph{leg-colength} are
respectively given by $A^t(s):= A_Y^t(s)=b-1$ and $L^t(s):= L_Y^t(s)=a-1$. The
\emph{hook} of $s$ is the set $H_s:=H_s(Y)=\{(c,d)\in Y\ | \ c=a, d \geq
b\} \cup \{(c,d)\in Y\ | \ c>a, d=b\}$; then $h(s)=\# H_s$ and we
say that $H_s$ is an \emph{$r$-hook} if $h(s)=r$. For
$\nu\in\IZ$, we define $N_\nu(Y)$ to be the number of nodes of $Y$ along
the line $b=a-\nu$ .

Fix an integer $k\geq 2$ and let $i\in\{0,1,\dots,k-1\}$. A node of a Young diagram $Y$ is called an $i$\emph{-node} if its content equals $i$ modulo $k$; this gives a \emph{$k$-colouring} of $Y.$ We define 
\begin{eqnarray*}
n_j(Y)&=&\#\big\{(a,b)\in Y\,\big\vert\, a-b=j\big\}\ ,\\[4pt]
\nu_i(Y)&=&\#\big\{(a,b)\in Y\, \big\vert\, a-b\equiv i \ \mathrm{mod}\, k
            \big\} \ , \qquad \vec{\nu}(Y) \ = \ \big(\nu_0(Y), \nu_1(Y),
            \ldots, \nu_{k-1}(Y) \big)\in \mathbb{Z}_{\geq0}^{k} \ .
\end{eqnarray*}

\subsection{Maya diagrams\label{app:Maya}}

A \emph{Maya diagram} is an increasing sequence of half-integers $\boldsymbol{m}:=(h_j)_{j\geq 1}$ such that $h_{j+1}=h_j+1$ for sufficiently large $j.$ Let $\mathcal{M}$ denote the set of all Maya diagrams. Any $\mbf m\in\cM$ can be identified with a map $\mbf m:\mathbb{Z}+\frac{1}{2}\rightarrow \{\pm\, 1\}$ such that
\begin{equation*}
\boldsymbol{m}(h)=\left\{
\begin{array}{rl}
1 & \mbox{for} \quad h\gg 0\ ,\\
-1 & \mbox{for} \quad h\ll 0\ .
\end{array}\right.
\end{equation*}
We define the \emph{charge} of $\boldsymbol{m}$ by $c(\boldsymbol{m}):=\#\{h<0\,\vert\, \boldsymbol{m}(h)=1\}-\#\{h>0\,\vert\,\boldsymbol{m}(h)=-1\}.$ For $z\in \mathbb{Z}$, we define $\boldsymbol{m}^{(z)}$ by $\boldsymbol{m}^{(z)}(h):=\boldsymbol{m}(h+z)$; then $c(\boldsymbol{m}^{(z)})=c(\boldsymbol{m})-z.$

We shall now explain how Maya diagrams can be identified with Young diagrams. Let $Y$ be a Young diagram, and note that for any half-integer $h$ one has
\begin{equation*}
n_{h-\frac{1}{2}}(Y)-n_{h+\frac{1}{2}}(Y)=\left\{
\begin{array}{rl}
-1 \ \mbox{ or } \ 0 & \mbox{for} \quad h<0\ ,\\
0 \ \mbox{ or } \ 1 & \mbox{for} \quad h>0\ .
\end{array}\right.
\end{equation*}
We then define
\begin{equation*}
\boldsymbol{m}_Y(h)=\left\{
\begin{array}{rl}
-1 & \mbox{for} \quad h<0 \ \mbox{ and } \ n_{h-\frac{1}{2}}(Y)-n_{h+\frac{1}{2}}(Y)=0\ ,\\
1 & \mbox{for} \quad h<0 \ \mbox{ and } \ n_{h-\frac{1}{2}}(Y)-n_{h+\frac{1}{2}}(Y)=-1\ ,\\
-1 & \mbox{for} \quad h>0 \ \mbox{ and } \ n_{h-\frac{1}{2}}(Y)-n_{h+\frac{1}{2}}(Y)=1\ ,\\
1 & \mbox{for} \quad h>0 \ \mbox{ and } \ n_{h-\frac{1}{2}}(Y)-n_{h+\frac{1}{2}}(Y)=0\ .
\end{array}\right.
\end{equation*} 
Then $c(\boldsymbol{m}_Y)=0.$ Conversely, given a Maya diagram $\boldsymbol{m}$ of zero charge, there exists a unique Young diagram $Y$ such that $\boldsymbol{m}_Y=\boldsymbol{m}.$ Consequently there is a bijection
\begin{eqnarray*}
F\, \colon \, \mathbb{Z}\times \Pi \ \longrightarrow \ \mathcal{M} \ , \qquad
(z, Y) \ \longmapsto \ \boldsymbol{m}_Y^{(-z)}\ .
\end{eqnarray*}
Define $q(\boldsymbol{m})\in \Pi$ by $F^{-1}(\boldsymbol{m})=(c(\boldsymbol{m}), q(\boldsymbol{m})).$

Let $\tilde{I}=\left\{\frac{1}{2}, \frac{3}{2}, \ldots, k-\frac{1}{2}\right\}.$ For $h\in \tilde{I}$ and $\boldsymbol{m}\in\mathcal{M}$, we define a Maya diagram $\boldsymbol{m}_h$ by $\boldsymbol{m}_h(l)=\boldsymbol{m}\big(k\,(l-\frac{1}{2})+h\big).$ Then $\boldsymbol{m}$ can be recovered from $\{\boldsymbol{m}_h\}_{h\in \tilde{I}}$ and $c(\boldsymbol{m})=\sum_{h\in\tilde I}\, c(\boldsymbol{m}_h).$ For a Young diagram $Y$, we set $c_h(Y)=c(\boldsymbol{m}_{Y,h})$ and $q_h(Y)=q(\boldsymbol{m}_{Y,h}).$ By~\cite[Lem.~2.5.2]{art:nagao2009}, one has explicitly $c_h(Y)=\nu_{h-\frac{1}{2}}(Y)-\nu_{h+\frac{1}{2}}(Y).$ We define the $k$\emph{-core} of $Y$ by 
\begin{equation}
\vec{c}\, (Y)=\big(c_h(Y) \big)_{h\in\tilde{I}} \ \in \ 
\big(\mathbb{Z}^{\tilde{I}}\,\big)_0:=\Big\{\big(c_{\frac{1}{2}},
    \ldots, c_{k-\frac{1}{2}}\big)\in
  \mathbb{Z}^{\tilde{I}}\,\Big\vert\, \mbox{$\sum\limits_{h\in\tilde I}\, c_h=0$} \Big\}\ .
\nonumber \end{equation}
The set $\big(\IZ^{\tilde I}\,\big)_0$ of $k$-cores is in bijection
with the set of Young diagrams which do not contain any $k$-hooks.
We also define the $k$\emph{-quotient} of $Y$ by $\vec{q}\,(Y)=\big(q_h(Y) \big)_{h\in\tilde{I}}\in \Pi^{\tilde{I}}.$ There is a bijection
\begin{eqnarray*}
{\tt CQ} \,\colon \, \Pi \ \longrightarrow \ \big(\mathbb{Z}^{\tilde{I}}\, \big)_0\times \Pi^{\tilde{I}}\ , \qquad 
Y \ \longmapsto \ \big(\vec{c}\,(Y)\,,\, \vec{q}\,(Y) \big)\ ,
\end{eqnarray*}
which is obtained by removing all $k$-hooks from a Young
diagram $Y$; the set of removed $k$-hooks uniquely constitutes a $k$-tuple of
partitions $\big(\lambda^{(0)}, \lambda^{(1)},\dots,\lambda^{(k)} \big)$ which corresponds to
the $k$-quotient $\vec q\,(Y)$.
In particular, for any element $\big((c_h)_{h\in \tilde{I}}, (Y_h)_{h\in \tilde{I}}\big)\in \big(\mathbb{Z}^{\tilde{I}}\, \big)_0\times \Pi^{\tilde{I}}$ we have
\begin{equation*}
{\tt CQ}^{-1}\left((c_h)_{h\in \tilde{I}} \,,\, (Y_h)_{h\in \tilde{I}}\right)=q(\boldsymbol{m})\ ,
\end{equation*}
where $\boldsymbol{m}$ is the Maya diagram recovered from the Maya
diagrams $\boldsymbol{m}_{Y_h}$ associated to the Young diagrams $Y_h$ for
$h\in \tilde{I}.$ The weight of $Y\in\Pi$ is given by
$$
|Y|=\big|\,\underline{Y}\, \big|+k\, \big|\vec q\,(Y)\big|
$$
where $\underline{Y}={\tt CQ}^{-1}\big(\vec c\,(Y),\vec\emptyset\ \big)$. If $\vec
q(Y)=\vec\emptyset$ then
$$
|Y| = \sum_{h\in\tilde I} \ \sum_{l\in\IZ}\, N_{k\,l+h-\frac12}(Y) =
\sum_{h\in\tilde I}\, \Big(\, \mbox{$\frac12$} \, k\, c_h(Y)^2+
\big(h-\mbox{$\frac12$}\big)\, c_h(Y)\Big) \ .
$$
On the other hand, if $\vec c\, (Y)=\vec0$ then
$\underline{Y}=\emptyset$ and the $k$-quotient
$\big(\lambda^{(0)}, \lambda^{(1)},\dots,\lambda^{(k)} \big)$ can be read off from
the relation
$$
(\lambda_i-i)_{i\geq1} = \bigcup_{r=0}^{k-1} \,
\Big(k\,\big(\lambda_{i_r}^{(r)}-i_r \big)+r\Big)_{i_r\geq1} \ ,
$$
where the left-hand side is the corresponding \emph{blended} partition
while the right-hand side contains the corresponding \emph{coloured} Young
diagrams.

\subsection{Three-dimensional Young diagrams\label{app:planepart}}

A \emph{three-dimensional Young diagram} can be regarded as a stack of ordinary (two-dimensional) Young diagrams, with boxes piled in the
positive octant $(x,y,z)\in \zed_{\geq0}^3$. We can construct an array
of non-negative decreasing integers $\pi=(\pi_{i,j})_{i,j\geq1}$, $\pi_{i,j}\geq\pi_{i+1,j}$, $\pi_{i,j}\geq\pi_{i,j+1}$, called a \emph{plane partition}, as the height function of the stack of cubes
defined on the $(x, y)$ plane, i.e. the three-dimensional Young diagram (that we also denote by $\pi$ for brevity) is the set $\pi=\{(i,j,k)\in\IZ_{>0}^3\,|\, 1\leq k\leq \pi_{i,j}\}$. Its projection to the $(x,y)$ plane defines an ordinary partition $\pi^z$. More generally, for fixed $a,b,p,r\geq0$ the sequence $\lambda=(\lambda_i):= (\pi_{a+p\,i, b+r\, i})$ defines a partition. The \emph{size} of $\pi$ is
\bea
|\pi|:= \sum_{(i,j)\in\pi^z}\, \pi_{i,j} \ .
\nonumber \eea

We also obtain different arrays of non-negative integers
$\pi^t=(\pi^t_{j,k})_{j,k\geq1}$ and $\pi^{tt}=(\pi^{tt}_{i,k})_{i,k\geq1}$ by considering
the height functions of the stack relative to the $(y,z)$ and $(x,z)$
planes, respectively; these
transformations are the analogs of the transpose operation for
two-dimensional Young diagrams. Again their respective projections to the $(y,z)$ and $(x,z)$ coordinate planes defines ordinary partitions $\pi^x$ and $\pi^y$. For each fixed integer $k\geq1$, the
sequences $\pi^t_{(k)}:=(\pi^t_{j,k})_{j\geq1}$ and $\pi^{tt}_{(k)}:=(\pi^{tt}_{i,k})_{i\geq1}$ are two-dimensional Young
diagrams obtained by cutting the three-dimensional Young diagram $\pi$
by the plane $z=k$ perpendicular to the $z$-axis.

\bigskip

\section{Symmetric functions}\label{sec:symmetric}

\subsection{Monomial symmetric functions}

The {algebra of symmetric polynomials in} $N$ {variables} is the subspace $\Lambda_{N}:= \IC[x_1,\ldots,x_N]^{S_N}$ of polynomials invariant under the action of the symmetric group of permutations $S_N.$ It is a graded ring $\Lambda_{N}=\bigoplus_{n\geq0}\,  \Lambda_{N}^n$, where $\Lambda_{N}^n$ is the ring of homogeneous symmetric polynomials in $N$ variables of degree $n$. There are graded inclusion morphisms $\Lambda_{N+1}\hookrightarrow\Lambda_N$ by setting $x_{N+1}=0$, and the corresponding inverse inductive limit is the {algebra of symmetric functions in infinitely many variables} $\Lambda:=\bigoplus_{n\geq0}\, \Lambda^n.$

For a partition $\mu=(\mu_1,\ldots,\mu_t)$ with $t\leq N$, we define the polynomial
\begin{equation*}
m_\mu(x_1,\ldots,x_N) =\sum_{\sigma\in S_N}\, x_1^{\mu_{\sigma_1}}\cdots x_N^{\mu_{\sigma_N}}\ ,
\end{equation*}
where we set $\mu_j=0$ for $j=t+1,\ldots, N.$ The polynomial $m_\mu$ is symmetric, and the set of all $m_\mu$ for all partitions $\mu$ with $\vert \mu \vert\leq N$ is a basis of $\Lambda_N.$ Then the collection of $m_\mu$, for all partitions $\mu$ with $\vert \mu \vert\leq N$ and $\sum_i\, \mu_i=n$, is a basis of $\Lambda^n_N$. Using the definition of inverse limit we can define the \emph{monomial symmetric functions} $m_\mu$; by varying over partitions $\mu$ of $n$, these functions form a basis for $\Lambda^n.$

The $n$\emph{-th power sum symmetric function} is $p_n := m_{(n)}= \sum_{i}\, x_i^n.$ The collection of symmetric functions $p_\mu:=p_{\mu_1}\cdots p_{\mu_t}$, for all partitions $\mu=(\mu_1,\ldots,\mu_t)$, is another basis of $\Lambda.$ 

\subsection{Macdonald functions\label{app:Macdonald}}

Fix parameters $q,t\in\IC$ with $|q|<1$. For $a\in\IC$, we use throughout the standard hypergeometric notation for the infinite $q$-shifted factorial
\begin{equation*}
(a;q)_\infty:= \prod_{n=0}^\infty\, \big(1-a\, q^n\big) \ .
\end{equation*}
Define an inner product on the vector space $\Lambda\otimes\IQ(q,t)$ with respect to which the basis of power sum symmetric functions $p_\lambda(x)$ are orthogonal with the normalization
\begin{equation*}
\langle p_\lambda ,p_\mu\rangle_{q,t}=\delta_{\lambda,\mu} \ z_\lambda
\ \prod_{i=1}^{\ell(\lambda)}\,
\frac{1-q^{\lambda_i}}{1-t^{\lambda_i}} \ ,
\end{equation*}
where $\delta_{\lambda,\mu}:=\prod_i\, \delta_{\lambda_i,\mu_i}$ and
\begin{equation*}
z_\lambda:= \prod_{j\geq1}\, j^{m_j}\, m_j! \ .
\end{equation*}
This is called the \emph{Macdonald inner product}.

The monic form of the \emph{Macdonald functions} $M_\lambda(x;q,t)\in\Lambda\otimes\IQ(q,t)$ for $x=(x_1,x_2,\dots)$ are uniquely defined by the following two conditions~\cite[Ch.~VI]{book:macdonald1995}:
\begin{itemize}
\item[(i)] Triangular expansion in the basis $m_\mu(x)$ of monomial symmetric
  functions:
\begin{equation}\label{eq:triangular-macdonald}
M_\lambda(x;q,t) = m_\lambda(x)+ \sum_{\mu<\lambda}\,
v_{\lambda,\mu}(q,t)\, m_\mu(x) \qquad \mbox{with} \quad
v_{\lambda,\mu}(q,t)\in\IC \ .
\end{equation}
\\
\item[(ii)] Orthogonality:
\begin{equation}\label{eq:orthogonality-macdonald}
\langle M_\lambda , M_\mu\rangle_{q,t} = \delta_{\lambda,\mu} 
\ \prod_{s\in Y_\lambda} \ \frac{1-q^{A(s)+1} \,
  t^{L(s)}}{1-q^{A(s)} \,
  t^{L(s)+1}}\ .
\end{equation}
\end{itemize}
For $t=1$ these functions coincide with the monomial symmetric functions, $M_\lambda(x;q,1)=m_\lambda(x).$

By their definition the Macdonald functions are homogeneous:
\begin{equation}\label{eq:homogeneity-macdonald}
M_\lambda(\zeta\,x;q,t) = \zeta^{|\lambda|}\, M_\lambda(x;q,t) \qquad
\mbox{for} \quad \zeta\in\IC \ ,
\end{equation}
and they satisfy the \emph{Macdonald specialization identity}
\begin{equation}\label{eq:specialization1-macdonald}
M_\lambda(t^\rho;q,t)=q^{\frac14\,\|\lambda\|^2}\, t^{-\frac14\,
  \|\lambda^t\|^2}\, \Big(\, \frac tq\, \Big)^{\frac{|\lambda|}4} \
\prod_{s\in Y_\lambda}\, \Big(q^{\frac{A(s)}2} \,
t^{\frac{L(s)+1}2} - q^{-\frac{A(s)}2}\,
t^{-\frac{L(s)+1}2}\Big)^{-1} \ ,
\end{equation}
where $t^\rho:=(t^{\rho_1},t^{\rho_2},\dots)$ with $\rho_i=-i+\frac12$. If we take $x=(t^{-1},\dots,t^{-N},0,0,\dots)$ for $N\in\IZ_{>0}$ then we can also write the special value
\begin{equation}\label{eq:specialization2-macdonald}
M_\lambda(t^{-1},\dots,t^{-N};q,t) = \prod_{s\in Y_\lambda}\,
\frac{t^{L^t(s)-N}-q^{A^t(s)}}{1-
      q^{A(s)}\, t^{L(s)+ 1}}
\end{equation}
for the Macdonald polynomials $M_\lambda(x_1,\dots,x_N;q,t)$ in $\Lambda_N\otimes\IQ(q,t).$

Macdonald functions also satisfy the \emph{generalized Cauchy--Binet formula}
\begin{equation}\label{eq:cauchybinet-macdonald}
\sum_\lambda\, \frac1{\langle M_\lambda,M_\lambda\rangle_{q,t}} \
M_\lambda(x;q,t)\, M_\lambda(y;q,t) = \prod_{i,j\geq1}\, \frac{(t\,
  x_i\, y_j;q)_\infty}{(x_i\, y_j;q)_\infty}\ ,
\end{equation}
where the sum runs over all partitions $\lambda.$ By taking the logarithm of this expression and resumming, we can rewrite it in the form of a \emph{generalized Cauchy--Stanley identity}
\begin{equation}\label{eq:cauchystanley-macdonald}
\sum_\lambda\, \frac1{\langle M_\lambda,M_\lambda\rangle_{q,t}} \
M_\lambda(x;q,t)\, M_\lambda(y;q,t) = \exp\Big(\, \sum_{n=1}^\infty\,
\frac1n\, \frac{1-t^n}{1-q^n}\ p_n(x)\, p_n(y)\, \Big) \ .
\end{equation}

\subsection{Uglov functions}\label{sec:uglovfunctions}

Let us consider now the limit
\begin{equation*}
q=\omega\, p \ , \qquad t=\omega\, p^{\beta} \qquad \mbox{with} \quad
p\to1  \ ,
\end{equation*}
where $\omega:=\e^{2\pi\ii/k}$ is a primitive $k$-th root of unity with $k\in\IZ_{>0}$ and $\beta\in\IC.$ The resulting symmetric functions are called the \emph{rank $k$ Uglov functions} or \emph{$\frgl_k$-Jack functions} and are denoted by
\begin{equation*}
U_\lambda(x;\beta,k):= \lim_{p\to1}\, M_\lambda\big(x;\omega\, p,\omega\,
p^\beta\big) \ .
\end{equation*}
They were first introduced in~\cite{art:uglov1998}. 

The Uglov functions of rank $k=1$ are just the usual \emph{monic Jack functions}
\begin{equation*}
J_\lambda(x;\beta )= U_\lambda(x;\beta,1)
\end{equation*}
in $\Lambda\otimes\IQ(\beta).$ Taking the limit $p\to1$ in the inner product $\langle-,-\rangle_{p,p^\beta}$ yields an inner product $\langle-,-\rangle_{\beta}$ on $\Lambda\otimes\IQ(\beta)$ with
\begin{equation*}
\langle p_\lambda,p_\mu\rangle_{\beta} = \delta_{\lambda,\mu}\
z_\lambda\, \beta^{-\ell(\lambda)} \ .
\end{equation*}
The orthogonality relation for the Jack functions thus reads
\begin{equation*}
\langle J_\lambda, J_\mu\rangle_{\beta} = \delta_{\lambda,\mu} \
\prod_{s\in Y_\lambda} \
\frac{A(s)+1 + \beta\,L(s)}{
  A(s)+\beta\,\big(L(s)+1 \big)}
\ .
\end{equation*}
The homogeneity property~\eqref{eq:homogeneity-macdonald} in this case becomes
\begin{equation} \nonumber
J_\lambda(\zeta\, y;\beta) = \zeta^{\vert\lambda\vert}\, J_\lambda(y;\beta) \qquad\mbox{for} \quad \zeta\in\IC \ .
\end{equation}
Moreover, the Macdonald specialisation identities~\eqref{eq:specialization1-macdonald} and \eqref{eq:specialization2-macdonald} in this limit become
\begin{eqnarray}
J_\lambda(1,1,\dots;\beta) &=& \prod_{s\in Y_\lambda}\,
\frac1{A(s)+\beta\, \big(L(s)+1\big)} \ ,
\nonumber\\[4pt]\nonumber J_\lambda(\, \underbrace{1,\dots,1}_{N\, {\rm
    times}}\,;\beta) &=& \prod_{s\in Y_\lambda}\,
\frac{A^t(s)-\beta\,
  \big(L^t(s)-N\big)}{A(s)+\beta\,
  \big(L(s)+1\big)} \ ,
\end{eqnarray}
while the generalised Cauchy--Stanley identities read (cf.~\cite[Prop.~2.1]{art:stanley1989})
\begin{equation} \nonumber
\sum_\lambda\, \frac1{\langle J_\lambda,J_\lambda\rangle_\beta} \
J_\lambda(x;\beta)\, J_\lambda(y;\beta) =
\prod_{i,j\geq1}\, \frac1{(1-x_i\, y_j)^\beta} = \exp\Big(\beta\,
\sum_{n=1}^\infty\, \frac1n\ p_n(x)\, p_n(y)\Big) \ .
\end{equation}

For $\beta=1$, the Jack functions reduce to the \emph{Schur functions} (cf.~\cite[Ch.~I, \S3]{book:macdonald1995})
\begin{equation*}
s_\lambda(x) = J_\lambda(x;1) \ ,
\end{equation*}
which are the characters of $\frgl_\infty$-representations associated to partitions $\lambda$, with the orthogonality relation $\langle
s_\lambda,s_\mu\rangle=\delta_{\lambda,\mu}.$ In this case the limiting inner product $\langle-,-\rangle$ is the usual Hall inner product on $\Lambda$~\cite[Ch.~I, \S4]{book:macdonald1995}. The specialisation formula
\begin{equation*}
s_\lambda(1,1,\dots)= \prod_{s\in Y_\lambda} \,
  \frac1{h(s)} = \dim\lambda
\end{equation*}
is the dimension of the irreducible $\frgl_\infty$-representation with highest weight $\lambda.$

Now we turn to the general case. Since $M_\lambda(x;q,q)=s_\lambda(x)$, for $\beta=1$ we obtain again the expected basis of Schur functions $U_\lambda(s;1,k)=s_\lambda(x)$ for all $k\geq1.$ For generic $\beta$, we have
\begin{proposition}
The rank $k$ Uglov functions $U_\lambda(x;\beta,k)$ have a unitriangular expansion in the basis of Schur functions $s_\lambda(x)$ given by
\begin{equation*}
U_\lambda(x;\beta,k)= s_\lambda(x)+ \sum_{\mu<\lambda}\,
u_{\lambda,\mu}(\beta,k)\, s_\mu(x) \qquad \mbox{with} \quad
u_{\lambda,\mu}(\beta,k)\in\IC \ .
\end{equation*}
\end{proposition}
\Proof{
Both the Schur and Uglov functions are limits of Macdonald functions and so can be expanded in unitriangular form by monomial symmetric functions. On the other hand, the bases $\{m_\lambda\}$ and $\{s_\lambda\}$ of $\Lambda$ are related by a unimodular transformation (cf.\ \eqref{eq:triangular-macdonald} for $t=q$) and so the monomial symmetric functions can be written in the form
\begin{equation*}
m_\lambda(x) = s_\lambda(x)+ \sum_{\mu<\lambda}\, a_{\lambda,\mu}\,
s_\mu(x) \ .
\end{equation*}
With $v_{\lambda,\mu}(\beta,k):= \lim_{p\to1}\, v_{\lambda,\mu}(\omega\, p,\omega\, p^\beta)$, this gives
\begin{eqnarray*}
U_\lambda(x;\beta,k)&=& m_\lambda(x)+ \sum_{\mu<\lambda}\,
v_{\lambda,\mu}(\beta,k)\, m_\mu (x) \nonumber \\[4pt]
&=& s_\lambda(x)+ \sum_{\mu<\lambda}\, a_{\lambda,\mu}\, s_\mu(x) +
\sum_{\mu<\lambda}\, v_{\lambda,\mu}(\beta,k)\,
\Big(s_\mu(x)+\sum_{\nu<\mu}\, a_{\mu,\nu}\, s_\nu(x)\Big) \nonumber
\\[4pt] 
&=& s_\lambda(x)+ \sum_{\mu<\lambda}\, u_{\lambda,\mu}(\beta,k)\,
  s_\mu(x) 
\end{eqnarray*}
with
\begin{equation*}
u_{\lambda,\mu}(\beta,k):=
a_{\lambda,\mu}+v_{\lambda,\mu}(\beta,k)+\sum_{\nu>\mu}\,
v_{\lambda,\nu}(\beta,k)\, a_{\nu,\mu} \ ,
\end{equation*}
as required.
}

The orthogonality relation for the Uglov functions follows from that of the Macdonald functions \eqref{eq:orthogonality-macdonald} and one has
\begin{equation*}
\langle U_\lambda, U_\mu\rangle_{\beta,k} = \delta_{\lambda,\mu} \
\prod_{\stackrel{\scriptstyle s\in
    Y_\lambda}{\scriptstyle h(s)\equiv0\ {\rm mod}\, k}} \
\frac{A(s)+1 + \beta\,L(s)}{
  A(s)+\beta\,\big(L(s)+1 \big)} 
\ , 
\end{equation*}
where the limit of the Macdonald inner product is given by
\begin{equation*}
\langle p_\lambda,p_\mu\rangle_{\beta,k}=\delta_{\lambda,\mu}\,
z_\lambda\, \beta^{-\#\{\lambda_i\equiv0\ {\rm mod}\, k \} } \ .
\end{equation*}
\begin{proposition}
The rank $k$ Uglov functions satisfy the bilinear sum relations
\begin{eqnarray*}
&& \sum_\lambda\, \frac1{\langle U_\lambda,U_\lambda\rangle_{\beta,k}} \
U_\lambda(x;\beta,k)\, U_\lambda(y;\beta,k) \ = \ \prod_{i,j\geq1}\,
\frac1{(1-x_i\, y_j)^{\frac{\beta-1+k}k}} \ \prod_{a=1}^{k-1}\,
\frac1{\big(1-\omega^a\,x_i\, y_j \big)^{\frac{\beta-1}k}} \\[4pt] && \qquad
\qquad \qquad \qquad \ = \ \exp\Big(\, \sum_{a=1}^{k-1} \ \sum_{n\equiv a\ {\rm mod}\, k} \
\frac1n\ p_n(x)\, p_n(y)+ \frac\beta k\, \sum_{n=1}^\infty\, \frac1n
\ p_{n\,k}(x)\, p_{n\,k}(y) \, \Big) \ . 
\end{eqnarray*}
\end{proposition}
\Proof{
For the first equality, we first prove it when $\beta\equiv 1\ {\rm  mod}\, k$ and then analytically continue to arbitrary values. To this end we set $\beta=s\, k+1$ for $s\in\IZ_{>0}.$ Setting $q=\omega\, p$ and $t=\omega\, p^\beta$ in the infinite product appearing in the generalized Cauchy--Binet formula \eqref{eq:cauchybinet-macdonald} for the Macdonald functions and then taking the limit $p\to1$, we obtain
\begin{eqnarray*}
\frac{(t\,z;q)_\infty}{(z;q)_\infty} &=& \prod_{n=0}^\infty\,
\frac{1-t\, z\, q^n}{1-z\, q^n} \\[4pt] &=& \prod_{n=0}^\infty\,
\frac{1-z\, \omega^{n+1}\, p^{n+\beta}}{1-z\, \omega^n\, p^n}
\\[4pt] &=& \prod_{a=0}^{k-1} \ \prod_{m=0}^\infty \,
\frac{1-z\, \omega^{a+1}\, p^{m\,k+a+\beta}}{1-z\, \omega^a\, p^{m\,
    k+a}}  \\[4pt] &=& \prod_{m=0}^\infty\,
\frac{1-p^{(m+s+1)\, k}\, z}{1-p^{m\,k}\, z} \ \prod_{a=1}^{k-1}\,
\frac{1-\omega^a\, p^{(m+s)\,k+a}\, z}{1-\omega^a\, p^{m\,k+a}\, z}
\ = \ \frac1{(1-z)^{s+1}}\, \prod_{a=1}^{k-1}\,
\frac1{\big(1-\omega^a\, z\big)^s} 
\end{eqnarray*}
and the result now follows. For the second equality, we proceed similarly with the argument of the exponential function appearing in the generalized Cauchy--Stanley identity \eqref{eq:cauchystanley-macdonald} for the Macdonald functions to obtain
\begin{eqnarray*}
\sum_{n=1}^\infty\, \frac1n\, \frac{1-t^n}{1-q^n}\ p_n(x)\, p_n(y) &=&
\sum_{n=1}^\infty\, \frac1n\, \frac{1-\omega^n\,
  p^{\beta\,n}}{1-\omega^n\, p^n}\ p_n(x)\, p_n(y) \\[4pt] &=&
\sum_{a=1}^{k} \ \sum_{m=0}^\infty\, \frac1{m\,k+a}\,
\frac{1-\omega^a\, p^{\beta\,(m\,k+a)}}{1- \omega^a\, p^{m\,k+a}}\
p_{m\,k+a}(x)\, p_{m\, k+a}(y) \\[4pt] &=& \sum_{a=1}^{k-1}
\ \sum_{m=0}^\infty\, \frac1{m\, k+a}\ p_{m\,k+a}(x)\, p_{m\,k+a}(y) \\ && +\,
\frac\beta k\, \sum_{m=0}^\infty\, \frac1{m+1}\ p_{(m+1)\,k}(x)\,
p_{(m+1)\, k}(y)
\end{eqnarray*}
and the result follows.
}

\bigskip

\section{$\cW$-algebras\label{app:Walgebras}}

Let $\frg$ be a finite-dimensional simply-laced Lie algebra of rank
$N$, and denote its Weyl group by $\sfW(\frg)$. The
\emph{affine Lie algebra} $\widehat{\frg}$
associated to $\frg$ is defined as a one-dimensional central
extension
\bea
0 \ \longrightarrow \ \IC\, c \ \longrightarrow \ \widehat{\frg} \
\longrightarrow \ \frg\otimes\IC\big[t,t^{-1}\big] \ \longrightarrow \ 0
\ ,
\nonumber \eea
where $c$ is the \emph{central charge} and $\frg\otimes\IC\big[t,t^{-1}\big]$ is the associated infinite-dimensional \emph{loop
  algebra}.

Let $e$ be a nilpotent element in the Lie algebra $\frg$. By the
Jacobson--Morozov theorem, it can be completed to an \emph{$\frsl_2$-triple} $(e,f,h)$, i.e. there exist elements
$f,h\in\frg$ with the Lie brackets
\bea
[h,e]=2e \ , \qquad [h,f]=-2f \qquad \mbox{and} \qquad [e,f]=h \ .
\nonumber \eea
The $\frsl_2$-triples correspond to embeddings of the Lie algebra $\frsu_2$ into $\frg$. For example, nilpotent elements $e$ of $\frg=\frgl_N$ are classified (up to conjugation) by their traceless Jordan normal form, which corresponds to a partition $\lambda=(\lambda_1,\dots,\lambda_\ell)=(1^{m_1}\ 2^{m_2}\ \ldots)$ of $N$ where $m_i$ is the number of regular $i\times i$ nilpotent block matrices
\bea
\begin{pmatrix}
   0&1&0&\cdots&0\\0&0&1&\ddots&\vdots\\
   \vdots&\vdots&\ddots&\ddots&0\\0&0& \cdots &0&1 \\
   0&0& \cdots &0&0 \end{pmatrix} 
\label{eq:renilpotent} \eea
in the Jordan form; alternatively, $m_i$ is the multiplicity of the
irreducible $i$-dimensional representation in the decomposition of a
representation of $\frsu_2$ on $\IC^N$. The two extreme cases are the
trivial embedding $\frsu_2\to0\in\frg$, where $e=0$ corresponds to the
maximal partition $\lambda=\big(1^N \big)$ of length $N$, and the
principal embedding $\frsu_2\subset\frg$, where $e=e_{\rm pr}$ is the
regular nilpotent matrix \eqref{eq:renilpotent} in the $N$-dimensional
irreducible representation of $\frsu_2$ corresponding to the trivial
partition $\lambda=(N)$ of length~one.

Given a nilpotent element $e\in\frg$, one constructs from the affine algebra $\widehat{\frg}$ a vertex operator algebra $\cW(\frg,e)$ by a method called \emph{Drinfeld--Sokolov reduction}. The two limiting extremes
\bea
\cW(\frg,e=0) = \widehat{\frg} \qquad \mbox{and} \qquad \cW(\frg,e_{\rm pr}) =: \cW({\frg})
\nonumber \eea
are of particular interest. The $\cW({\frg})$-algebra has Virasoro quasi-primary fields
\bea
W_{d_a}(z) = \sum_{n\in\IZ}\, \frac{W_{d_a,n}}{z^{d_a+n}} \ \dd z^{d_a}
\nonumber \eea
of dimension $d_a$ for $a=1,\dots,N$, where $z\in\IC$, $W_{d_a,n}$ are
certain operators and $d_a-1$ is the \emph{$a$-th exponent} of the Lie
algebra $\frg$ which satisfies the dimension formulas
\bea
\#\sfW(\frg)= \prod_{a=1}^N\, d_a \qquad \mbox{and} \qquad \dim \frg = \sum_{a=1}^N\, (2d_a-1) \ .
\nonumber \eea
Note that $d_a=a$ for $\frg=\frgl_N$.

The $\cW({\frg} )$-algebras are generalizations of the
Heisenberg and Virasoro algebras: $\cW({\frgl}_1)=\widehat{\frgl}_1$ is the Heisenberg algebra $\frh$ which is defined by generators $\alpha_m$, $m\in\IZ\setminus\{0\}$ with the Heisenberg commutation relations
\bea
[\alpha_m,\alpha_n] = m\, \delta_{m+n,0}\ c \ .
\nonumber \eea
The $\cW({\frsl}_2)$-algebra is the Virasoro algebra which is defined by generators $L_m$, $m\in\IZ$ with the relations
\bea
[L_m,L_n] = (m-n)\, L_{m+n}+\mbox{$\frac m{12}$}\, \big(m^2-1\big)\, \delta_{m+n,0}\ c \ .
\nonumber \eea
In the general case, $\cW({\frg} )$ is a vertex algebra which does not admit a presentation in terms of generators and relations.

There is a surjective functor from the category of irreducible highest weight
representations $\Line_\lambda$ of the affine Lie algebra
$\widehat{\frg}$, which are parameterized by highest weights $\lambda$
in the Cartan subalgebra of $\frg$, to the category of irreducible
highest weight representations $\cW_\lambda$ of $\cW(\frg,e)$, which are called \emph{Verma modules}. In particular, this functor sends the vacuum representation $\Line_0$ to the vacuum representation~$\cW_0$.

Let $h$ denote the dual Coxeter number of $\frg$ ($h=N$ for $\frg=\frgl_N$).
\begin{definition}
A vector $\psi\in\cW_\lambda$ is called a \emph{Whittaker vector} if
\bea
W_{d_a,n}\psi=0 = W_{h,m}\psi \qquad \mbox{and} \qquad W_{h,1}\psi=\psi
\nonumber \eea
for all $d_a\neq h$, $a=1,\dots,N$, $n\geq1$ and $m>1$.
\end{definition}


\bigskip

\begin{thebibliography}{99}

\bibitem{Alba:2010qc}
  {\sc V.~A.~Alba, V.~A.~Fateev, A.~V.~Litvinov, G.~M.~Tarnopolsky}:
  \emph{On combinatorial expansion of the conformal blocks arising from AGT conjecture},
  Lett.\ Math.\ Phys.\ {98} (2011), 33--64.
  
\bibitem{art:aldaytachikawa2010}
{\sc L.~F. Alday, Y.~Tachikawa}: {\em Affine {$SL(2)$} conformal blocks from
  $4D$ gauge theories}, Lett. Math. Phys. 94 (2010), 87--114.

\bibitem{art:aldaygaiottotachikawa2010}
{\sc L.~F. Alday, D.~Gaiotto, Y.~Tachikawa}: {\em Liouville correlation
  functions from four-dimensional gauge theories}, Lett. Math. Phys. 91
  (2010), 167--197.
  
\bibitem{Alfimov:2011ju}
  {\sc M.~N.~Alfimov, G.~M.~Tarnopolsky}:
  \emph{Parafermionic Liouville field theory and instantons on ALE spaces},
  J. High Energy Phys. {1202} (2012), 036.
  
\bibitem{Alfimov:2013cqa}
  {\sc M.~N.~Alfimov, A.~A.~Belavin, G.~M.~Tarnopolsky}:
  \emph{Coset conformal field theory and instanton counting on $\IC^{2}/\IZ_{p}$},
  J. High Energy Phys. {1308} (2013), 134.

\bibitem{Andrews}
  {\sc G.~E.~Andrews}: 
  {\sl The Theory of Partitions} (Cambridge University Press, 1998).

\bibitem{Awata:2009dd}
 {\sc H.~Awata, H.~Kanno}:
  \emph{Quiver matrix model and topological partition function in six dimensions},
  J. High Energy Phys. {0907} (2009), 076.
  
\bibitem{Awata:2011dc}
 {\sc H.~Awata, B.~L.~Feigin, A.~Hoshino, M.~Kanai, J.~Shiraishi, S.~Yanagida}:
  \emph{Notes on Ding--Iohara algebra and AGT conjecture},
  RIMS Kokyuroku 1765 (2011), 12--32.
  
\bibitem{art:belavinbelavinbershtein2011}
{\sc A.~A.~Belavin, V.~A.~Belavin, M.~A.~Bershtein}: {\em {Instantons and $2D$
  superconformal field theory}}, J. High Energy Phys. 1109 (2011),
117.

\bibitem{art:belavinbershteintarnopolsky2013}
{\sc A.~A.~Belavin, M.~A.~Bershtein, G.~M.~Tarnopolsky}: {\em {Bases in coset
  conformal field theory from AGT correspondence and Macdonald polynomials at
  the roots of unity}}, J. High Energy Phys. 1303 (2013), 019.

\bibitem{Belavin:2011sw}
 {\sc A.~A.~Belavin, M.~A.~Bershtein, B.~L.~Feigin, A.~V.~Litvinov, G.~M.~Tarnopolsky}:
  \emph{Instanton moduli spaces and bases in coset conformal field theory},
  Commun.\ Math.\ Phys.\  {319} (2013), 269--301.
  
\bibitem{art:belavinfeigin2011}
{\sc V.~A.~Belavin, B.~L.~Feigin}: {\em {Super Liouville conformal blocks from
  $\cN=2$ $SU(2)$ quiver gauge theories}}, J. High Energy Phys. 1107 (2011),
  079.
  
\bibitem{BK}
{\sc G.~Bellamy, T.~Kuwabara}:
\emph{On deformation quantizations of hypertoric varieties},
Pacific J. Math 260 (2012), 89--127.

\bibitem{Bershtein:2014yia}
  {\sc M.~A.~Bershtein, A.~I.~Shchechkin}:
  \emph{Bilinear equations on Painlev\'e $\tau$-functions from CFT},
  Preprint arXiv:1406.3008 [math-ph].
  
\bibitem{Bershtein:2013oka}
  {\sc M.~A.~Bershtein, B.~L.~Feigin, A.~V.~Litvinov}:
  \emph{Coupling of two conformal field theories and Nakajima--Yoshioka blow-up equations},
  Preprint arXiv:1310.7281 [math.QA].
  
\bibitem{Bonelli:2007hb}
  {\sc G.~Bonelli, A.~Tanzini}:
  \emph{Topological gauge theories on local spaces and black hole entropy countings},
  Adv.\ Theor.\ Math.\ Phys.\ {12} (2008), 1429--1446.

\bibitem{art:bonellimaruyoshitanzini2011}
{\sc G.~Bonelli, K.~Maruyoshi, A.~Tanzini}: {\em {Instantons on ALE spaces
  and super Liouville conformal field theories}}, J. High Energy Phys. 1108
  (2011), 056.

\bibitem{art:bonellimaruyoshitanzini2012}
{\sc G.~Bonelli, K.~Maruyoshi, A.~Tanzini}: {\em {Gauge theories on
  ALE space and super Liouville correlation functions}}, Lett. Math. Phys. 101
  (2012), 103--124.

\bibitem{art:bonellimaruyoshitanziniyagi2012}
{\sc G.~Bonelli, K.~Maruyoshi, A.~Tanzini, F.~Yagi}: {\em {$\cN=2$ gauge
  theories on toric singularities, blow-up formulae and $\cW$-algebrae}}, J. High
  Energy Phys. 1301 (2013), 014.

\bibitem{BCS}
{\sc L.~Borisov, L.~Chen, G.~Smith}:
\emph{The orbifold Chow ring of a toric Deligne--Mumford stack},
J. Amer. Math. Soc. 18 (2005), 193--215.

\bibitem{Bouwknegt:1992wg}
  {\sc P.~Bouwknegt, K.~Schoutens}:
  \emph{$\cW$-symmetry in conformal field theory},
  Phys.\ Rept.\ {223} (1993), 183--276.

\bibitem{Boyarchenko}
{\sc M.~Boyarchenko}:
\emph{Quantization of minimal resolutions of Kleinian singularities},
Adv. Math. 211 (2007), 244--265.

\bibitem{BruzzoSala13}
{\sc U. Bruzzo, F. Sala}: \emph{Framed sheaves on projective stacks},
Adv. Math. 272 (2015), 20--95.

\bibitem{Bruzzo:2014jza}
  {\sc U.~Bruzzo, F.~Sala, R.~J.~Szabo}:
  \emph{${\mathcal{N} = 2}$ quiver gauge theories on A-type ALE spaces},
  Lett.\ Math.\ Phys.\ {105} (2015), 401--445.
  
\bibitem{Bruzzo:2002xf}
  {\sc U.~Bruzzo, F.~Fucito, J.~F.~Morales, A.~Tanzini}:
  \emph{Multi-instanton calculus and equivariant cohomology},
  J. High Energy Phys. {0305} (2003), 054.
  
\bibitem{Bruzzo:2013daa}
  {\sc U.~Bruzzo, M.~Pedrini, F.~Sala, R.~J.~Szabo}:
  \emph{Framed sheaves on root stacks and supersymmetric gauge theories on ALE spaces},
  Preprint arXiv:1312.5554 [math.AG].
  
\bibitem{art:carlssonokounkov2012}
{\sc E.~{Carlsson}, A.~{Okounkov}}: {\em {Exts and vertex operators}}, Duke
  Math. J. 161 (2012), 1797--1815.
  
\bibitem{Carlsson:2013jka}
  {\sc E.~Carlsson, N.~A.~Nekrasov, A.~Okounkov}:
  \emph{Five-dimensional gauge theories and vertex operators},
  Moscow Math. J. 14 (2014), 39--61.

\bibitem{art:ciraficiszabo2012}
{\sc M.~Cirafici, R.~J. Szabo}: {\em {Curve counting, instantons and McKay
  correspondences}}, J. Geom. Phys. 72 (2013), 54--109.

\bibitem{CA-KS}
  {\sc M.~Cirafici, A.-K.~Kashani-Poor, R.~J.~Szabo}:
  \emph{Crystal melting on toric surfaces},
  J. Geom. Phys. {61} (2011), 2199--2218.

\bibitem{Cirafici:2008sn}
  {\sc M.~Cirafici, A.~Sinkovics, R.~J.~Szabo}:
  \emph{Cohomological gauge theory, quiver matrix models and Donaldson--Thomas
  theory},
  Nucl.\ Phys.\ B {809} (2009), 452--518.

\bibitem{Cirafici:2010bd}
  {\sc M.~Cirafici, A.~Sinkovics, R.~J.~Szabo}:
  \emph{Instantons, quivers and noncommutative Donaldson--Thomas theory},
  Nucl.\ Phys.\  B {853} (2011), 508--605.

\bibitem{CLSI}
  {\sc L.~S.~Cirio, G.~Landi, R.~J.~Szabo}:
  \emph{Algebraic deformations of toric varieties I. General constructions}, Adv. Math. 246 (2013), 33--88.

\bibitem{CLSII}
  {\sc L.~S.~Cirio, G.~Landi, R.~J.~Szabo}:
  \emph{Algebraic deformations of toric varieties II. Noncommutative instantons}, Adv. Theor. Math. Phys. 15 (2011), 1817--1907.

\bibitem{CLSIII}
  {\sc L.~S.~Cirio, G.~Landi, R.~J.~Szabo}:
  \emph{Instantons and vortices on noncommutative toric varieties},
  Rev.\ Math.\ Phys.\ {26} (2014), 1430008.

\bibitem{Cordes:1994fc}
  {\sc S.~Cordes, G.~W.~Moore, S.~Ramgoolam}:
  \emph{Lectures on $2D$ Yang--Mills theory, equivariant cohomology and topological field theories},
  Nucl.\ Phys.\ Proc.\ Suppl.\  {41} (1995), 184--244.
  
\bibitem{Dijkgraaf:2007fe}
  {\sc R.~Dijkgraaf, P.~Sulkowski}:
  \emph{Instantons on ALE spaces and orbifold partitions},
  J. High Energy Phys. {0803} (2008), 013.
  
\bibitem{Dijkgraaf:2007sw}
  {\sc R.~Dijkgraaf, L.~Hollands, P.~Sulkowski, C.~Vafa}:
  \emph{Supersymmetric gauge theories, intersecting branes and free fermions},
  J. High Energy Phys. {0802} (2008), 106.
  
\bibitem{Donagi:1995cf}
  {\sc R.~Donagi, E.~Witten}:
  \emph{Supersymmetric Yang--Mills theory and integrable systems},
  Nucl.\ Phys.\ B {460} (1996), 299--334.

\bibitem{Donaldson84}
{\sc S. K. Donaldson}:
\emph{Instantons and geometric invariant theory},
Commun. Math. Phys. 93 (1984), 453--460.

\bibitem{Douglas:1995bn}
  {\sc M.~R.~Douglas}:
  \emph{Branes within branes},
  in: {\sl Strings, Branes and Dualities}, L.~Baulieu, P.~Di~Francesco, M.~R.~Douglas,
  V.~A.~Kazakov, M.~Picco, P.~Windey, eds. 
  (Dordrecht, 1997), 267--275.

\bibitem{EyssidieuxSala14}
{\sc P. Eyssidieux, F. Sala}: \emph{Instantons and framed sheaves on K\"ahler Deligne--Mumford stacks}, Preprint arXiv:1404.3504 [math.AG].

\bibitem{Fateev:2009aw}
  {\sc V.~A.~Fateev, A.~V.~Litvinov}:
  \emph{On AGT conjecture},
  J. High Energy Phys. {1002} (2010), 014.
  
\bibitem{Feigin}
{\sc B.~L.~Feigin, A.~I.~Tsymbaliuk}:
\emph{Equivariant K-theory of Hilbert schemes via shuffle algebra},
Kyoto J. Math. 51 (2011), 831--854.

\bibitem{Flume:2002az}
  {\sc R.~Flume, R.~Poghossian}:
  \emph{An algorithm for the microscopic evaluation of the coefficients of the Seiberg--Witten prepotential},
  Int.\ J.\ Mod.\ Phys.\ A {18} (2003), 2541--2563.
  
\bibitem{art:fucitomoralespoghossian2004}
{\sc F.~Fucito, J.~F. Morales, R.~Poghossian}: {\em {Multi-instanton
  calculus on ALE spaces}}, Nucl. Phys. B 703 (2004), 518--536.
  
\bibitem{art:fujii2005}
{\sc S.~Fujii, S.~Minabe}:
\emph{A combinatorial study on quiver varieties},
Preprint arXiv:math.AG/0510455.

\bibitem{Gaiotto:2009we}
  {\sc D.~Gaiotto}:
  \emph{$\cN=2$ dualities},
  J. High Energy Phys. {1208} (2012), 034.
  
\bibitem{art:gaiotto2009}
{\sc D.~{Gaiotto}}: {\em {Asymptotically free $\cN=2$ theories and irregular
  conformal blocks}}, J. Phys. Conf. Ser. 462 (2013), 012014.
  
\bibitem{Gaiotto:2009hg}
  {\sc D.~Gaiotto, G.~W.~Moore, A.~Neitzke}:
  \emph{Wall-crossing, Hitchin systems, and the WKB approximation},
  Adv. Math. 234 (2013), 239--403.

\bibitem{GL1}
  {\sc E.~Gasparim, C.-C.~M.~Liu}:
  \emph{The Nekrasov conjecture for toric surfaces},
  Commun.\ Math.\ Phys.\ {293} (2010), 661--700.

\bibitem{GS1}
{\sc I.~Gordon, J.~T.~Stafford}:
\emph{Rational Cherednik algebras and Hilbert schemes}, 
Adv. Math. 198 (2005), 222--274.

\bibitem{GS2}
{\sc I.~Gordon, J.~T.~Stafford}:
\emph{Rational Cherednik algebras and Hilbert schemes
  II. Representations and sheaves}, 
Duke Math. J. 132 (2006), 73--135.

\bibitem{GSST}
  {\sc L.~Griguolo, D.~Seminara, R.~J.~Szabo, A.~Tanzini}:
  \emph{Black holes, instanton counting on toric singularities and $q$-deformed two-dimensional Yang--Mills theory},
  Nucl.\ Phys.\ B {772} (2007), 1--24.

\bibitem{Hadasz:2010xp}
  {\sc L.~Hadasz, Z.~Jaskolski, P.~Suchanek}:
  \emph{Proving the AGT relation for $N_f = 0,1,2$ antifundamentals},
  J. High Energy Phys. {1006} (2010), 046.
  
\bibitem{Hamanaka}
  {\sc M.~Hamanaka, Y.~Imaizumi, N.~Ohta}:
  \emph{Moduli space and scattering of D0-branes in noncommutative super Yang--Mills theory},
  Phys.\ Lett.\ B {529} (2002), 163--170.
  
\bibitem{Hanany:2003hp}
  {\sc A.~Hanany, D.~Tong}:
  \emph{Vortices, instantons and branes},
  J. High Energy Phys. {0307} (2003), 037.

\bibitem{HW}
{\sc P.-M.~Ho, Y.-S.~Wu}:
\emph{Noncommutative gauge theories in Matrix theory},
Phys. Rev. D 58 (1998), 066003.

\bibitem{Iqbal:2003ds}
  {\sc A.~Iqbal, N.~A.~Nekrasov, A.~Okounkov, C.~Vafa}:
  \emph{Quantum foam and topological strings},
  J. High Energy Phys. {0804} (2008), 011.
  
\bibitem{Ito:2011mw}
  {\sc Y.~Ito}:
  \emph{Ramond sector of super Liouville theory from instantons on an ALE space},
  Nucl.\ Phys.\ B {861} (2012), 387--402.

\bibitem{art:itomaruyoshiokuda2013}
{\sc Y.~Ito, K.~Maruyoshi, T.~Okuda}: {\em Scheme dependence of instanton
  counting in ALE spaces}, J. High Energy Phys. 1305 (2013), 045.
  
\bibitem{Itoyama:2015xia}
  {\sc H.~Itoyama, R.~Yoshioka},
  \emph{Developments of theory of effective prepotential from extended Seiberg--Witten system and matrix models},
  Preprint arXiv:1507.00260 [hep-th].
  
\bibitem{Itoyama:2013mca}
  {\sc H.~Itoyama, T.~Oota, R.~Yoshioka}:
  \emph{2d/4d connection between $q$-Virasoro/$\cW$ block at root of unity limit and instanton partition function on ALE space},
  Nucl.\ Phys.\ B {877} (2013), 506--537.

\bibitem{Kapustin:2000ek}
  {\sc A.~Kapustin, A.~Kuznetsov, D.~Orlov}:
  \emph{Noncommutative instantons and twistor transform},
  Commun.\ Math.\ Phys.\  {221} (2001), 385--432.

\bibitem{KR}
{\sc M.~Kashiwara, R.~Rouquier}:
\emph{Microlocalization of rational Cherednik algebras},
Duke Math. J. 144 (2008), 525--573.

\bibitem{Kimura:2011zf}
  {\sc T.~Kimura}:
  \emph{Matrix model from $\cN = 2$ orbifold partition function},
  J. High Energy Phys. {1109} (2011), 015.

\bibitem{art:kronheimer1989}
{\sc P.~B. Kronheimer}: {\em The construction of {ALE} spaces as
  hyper-{K}\"ahler quotients}, J. Diff. Geom. 29 (1989), 665--683.

\bibitem{KN}
{\sc P.~B.~Kronheimer, H.~Nakajima}:
\emph{Yang--Mills instantons on ALE gravitational instantons},
Math. Ann. 288 (1990), 263--307.

\bibitem{art:kuznetsov2007}
{\sc A.~Kuznetsov}: {\em Quiver varieties and {H}ilbert schemes}, Moscow Math.
  J. 7 (2007), 673--697.

\bibitem{Lazaroiu}
{\sc C.~I.~Lazaroiu}:
\emph{A noncommutative geometric interpretation of the resolution of
  equivariant instanton moduli spaces},
Preprint arXiv:hep-th/9805132.

\bibitem{book:macdonald1995}
{\sc I.~G. Macdonald}:
{\sl Symmetric Functions and {H}all Polynomials} (Oxford University
  Press, 1995).

\bibitem{MM}
{\sc E.~J.~Martinec, G.~W.~Moore}:
\emph{Noncommutative solitons on orbifolds},
Preprint arXiv:hep-th/0101199.

\bibitem{Martinec:1995by}
  {\sc E.~J.~Martinec, N.~P.~Warner}:
  \emph{Integrable systems and supersymmetric gauge theory},
  Nucl.\ Phys.\ B {459} (1996), 97--112.

\bibitem{art:maulikokounkov2012}
{\sc D.~{Maulik}, A.~{Okounkov}}: {\em {Quantum groups and quantum
  cohomology}}, Preprint arXiv:1211.1287 [math.AG].

\bibitem{MNOPII}
    {\sc D.~Maulik, N.~A.~Nekrasov, A.~Okounkov, R.~Pandharipande}:
    \emph{Gromov--Witten theory and Donaldson--Thomas theory II},
    Compos. Math. {142} (2006), 1286--1304.
    
\bibitem{Morozov:2013rma}
  {\sc A.~Morozov, A.~Smirnov}:
  \emph{Towards the proof of AGT relations with the help of the generalized Jack polynomials},
  Lett.\ Math.\ Phys.\ {104} (2014), 585--612.

\bibitem{Musson}
{\sc I.~M.~Musson}:
\emph{Noncommutative deformations of type $A$ Kleinian singularities and
  Hilbert schemes},
J. Algebra 293 (2005), 102--129.

\bibitem{art:nagao2009}
{\sc K.~Nagao}:
\emph{Quiver varieties and {F}renkel--{K}ac construction},
J. Algebra 321 (2009), 3764--3789.

\bibitem{Nakajima1}
{\sc H.~Nakajima}:
\emph{Instantons on ALE spaces, quiver varieties and Kac--Moody
  algebras},
Duke Math. J. 76 (1994), 365--416.

\bibitem{art:nakajima1997}
{\sc H.~Nakajima}: {\em Heisenberg algebra
  and {H}ilbert schemes of points on projective surfaces}, Ann. Math. 145
  (1997), 379--388.

\bibitem{book:nakajima1999}
{\sc H.~Nakajima}: {\sl Lectures on
  {H}ilbert Schemes of Points on Surfaces} (American Mathematical Society, 1999).

\bibitem{nakajima2}
  {\sc H.~Nakajima, K.~Yoshioka}:
  \emph{Lectures on instanton counting},
  CRM Proc. Lect. Notes {38} (2004), 31--102.

\bibitem{NakYosh}
{\sc H.~Nakajima, K.~Yoshioka}:
\emph{Instanton counting on blowup I. $4$-dimensional pure gauge
  theory},
Invent. Math. {162} (2005), 313--355.

\bibitem{art:nekrasov2003}
{\sc N.~A. Nekrasov}:
{\em Seiberg--{W}itten prepotential from instanton
  counting}, Adv. Theor. Math. Phys. 7 (2003), 831--864.

\bibitem{NekJapan}
{\sc N.~A.~Nekrasov}:
\emph{Instanton partition functions and M-theory},
Japan. J. Math. {4} (2009), 63--93.

\bibitem{art:nekrasovokounkov2006}
{\sc N.~A. Nekrasov, A.~Okounkov}: {\em Seiberg--{W}itten theory and random
  partitions}, Progr. Math. 244 (2006), 525--596.
  
\bibitem{Nekrasov:2014nea}
  {\sc N.~A.~Nekrasov, A.~Okounkov}:
  \emph{Membranes and sheaves},
  Preprint arXiv:1404.2323 [math.AG].

\bibitem{NS}
{\sc N.~A.~Nekrasov, A.~S.~Schwarz}:
{\it Instantons on noncommutative $\real^4$ and $(2,0)$ superconformal six-dimensional theory},
Commun. Math. Phys. 198 (1998), 689--703.

\bibitem{Nekrasov:2009rc}
  {\sc N.~A.~Nekrasov, S.~L.~Shatashvili}:
  \emph{Quantization of integrable systems and four-dimensional gauge theories},
  in: {\sl 16th International Congress on Mathematical Physics},
  P. Exner, ed. (World Scientific, 2010), 265--289.

\bibitem{Nishinaka:2013mba}
  {\sc T.~Nishinaka, S.~Yamaguchi, Y.~Yoshida}:
  \emph{Two-dimensional crystal melting and D4--D2--D0 on toric Calabi--Yau singularities},
  J. High Energy Phys. {1405} (2014), 139.
  
\bibitem{art:nishiokatachikawa2011}
{\sc T.~Nishioka, Y.~Tachikawa}: {\em {Central charges of para-Liouville and
  Toda theories from M5-branes}}, Phys. Rev. D 84 (2011), 046009.
  
\bibitem{Pedrini:2014yoa}
  {\sc M.~Pedrini, F.~Sala, R.~J.~Szabo}:
  \emph{AGT relations for abelian quiver gauge theories on ALE spaces},
  Preprint arXiv:1405.6992 [math.RT].
  
\bibitem{SVHall}
{\sc O.~{Schiffmann}, E.~{Vasserot}}:
\emph{The elliptic Hall algebra and the K-theory of the Hilbert scheme of $\mathbb{A}^2$},
Duke Math. J. 162 (2013), 279--366.
  
\bibitem{art:schiffmannvasserot2013}
{\sc O.~{Schiffmann}, E.~{Vasserot}}: {\em {Cherednik algebras, $\cW$-algebras
  and the equivariant cohomology of the moduli space of instantons on
  $\mathbb{A}^2$}}, Publ. Math. IH\'{E}S 118 (2013), 213--342.
  
\bibitem{art:seibergwitten1994-I}
{\sc N.~Seiberg, E.~Witten}: {\em Electric-magnetic duality, monopole
  condensation, and confinement in {$\cN=2$} supersymmetric {Y}ang--{M}ills
  theory}, Nucl. Phys. B 426 (1994), 19--52.
  
\bibitem{Smirnov:2013hh}
  {\sc A.~Smirnov}:
  \emph{On the instanton $R$-matrix},
  Preprint arXiv:1302.0799 [math.AG].
  
\bibitem{Smirnov:2014npa}
  {\sc A.~Smirnov}:
  \emph{Polynomials associated with fixed points on the instanton moduli space},
  Preprint arXiv:1404.5304 [math-ph].
  
\bibitem{Smith} 
{\sc S. P. Smith}: \emph{Twisted homogeneous coordinate rings and
three superpotential algebras}, Talk at the Simons Center for Geometry and Physics, December 2011. Available at {\tt http://media.scgp.stonybrook.edu/video/video.php?f=20111213\_6\_qtp.mp4}

\bibitem{Spodyneiko:2014qsa}
  {\sc L.~Spodyneiko}:
  \emph{AGT correspondence, Ding--Iohara algebra at roots of unity and Lepowsky--Wilson construction},
  J.\ Phys.\ A {48} (2015), 275404.

\bibitem{art:stanley1989}
{\sc R.~P. Stanley}:
\emph{Some combinatorial properties of {J}ack symmetric functions},
Adv. Math. 77 (1989), 76--115.

\bibitem{Strominger:1995ac}
  {\sc A.~Strominger}:
  \emph{Open $p$-branes},
  Phys.\ Lett.\ B {383} (1996), 44--47.

\bibitem{Szabo:2009vw}
  {\sc R.~J.~Szabo}:
  \emph{Instantons, topological strings and enumerative geometry},
  Adv.\ Math.\ Phys.\  {2010} (2010), 107857.

\bibitem{Szabo:2011mj}
  {\sc R.~J.~Szabo}:
  \emph{Crystals, instantons and quantum toric geometry},
  Acta Phys.\ Polon.\ Suppl.\ {4} (2011), 461--494.
  
\bibitem{Tachikawa:2014dja}
  {\sc Y.~Tachikawa}:
  \emph{A review on instanton counting and $\cW$-algebras},
  Preprint arXiv:1412.7121 [hep-th].
  
\bibitem{Tan:2013tq}
  {\sc M.-C.~Tan}:
  \emph{M-theoretic derivations of 4d/2d dualities: From a geometric Langlands duality for surfaces, to the AGT correspondence, to integrable systems},
  J. High Energy Phys. {1307} (2013), 171.

\bibitem{art:uglov1998}
{\sc D.~Uglov}:
\emph{Yangian Gelfand--Zetlin bases, $\frgl_N$-Jack polynomials and
  computation of dynamical correlation functions in the spin
  Calogero--Sutherland model},
Commun. Math. Phys.
 191 (1998), 663--696.
 
\bibitem{art:vafawitten1994}
{\sc C.~Vafa, E.~Witten}:
 \emph {A strong coupling test of S-duality}, Nucl.
  Phys. B 431 (1994), 3--77.
  
\bibitem{art:vasserot2001}
{\sc E.~Vasserot}: {\em Sur l'anneau de cohomologie du sch\'ema de {H}ilbert de
  {$\mathbb{C}^2$}}, C. R. Acad. Sci. Paris S\'er. I Math. 332 (2001),
  7--12.
  
\bibitem{Witten:1988ze}
  {\sc E.~Witten}:
  \emph{Topological quantum field theory},
  Commun.\ Math.\ Phys.\ {117} (1988), 353--386.
  
\bibitem{Witten:1995zh}
  {\sc E.~Witten}:
  \emph{Some comments on string dynamics},
  in: {\sl Future Perspectives in String Theory}, I.~Bars, P.~Bouwknegt, J.~A.~Minahan, D.~Nemeschansky, K.~Pilch, H.~Saleur, N.~P.~Warner, eds. (World Scientific, 1996), 501--523.
  
\bibitem{Witten:2000mf}
  {\sc E.~Witten}:
  \emph{BPS bound states of D0--D6 and D0--D8 systems in a $B$-field},
  J. High Energy
Phys. {0204} (2002), 012.
 
\bibitem{Witten:2009at}
  {\sc E.~Witten}:
  \emph{Geometric Langlands from six dimensions},
  Preprint arXiv:0905.2720 [hep-th].
  
 \bibitem{art:wyllard2009}
{\sc N.~Wyllard}: {\em {$A_{N-1}$}
  conformal {T}oda field theory correlation functions from conformal
  {$\mathcal{N}=2$} {$SU(N)$} quiver gauge theories}, J. High Energy
Phys. 0911
  (2009), 002.
  
\bibitem{art:wyllard2011}
{\sc N.~Wyllard}: {\em Coset conformal blocks and $\cN=2$ gauge theories},
Preprint arXiv:1109.4264 [hep-th].

\bibitem{Yang}
{\sc C.~Yang}:
\emph{Isospectral deformations of Eguchi--Hanson spaces as nonunital
  spectral triples},
Commun. Math. Phys. 288 (2009), 615--652.

\end{thebibliography}
\end{document}